\documentclass[10pt,final,twocolumn]{IEEEtran}
\usepackage{cite}
\usepackage{float}
\usepackage[cmex10]{amsmath}
\usepackage{amssymb}
\usepackage{amsfonts}
\usepackage{psfrag}
\usepackage{subfigure}
\usepackage{epsfig}
\usepackage{algorithm}
\usepackage{algorithmic}

\usepackage[normalem]{ulem}
\usepackage[usenames,dvipsnames]{color}
\newcommand{\rd}{\textcolor{Red}}
\newcommand{\bl}{\textcolor{Blue}}

\begin{document}

\if 0
\title{Practical Limitations of Joint Reconstruction 
of Absorbed Optical Energy Density and Speed 
of Sound Distributions in Photoacoustic 
Computed Tomography 
}
\fi

\title{Joint Reconstruction 
%of Absorbed Optical Energy Density and Speed
%of Sound Distributions in Photoacoustic
of Absorbed Optical Energy Density and Sound Speed Distributions in Photoacoustic
Computed Tomography:\\ A Numerical Investigation}

\author{Chao~Huang, Kun~Wang,~\IEEEmembership{Member,~IEEE,} Robert W. Schoonover, Lihong~V.~Wang,~\IEEEmembership{Fellow,~IEEE,}
 and~Mark~A.~Anastasio,~\IEEEmembership{Senior Member,~IEEE}
\thanks{
Department of Biomedical Engineering, Washington University in St. Louis, 
St. Louis, MO 63130
}}
%\date{\today}
\maketitle

% $\,$ \vspace{-2cm}
\begin{abstract}
Photoacoustic computed tomography (PACT) is a rapidly emerging bioimaging modality that seeks to reconstruct an estimate of the
absorbed optical energy density within an object.  
Conventional PACT image reconstruction methods assume a constant
speed-of-sound (SOS), which can result in image artifacts when
acoustic aberrations are significant.
It has been demonstrated that incorporating knowledge 
of an object's SOS distribution into a PACT image
 reconstruction method can improve image quality.
 However, in many cases, the SOS distribution cannot be accurately
and/or conveniently estimated prior to the PACT experiment.
Because variations in the  SOS distribution
induce aberrations in the measured photoacoustic wavefields,
certain information regarding an object's SOS distribution
is encoded in the PACT measurement data.
Based on this observation,  a joint reconstruction (JR) problem
has been proposed in which
 the SOS distribution is concurrently estimated along with the sought-after
absorbed optical energy density from the photoacoustic measurement data.
A broad understanding of the extent to which the JR problem 
  can be accurately and reliably solved has not been reported.
 In this work, a series of numerical experiments is described that elucidate some
important properties of the JR problem that pertain to its practical feasibility.
To accomplish this, an optimization-based formulation of the JR
problem is developed that yields a non-linear iterative algorithm that 
alternatingly
 updates the two image estimates.
Heuristic analytic insights into the reconstruction problem are also provided.
These results confirm the ill-conditioned nature of the joint reconstruction problem
that will present significant challenges for practical applications.
%We also give a heuristic necessary condition for the accurate reconstruction of c(r).
% Computer simulations are employed to assess the accuracy and robustness of the alternating optimization method, as well as verify the heuristic argument.
% The numerical results show that, in some cases, it is possible to achieve accurate simultaneous reconstruction of both properties when the heuristic condition is met by incorporating a regularization term into the reconstruction algorithm.
%  Examples of cases in which accurate simultaneous reconstruction is not possible are also provided. Even when c(r) cannot be accurately reconstructed, we demonstrate that A(r) can be more accurately reconstructed by this method than when c(r) is assumed homogeneous.
\end{abstract}


\section{Introduction}
\label{sect:intro}

Photoacoustic computed tomography (PACT), also known as 
optoacoustic or thermoacoustic tomography, is a rapidly 
emerging hybrid imaging modality that combines optical 
image contrast with ultrasound detection 
\cite{WanglihongBook2009,OraevskyBook2003,XuminghuaRSI2006}.
In PACT, the to-be-imaged object is illuminated with a 
short laser pulse that results in the generation of internal acoustic 
wavefields via the photoacoustic effect. The % initial 
amplitudes of the induced acoustic wavefields are 
proportional to the spatially variant absorbed optical 
energy density distribution within the object, which will 
be denoted by the object function $A(\mathbf r)$. The acoustic
waves propagate out of the object and are subsequently
measured by use of wide-band ultrasonic transducers.
The goal of image reconstruction in PACT is to estimate the
object function $A(\mathbf r)$ from these measurements.
% From these measurements, a reconstruction 
%algorithm is employed to estimate the object function 
%$A(\mathbf r)$.
% Because the optical absorption properties 
%of biological tissue are highly related to its hemoglobin 
%concentration and molecular constitution, PACT can reveal
%the pathological condition of the tissue, and therefore
%holds great potential for a wide-range of anatomical, 
%functional, and molecular imaging tasks in preclinical 
%and clinical medicine \cite{XuminghuaRSI2006,KrugerMP1999,
%HaltmeierIP2004,CoxOSA2006,XuzhunJBO2010,XuzhunJBO2011}.

Image reconstruction methods for PACT
are often based on idealized imaging models that assume an 
acoustically homogeneous medium \cite{Xuplanar,Xu2005bp,FinchFBP3D,Kunyansky07,KunFBP}. 
However, these assumptions are not warranted 
in certain biomedical applications of PACT \cite{HuangchaoJBObrain}.
Numerous image reconstruction methods have been proposed
\cite{XuyuanIEEE2003,DimpleJBO2010,JoseOE2011,Yuanzhen2007MP,
HristovaIP2008,StefanovIP2009,HuangchaoTMIiter,DeanBenAPL2011}
that compensate for aberrations of the measured
photoacoustic (PA) wavefields caused by an object's speed-of-sound (SOS) variations,  $c(\mathbf r)$,
 and hence improve PACT image quality.
It has been demonstrated that these methods can improve the
fidelity of reconstructed images by incorporating accurate knowledge of the SOS variations in the PACT imaging model. 
However, accurate estimation of  $c(\mathbf r)$ prior to the PACT study generally
requires solution of an ultrasound computed tomography (USCT) inverse
problem, which can present experimental and computational
challenges \cite{ZZ2012,Li09:ClinSOS,glover1979characterization,Schreiman84:USCT,wang2015waveform}.
%in many cases of practical interest, it may not be $c(\mathbf r)$

An important observation is that, because variations in the  SOS distribution
induce the PA wavefield aberrations,
certain information regarding an object's SOS distribution
is encoded in the PACT measurement data.
Based on this observation, it is natural to question
whether  $A(\mathbf r)$ and  $c(\mathbf r)$
%This brings up an important question in PACT: can
% the absorbed optical energy density distribution, $A(\mathbf r)$, and the speed of sound 
%distribution, $c(\mathbf r)$, both be accurately 
%determined from the measured PA data alone? 
can both be accurately determined \emph{from only the PACT measurement data.}
\cite{ZhangjinSPIE08,JiangJOSA06,YuanOE06,TRadjoint,kirsch2012simultaneous}, thereby circumventing
the need to perform a dedicated USCT study.
This will be referred to as the joint reconstruction (JR) problem
and is the subject of this article.
 
Theoretical and computational studies of the JR problem have
been conducted but all are limited in scope.
Theoretical work on the JR problem that neglects
discrete sampling effects
has established that $A(\mathbf r)$ and 
$c(\mathbf r)$ can be uniquely determined from
the measured PACT data only under certain restrictive 
assumptions regarding the forms of $A(\mathbf r)$ and
$c(\mathbf r)$ \cite{HickmannThesis, HristovaIP2008}
or the measurement surface\cite{Scherzer2012}.
However, the uniqueness of the JR problem for the general case has not been established.
Another study established that the solution of the linearized JR problem is  unstable
\cite{StefanovArxiv12} and suggested that the same conclusion would hold for the general case where wavefield
propagation modeling is based on the full wave equation.
 
Despite the lack of theoretical works,
others have moved forward and developed computational methods
for solving the JR problem by use of discretely sampled measurement data
\cite{ZhangjinSPIE08,JiangJOSA06,YuanOE06}. 
In \cite{ZhangjinSPIE08}, an iterative reconstruction 
method was proposed to jointly estimate both 
$A(\mathbf r)$ and $c(\mathbf r)$.  That study
employed a geometrical acoustics propagation model
and assumed \emph{a priori} information regarding
the singular support of $c(\mathbf r)$.
%described the geometry of the SOS distribution by 
%a small number of parameters.
 In \cite{JiangJOSA06,
YuanOE06},  a JR method
 based on the Helmholtz equation was proposed that
was solved by the finite element method (FEM). While
this method is grounded in an accurate model of the 
imaging physics, it suffers from an intensive 
computational burden.
A similar JR approach 
was proposed \cite{TRadjoint}
that employed a time-reversal
(TR) adjoint method.
All of these works are preliminary 
in the sense that they did not
systematically explore the numerical properties of the JR problem
or provide broad insights that allow one to predict when accurate JR may be possible. 
In combination with the scarcity of theoretical works, this indicates an important need
to further elucidate the practical feasibility of JR.

To address this, the primary objective of this work is to
investigate the numerical properties of the JR problem,
which will provide important insights into its practical feasibility.
A novel JR method is developed for this purpose.
The developed reconstruction method is based on an alternating optimization scheme, 
where $A(\mathbf r)$ is reconstructed by use of a 
previously-developed full-wave iterative method \cite{
HuangchaoTMIiter}, while $c(\mathbf r)$ is reconstructed 
by use of a nonlinear optimization algorithm based on the 
Fr\'echet derivative of an objective function with respect 
to $c(\mathbf r)$ \cite{BunksGP1995,NortonJASA1999}. 
Computer-simulation studies are conducted to
investigate the topology of the cost function
defined in the optimization-based approach to JR.  Additionally,
numerical experiments are conducted that reveal how the supports
and relative smoothness
 of $A(\mathbf r)$ and 
$c(\mathbf r)$ affect the numerical stability of the JR problem. 
%Specifically, we reveal that the numerical stablity of the JR problem is influenced
%by the relative extents of the spatial supports of $A(\mathbf r)$ and $c(\mathbf r)$. 
%Additionally, we demonstrate that the relative spatial frquency bandwidths of 
%$A(\mathbf r)$ and $c(\mathbf r)$ influence this.
 Demonstrations
of how errors in the imaging model associated with imperfect transducer modeling and acoustic
attenuation affect JR accuracy are also provided.

%The developed method is utilized to investigate the
%numerical properties of the JR problem 
%and its feasibility in practice.
%through computer simulations.
%We also investigate how the spatial structure of the
%two sought-after properties affect the JR problem.
%We also give two heuristic conditions for 
%the accurate reconstruction of $c(\mathbf r)$ in PACT. 

The paper is organized as follows. In Section 
\ref{sect:background}, the imaging physics of 
PACT in acoustically heterogeneous  media is reviewed briefly. 
The derivation of the Fr\'echet derivative with respect to $c(\mathbf r)$ of 
a pertinent objective function 
is also provided. Section \ref{sect:method} 
describes the alternating optimization
approach for solving the JR problem. 
In Sec.\ \ref{sect:supportTheory}, heuristic insights into how the relative
extents of the
spatial supports of $A(\mathbf r)$ and $c(\mathbf r)$ can 
affect the ability to perform accurate JR are provided.
The computer-simulation methodology and numerical studies are 
given in Secs.\ \ref{sect:studies} and \ref{sect:investigations}.
 The paper concludes with 
a summary and discussion in Section \ref{sect:summary}.

\section{Background}
\label{sect:background}

%\noindent 1. Quick review of PACT
%with variable SOS; give PA wave equation
\if 0
Below we first review the PA wavefield 
propagation model in heterogeneous, fluid,
lossless media. We then give the brief derivation
of the Fr\'echet derivative of an objective functional
with respect to $c(\mathbf r)$, which is the
key requirement of the optimization approach
to JR proposed in Section \ref{sect:method}.
\fi

\subsection{Photoacoustic wavefield propagation in heterogeneous media}

We consider PA wavefield propagation
in lossless fluid media having a constant mass density. 
Let $p(\mathbf r, t)$ denote 
the photoacoustically-induced pressure wavefield 
at location $\mathbf r \in \mathbb{R}^3$ and time $t \ge 0$.
The photoacoustic wavefield $p(\mathbf r, t)$ satisfies \cite{WanglihongBook2009}:
\begin{equation}
\label{eq:we}
\nabla^2 p(\mathbf{r},t) - \frac{1}{c(\mathbf{r})^2}\frac{\partial^2 p(\mathbf{r},t)}{\partial t^2} = 0,
\end{equation}
subject to initial conditions
\begin{equation}
\label{eq:ic}
p(\mathbf{r},0)=\Gamma(\mathbf r)A(\mathbf r), \quad \left.\frac{\partial p(\mathbf{r},t)}{\partial t}\right|_{t=0}=0,
\end{equation}
where $\Gamma(\mathbf r)$ is the Grueneisen parameter
that is assumed to be known.

%\noindent 2. Recontruction of $c(\mathbf r)$ given the measurement and $A(\mathbf r)$
%
%a. Formulate the reconstruction of $c(\mathbf r)$
%as an optimization problem; Give the cost functional
%
%b. Derive the Fr\'echet derivative of 
%the cost functional with respect to $c(\mathbf r)$

\subsection{Fr\'echet derivative with respect to $c(\mathbf r)$}
\label{sect:frechet}
%Here, for simplicity, we neglect the acousto-electrical
%impulse response (EIR) of the ultrasonic transducers and assume 
%each transducer is point-like.  With these assumptions,
%impulse response (EIR) and the spatial impulse response (SIR)
%of the ultrasonic transducers; i.e. assuming EIR to be Dirac
%delta function and transducers to be point-like. With these assumptions, 

Here, for simplicity, we neglect the 
acousto-electrical impulse response (EIR)
and the spatial impulse response (SIR)
of the ultrasonic transducers employed to record the PA signals. However,
the impact of these will be addressed
in Section \ref{sect:JR_real}.
The quantity $\hat{p}(\mathbf{r}^m,t)$ represents the  PA data
recorded by  the $m$-th transducer at location $\mathbf{r}^m$ 
($m=1,\cdots,M$).  For ease of description, we represent the measured PA data
as continuous functions of $t$, 
but the results below will be discretized for
numerical implementation as described in
Section \ref{sect:method}.

%For a given $A(\mathbf r)$,
 The problem
of reconstructing $c(\mathbf r)$ from PA data for a fixed $A(\mathbf r)$, defined as Sub-Problem \#2 below,  can be formulated as 
an optimization problem in which the following objective
functional is minimized with respect to $c(\mathbf r)$:
\begin{equation}
\label{eq:cost_fd}
\mathcal{E}[c(\mathbf r)] =
\sum_{m=1}^M\int_0^T \mathrm{d}t 
[p(\mathbf{r}^m,t)-\hat{p}(\mathbf{r}^m,t)]^2,
\end{equation}
subject to the constraint that $p(\mathbf{r}^m,t)$ satisfies Eqs.\ \eqref{eq:we} and \eqref{eq:ic}, 
where $T$ denotes the maximum time at which 
the PA data were recorded.
%  \rd{[What about Eq.\ (2)?]}

Gradient-based  algorithms can be 
utilized to minimize the nonlinear functional 
\eqref{eq:cost_fd}. Such methods require 
the functional gradient, or Fr\'echet derivative,
of $\mathcal{E}$ with respect to $c(\mathbf r)$, 
which can be calculated by use of an adjoint method
\cite{BunksGP1995,NortonJASA1999}.
In the adjoint method, the adjoint wave equation
is defined as 
\begin{equation}
\label{eq:adj_we}
\nabla^2 q(\mathbf{r},t) - \frac{1}{c(\mathbf{r})^2}\frac{\partial^2 q(\mathbf{r},t)}{\partial t^2} = -s(\mathbf{r},t),
\end{equation}
subject to terminal conditions
\begin{equation}
\label{eq:tc}
q(\mathbf{r},T)=0, \quad \left.\frac{\partial q(\mathbf{r},t)}{\partial t}\right|_{t=T}=0.
\end{equation}
The source term $s(\mathbf r,t)$ is defined as
\begin{equation}
\label{eq:s}
s(\mathbf{r},t) = \sum_{m=1}^M
[p(\mathbf{r}^m,t) - \hat{p}(\mathbf{r}^m,t)]
\delta(\mathbf{r} - \mathbf{r}^m).
\end{equation}

Upon solving \eqref{eq:we} and \eqref{eq:adj_we},
%(see Appendix), 
the Fr\'echet derivative of $\mathcal{E}$ 
with respect to $c(\mathbf r)$ can be
determined as \cite{BunksGP1995,NortonJASA1999},
\begin{equation}
\label{eq:fd}
\nabla_c \mathcal{E} = 
-\frac{4}{c(\mathbf r)^3} \int_0^T \mathrm{d}t
\frac{\partial p(\mathbf{r},t)}{\partial t}
\frac{\partial q(\mathbf{r},t)}{\partial t}.
\end{equation}
Once the Fr\'echet derivative is obtained,
it can be utilized by any gradient-based method
as the search direction to iteratively reduce 
the functional value of \eqref{eq:cost_fd}.
%The algorithm that was employed to compute
The computation of this Fr\'echet derivative 
%the right-hand side of Eq.\ (\ref{eq:fd})
 from discrete measurements
%compute this Fr\'echet derivative is described
is described in the Appendix.

\section{Joint Reconstruction of $A(\mathbf r)$ and $c(\mathbf r)$}
\label{sect:method}
\if 0
The Fr\'echet derivative described above
permits us to formulate the JR of
$A(\mathbf r)$ and $c(\mathbf r)$
as an optimization problem.
%which consists two sub-problems: reconstruction 
%of $A(\mathbf r)$ given $c(\mathbf r)$ 
%and reconstruction of $c(\mathbf r)$ given 
%$A(\mathbf r)$. 
To numerically solve the optimization problem, 
a discrete imaging model of PACT needs 
to be established first.
%, that is described below.
\fi

In this section, a JR method for concurrently estimating $A(\mathbf r)$ and $c(\mathbf r)$ is formulated based on an alternating optimization
strategy.

\subsection{Discrete imaging model}

Let the  $N \times 1$ vectors 
\begin{equation}
\label{eq:A}
\mathbf A \equiv [A(\mathbf{r}_1),\cdots, A(\mathbf{r}_N)]^{\rm T}
\end{equation}
and
\begin{equation}
\label{eq:c}
\mathbf c \equiv [c(\mathbf{r}_1),\cdots, c(\mathbf{r}_N)]^{\rm T}
\end{equation}
denote the finite-dimensional approximations of
$A(\mathbf r)$ and $c(\mathbf r)$ 
formed by sampling the functions at the $N$ vertices on a Cartesian grid
that correspond to locations $\mathbf r_n$, $n=1,\cdots, N$.
As introducted earlier, $\mathbf r^k$, $k=1,\cdots,M$ denote the transducer locations.

The quantity
\begin{equation}
\label{eq:tp_l}
\hat{\mathbf{p}}_l \equiv [\hat{p}(\mathbf{r}^1,l\Delta t), \cdots, \hat{p}(\mathbf{r}^M, l\Delta t)]^{\rm T}
\end{equation}
represents the measured PA data sampled at time 
$t=l\Delta t$ ($l=1,\cdots,L$) at
each transducer location.
Here, $\Delta t$ is the sampling time step, and
$L$ is the total number of time steps.
The complete set of measured PA data can be represented 
by the $LM \times 1$ vector
\begin{equation}
\label{eq:tp}
\hat{\mathbf{p}} \equiv [\hat{\mathbf{p}}_1, \cdots, \hat{\mathbf{p}}_{L}]^{\rm T}.
\end{equation}

By use of \eqref{eq:A}, \eqref{eq:c}, 
and \eqref{eq:tp}, a discrete PACT imaging model 
 can be expressed as \cite{HuangchaoTMIiter}
\begin{equation}
\label{eq:dim}
\hat{\mathbf p} =  \mathbf{H}(\mathbf c) \mathbf{A},
\end{equation}
where $\mathbf{H}(\mathbf c)$ is the $LM \times N$
system matrix that depends non-linearly on $\mathbf c$. 
A procedure to establish an explicit
matrix representation of $\mathbf{H}(\mathbf c)$
was provided in \cite{HuangchaoTMIiter}.

In conventional applications of PACT the SOS distribution $\mathbf c$
is assumed to be known.
Alternatively, the goal of the JR problem is to concurrently estimate $\mathbf A$ and $\mathbf c$
from the measured data $\hat{\mathbf p}$ by use of the model in Eq.\ (\ref{eq:dim}).

\subsection{Optimization-based joint image reconstruction}

Based on Eq.\
\eqref{eq:dim}, the JR problem can be 
formulated as 
\begin{equation} 
\label{eq:cost}
\hat{\mathbf{A}}, \hat{\mathbf{c}} =
\operatorname*{arg\,min}_{\mathbf{A} \ge 0, \mathbf{c} > 0}
\| \mathbf{H}(\mathbf c) \mathbf{A} - \hat{\mathbf p}\|^2
%+ \lambda_A |\mathbf{A}|_{\text{TV}},
+ \lambda_1 \rm{R}_A(\mathbf{A}) + \lambda_2 \rm{R}_c(\mathbf{c}),
\end{equation}
where $\rm{R}_A(\mathbf{A})$ and $\rm{R}_c(\mathbf{c})$
are penalty functions that impose regularity on the estimates 
of $\mathbf{A}$ and $\mathbf{c}$, respectively, and $\lambda_1 $,
$\lambda_2$ are the corresponding regularization parameters. 
%\rd{[Do we need to change to $\lambda_A$ and $\lambda_c$ in (13)?]}
%\bl{[I think $\lambda_1$ and $\lambda_2$ are fine here; they are the weights
%between 3 terms in (13), while $\lambda_A$ and $\lambda_c$ balance 
%the weights between two terms]}
\if0
Although the data fidelity
term is convex with respect to $\mathbf{A}$,
the convexity with respect to $\mathbf{c}$
has not been established to our knowledge. 
Even if it is convex with respect to $\mathbf{c}$, 
the biconvex data fidelity term is not convex 
in general \cite{biconvex}. 
\fi
%Since
As discussed in Sec.\ \ref{sect:cf}, the cost function in \eqref{eq:cost} is non-convex.
However, a heuristic alternating optimization approach can be 
employed to find solutions that approximately
satisfy \eqref{eq:cost}. This approach
consists of the two sub-problems described below: (1) reconstruction 
of $\mathbf A$ given $\mathbf c$, and (2) 
reconstruction of $\mathbf c$ given 
$\mathbf A$.

{\bf Sub-Problem \#1: \emph{Reconstruction of $\mathbf A$ 
given $\mathbf c$}}:  The  
problem of estimating $\mathbf{A}$ 
for a given  ({i.e.}, fixed) $\mathbf{c}$ can be 
formulated as the penalized least 
squares problem
\begin{equation} 
\label{eq:cost_A}
\hat{\mathbf{A}} =
\operatorname*{arg\,min}_{\mathbf{A} \ge 0}
\| \mathbf{H}(\mathbf c) \mathbf{A} - \hat{\mathbf p}\|^2
%+ \lambda_A |\mathbf{A}|_{\text{TV}},
+ \lambda_A \rm{R}_A(\mathbf{A}),
\end{equation}
%where $\rm{R}(\mathbf{A})$ is a regularization
where $\lambda_A$ is the regularization 
parameter, which is different from $\lambda_1$
%\bl{[since we already use a different notation here,
%we may not need to explicitly say there are different]}
in Eq.\ \eqref{eq:cost} in general.
Reconstruction  methods have been proposed for
solving problems of this form \cite{HuangchaoTMIiter,TreebyIP2010}. 
%and a non-negativity constraint was employed.
%Depending on the choice of $\rm{R}(\mathbf{A})$,
%there exist different optimization algorithms to
%efficiently solve \eqref{eq:cost_A} \cite{NocedalBook}.

{\bf Sub-Problem \#2: {Reconstruction of $\mathbf c$ 
given $\mathbf A$}}:  
For a given ${\mathbf{A}}$,
an estimate of ${\mathbf{c}}$ can be formed as
%Reconstruction of $\mathbf{c}$ for a given 
%$\mathbf{A}$ can be accomplished by solving 
%the following optimization problem
\begin{equation} 
\label{eq:cost_c}
\hat{\mathbf{c}} =
\operatorname*{arg\,min}_{\mathbf{c} > 0}
\| \mathbf{H}(\mathbf c) \mathbf{A} - \hat{\mathbf p}\|^2
+ \lambda_c \rm{R}_c(\mathbf{c}),
\end{equation}
where $\lambda_c$ is the regularization   
parameter, {which is different from $\lambda_2$} 
in Eq.\ \eqref{eq:cost}, in general.
Equation \eqref{eq:cost_c} can be solved
by use of gradient-based methods, which require
computation of the gradient of the objective function in
Eq.\ \eqref{eq:cost_c} with respect to $\mathbf c$.
Details regarding this gradient computation
are provided in Sec.\ \ref{sect:frechet} and the Appendix.

\begin{algorithm}[H]
\caption{\label{alg:alter}
Alternating optimization approach to
JR of $\mathbf A$ and $\mathbf c$
}
% \algsetup{indent=2em}
\begin{algorithmic}[1]
\REQUIRE $\hat{\mathbf p}$,
$\mathbf{A}^{(0)}$, $\mathbf{c}^{(0)}$,
$\epsilon_A$, $\epsilon_c$,
$\lambda_A$, $\lambda_c$
\ENSURE $\hat{\mathbf A}$, $\hat{\mathbf c}$
  \STATE $i = 0$
  \WHILE{$\epsilon_1 < \epsilon_A$ \textbf{and} $\epsilon_2 < \epsilon_c$}
    \STATE $\mathbf{A}^{(i+1)}
        \leftarrow \rm {F_A}
          \big( \mathbf{A}^{(i)}, \mathbf{c}^{(i)}, \hat{\mathbf{p}}, \lambda_A \big)$
% \hskip 0.8cm 
\COMMENT{Sub-Problem \#1}
	\STATE $\mathbf{c}^{(i+1)}
        \leftarrow \rm {F_c}
          \big( \mathbf{c}^{(i)}, \mathbf{A}^{(i+1)}, \hat{\mathbf{p}}, \lambda_c \big)$
% \hskip 0.75cm 
\COMMENT{Sub-Problem \#2}
    \STATE $\epsilon_1 \leftarrow \rm{Dist}(\mathbf{A}^{(i)}, \mathbf{A}^{(i+1)})$
    \STATE $\epsilon_2 \leftarrow \rm{Dist}(\mathbf{c}^{(i)}, \mathbf{c}^{(i+1)})$
    \STATE $i \leftarrow i+1$
  \ENDWHILE
  \STATE $\hat{\mathbf A} \leftarrow \mathbf{A}^{(i)}$
  \STATE $\hat{\mathbf c} \leftarrow \mathbf{c}^{(i)}$
\end{algorithmic}
\end{algorithm}

\underline{\emph{Alternating optimization algorithm}}:
%Based on \eqref{eq:cost_A} and \eqref{eq:cost_c},
As described in  Algorithm 1,
JR of $\mathbf A$ and $\mathbf c$ can be
accomplished by alternately solving  Sub-Problems \#1 and \#2.
%\eqref{eq:cost_A} and \eqref{eq:cost_c}. 
The quantities $\mathbf{A}^{(0)}$ and $\mathbf{c}^{(0)}$
are the initial estimates of $\mathbf A$ and $\mathbf c$,
respectively, and $\epsilon_A$ and $\epsilon_c$
are convergence tolerances. The functions
`$\rm{F_A}$' and `$\rm{F_c}$' compute the
solutions of Eqs.\ \eqref{eq:cost_A} and 
\eqref{eq:cost_c}, respectively, and are described
below in Section \ref{sect:studies}.
The function `Dist' measures the 
Euclidean distance between $\mathbf{A}^{(i)}$ and $\mathbf{A}^{(i+1)}$
(or between $\mathbf{c}^{(i)}$ and $\mathbf{c}^{(i+1)}$).

%\section{Conditions for accurate reconstruction of $\mathbf c$ given $\mathbf A$ (sub-problem (\ref{eq:cost_c}))}
\section{Heuristic insights into support conjecture  and Sub-problem \#2}
\label{sect:supportTheory}

As revealed throughout this work, difficulities in solving Sub-Problem \#2 represents
a significant challenge for JR.
\if 0
\bl{In this study, we propose two heuristic conditions that,
when satisfied, suggest that $\mathbf c$ can be accurately
estimated from PACT measurement data given known $\mathbf A$.
One condition is with respect to
to the support of $\mathbf A$, which will
be referred as the support condition
and is described in Appendix-B. Another condition
is related to the spatial specarum of $\mathbf A$,
which will be referred as the k-space condition
and discussed in Section \ref{sect:k-space}.
The $\mathbf A$ satisfying the support condition
will be called `adequate' in this paper, otherwise
it will be called `defective'. The $\mathbf A$ satisfying
both heuristic conditions will be called `sufficient',
otherwise it will be called `deficient'.
When $\mathbf A$ is sufficient, we will show that it is possible
to achieve accurate JR in Section \ref{sect:JR_ideal}.}
\bl{Note that the support condition is based on
two assumptions and is only verified by
computer simulations instead of rigorous
mathematical proofs, so the condition is
more an observation than a
necessary/sufficient condition in
mathematical sense; i.e. when the support
condition is satisfied, it is more likely
to achieve accurate reconstruction of
$\mathbf c$ or accurate JR. Therefore, we
call the support condition a heuristic condition.
This also applies to the heuristic k-space
condition described below.}

\fi
In this section, 
 heuristic insights into how the relative
extents of the
spatial supports of $A(\mathbf r)$ and $c(\mathbf r)$ can
affect the ability to accurately solve Sub-Problem \#2, and
hence the JR problem, are provided.

The supports of the functions $A(\mathbf r)$ and $c(\mathbf r)$
will be denoted as $\text{supp}(A)$ and $\text{supp}(c)$, respectively.
The supports are defined to be the regions where
$A(\mathbf r) \neq 0$ and
$c(\mathbf r) - c_0 \neq 0$,
respectively. Here $c_0$ is
the known SOS in the background
(i.e., in the coupling water-bath).
The functions $A(\mathbf r)$
and $c(\mathbf r)$ are assumed to be compactly supported,
reflecting the physical constraint that their extents are limited
and the object of interest resides within the imaging system.

The following assumptions will be made in order to permit a simple analysis of
the problem that yield insights.
First, it is assumed that acoustic scattering is weak. More specifically,
a geometrical acoustics model \cite{XuyuanIEEE2003} is employed to describe the propagation
of the PA wavefields.  Amplitude variations are assumed to
be negligible and wavefront aberrations are modeled by computing the
 time-of-flight along straight propagation paths.
%the SOS distribution must occur on
%length scales that are large
%compared to the effective
%acoustic wavelength. Under
%this assumption, a straight ray model
%is utilized to approximate the wave equation.
The second assumption exploits the fact that an arbitrary
function $A(\mathbf r)$ can be decomposed into a collection of
point sources.
 Specifically, it is assumed that the PA signal
generated by each of these point sources 
can be recorded independently by transducers.
In reality, of course, this is not the case. This assumption, in effect,
assumes that an `oracle' records the PA signals and is able to decompose
them into the individual components that were produced by each point source.  Below, we will
describe how the relative extents of  $\text{supp}(A)$ and $\text{supp}(c)$
affect the ability of the oracle to accurately solve Sub-Problem \#2.  Since
the oracle has access to more information than is actually recorded in an
experiment, inability of the oracle to perform accurate image reconstruction
implies that accurate image reconstruction by use of the actual recorded data will
also be unfeasible. This is the logical basis for the analysis below.
Third, the analysis is presented in 2D but can be
extended to the 3D case readily.
Without loss of generality, the measurement surface
is assumed to be a circle with radius $R$
that encloses $\text{supp}(A)$.

%These transducers are assumped to be densely distributed on a measurement surface
%that encloses $\text{supp}(A)$ and $\text{supp}(c)$.

\begin{figure}[h]
  \centering
    \resizebox{1.5in}{!}{
      \includegraphics{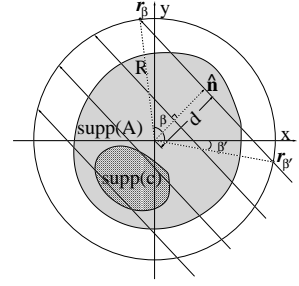}
  }
  \caption{\label{fig:radon}
%Projection of the slowness in direction $\alpha$
%and $\text{supp}(c) \subseteq \text{supp}(A)$.
Schematic of the 2D circular measurement geometry employed in
the heuristic analysis  of Sub-Problem \#2 described
in Sec.\ \ref{sect:supportTheory}. The coordinates
$\mathbf r_\beta$ and $\mathbf r_\beta^\prime$ denote
the transducer locations that correspond to the intersection
points of the line $\mathbf r \cdot \hat{n}=d$ with
the measurement circle.
In this example,  $\text{supp}(c) \subseteq \text{supp}(A)$.
%\rd{[Chao: Can you please update this figure?  All of the text and symbols
%are too small.  Also, we don't need $\mathbf r_0$
% or $\mathbf r_1$ and
%need to add $\mathbf r_\beta^\prime$.]}
%As an example, $\text{supp}(A)$ can be seen as
%the area occupied by a breast, where the SOS is
%approximately the same as the background SOS in
%water, and $\text{supp}(c)$ can be seen as the
%area occupied by a tumor, where the SOS is larger
%than the background SOS.
}
\end{figure}

\if 0
Under the above assumptions,
the reconstruction of the slowness
$s(\mathbf r) \equiv \frac{1}{c(\mathbf r)}$,
which is equivalent to reconstructing
$c(\mathbf r)$, is analogous to the
reconstruction of attenuation coefficients
in X-ray computed tomography (CT). The 2D analogy
is shown in Fig. \ref{fig:radon} and described below.
First, consider the case where $\text{supp}(c) \subseteq \text{supp}(A)$.
Without loss of generality, the measurement surface
is assumed to be a circle with radius $R$
that encloses $\text{supp}(A)$.
\fi

For convenience, we introduce the perturbed slowness distribution
 $s(\mathbf r)\equiv{1\over c_0}-{1\over c(\mathbf r)}$.
%Two cases will be considered to reveal the impact of the
%relative extents of $\text{supp}(A)$ and $\text{supp}(c)$
%on the ability of the oracle to reconstruct $s(\mathbf r)$,
%or equivalently, $c(\mathbf r)$,
%given $A(\mathbf r)$.
As assumed above, the oracle can resolve the PA signal generated
by each point source that comprises $A(\mathbf r)$.
As such, the time it takes for a PA pulse to propagate
from its emission  location 
 $\mathbf r_s \in \text{supp}(A)$
to each transducer location $\mathbf{r}_{\beta} \equiv
[R\cos(\beta), R\sin(\beta)]^{\rm{T}}$
% ($\beta \in [\alpha - \frac{\pi}{2}, \alpha + \frac{\pi}{2}]$)
 ($\beta \in [-\pi, \pi)$)
is also known to the oracle, where the geometry is defined in Fig.\ \ref{fig:radon}.
From these time-of-flight (TOF) data, a tomographic data function
can be defined as
\begin{equation}
g(\mathbf r_\beta,\mathbf r_\beta^\prime;\mathbf r_s)\equiv  \tau_{c_0}
(\mathbf r_\beta, \mathbf r_{\beta^\prime})-
\tau(\mathbf r_s, \mathbf r_\beta) - \tau(\mathbf r_s, \mathbf r_\beta^\prime),
\label{eq:datafun}
\end{equation}
where $\mathbf r_\beta^\prime$ is the transducer location defined by
the intersection of the line connecting $\mathbf r_\beta$ and $\mathbf r_s$
and the measurement circle.
% as shown in Fig.\ \ref{fig:radon}.
The quantity ${\tau_{c_0}(\mathbf r_\beta, \mathbf r_{\beta^\prime})}$
denotes the time it takes for a pulse to propagate between the
two transducer locations when only the background medium is present.
Since a straight-ray, geometrical acoustics, model is assumed,
 this data function is related to the sought-after
slowness distribution as
\if 0
Consider the 2D Radon transform of $s(\mathbf r)$
%($\mathbf r \in \text{supp}(A)$)
 corresponding to a
tomographic view angle specified by the direction
$\alpha$, as shown in Fig. \ref{fig:radon}.
\fi
\begin{equation}
\label{eq:tau}
%\tau(\alpha, \beta) \equiv t_1(\alpha , \beta) - t_0(\alpha , \beta)
%= \int_{L(\mathbf{r}_0 , \mathbf{r}_1)}
%\hskip -.5cm s(\mathbf r) \rm{d} \mathbf{r}, \quad
%\tau(\mathbf r_s, \mathbf r_\beta) + \tau(\mathbf r_s, \mathbf r_\beta^\prime)-{D(\mathbf r_\beta,
%\mathbf r_{\beta^\prime})\over c_0}
g(\mathbf r_\beta,\mathbf r_\beta^\prime;\mathbf r_s)
= \int_{L(\mathbf{r}_\beta^\prime , \mathbf{r}_\beta)}
\hskip -.5cm s(\mathbf r) \rm{d} \mathbf{r}, \quad 
\end{equation}
%Here, $\mathbf r_{\beta^\prime}$ denotes the transducer location that
%is conjugate to  $\mathbf r_{\beta}$ in the sense that it is intersected
%by the chord $\mathbf r \cdot \hat{n}=d$ that intersects the measurment
%circle, as shown in  Fig.\ \ref{fig:radon}.
%The length of this chord is denoted as $D(\mathbf r_\beta, \mathbf r_{\beta^\prime})$.
where the path of integration is along the line segment $L(\mathbf{r}_\beta^\prime , \mathbf{r}_\beta)$
that connects the two transducer locations.
\if 0
Here,
\begin{equation}
\label{eq:t0}
t_0(\alpha , \beta)=\inf\{t_f (\beta, \mathbf{r}):
\mathbf{r} \cdot \hat{\mathbf{n}}=d, \mathbf{r} \in \text{supp}(A)\}
\end{equation}
and
\begin{equation}
\label{eq:t1}
t_1(\alpha , \beta)=\sup\{t_f (\beta, \mathbf{r}):
\mathbf{r} \cdot \hat{\mathbf{n}}=d, \mathbf{r} \in \text{supp}(A)\},
\end{equation}
where $\hat{\mathbf{n}} = [\cos(\alpha-\frac{\pi}{2}),
\sin(\alpha-\frac{\pi}{2})]^{\rm{T}}$ is the unit vector
in direction $\alpha-\frac{\pi}{2}$, and $d=R\sin(\beta-\alpha)$,
as shown in Fig. \ref{fig:radon}. Geometrically, $t_0(\alpha , \beta)$
(resp. $t_1(\alpha , \beta)$) is the TOF of the signal
traveling from point $\mathbf{r}_0$ (resp. point $\mathbf{r}_1$)
to $\mathbf{r}_{\beta}$, where the line segment $\mathbf{r}_0\mathbf{r}_1$
is the intersection of the line $\mathbf{r} \cdot \hat{\mathbf{n}}=d$
and $\text{supp}(A)$. If the line $\mathbf{r} \cdot \hat{\mathbf{n}}=d$
does not intersect with $\text{supp}(A)$,
we define $t_0(\alpha , \beta)=t_1(\alpha , \beta)=0$
($\beta \in [\alpha - \frac{\pi}{2}, \alpha + \frac{\pi}{2}]$).
\fi

Two cases will be considered that reveal the impact of the
relative extents of $\text{supp}(A)$ and $\text{supp}(c)$
on the ability of the oracle to reconstruct $s(\mathbf r)$,
or equivalently, $c(\mathbf r)$,
given $A(\mathbf r)$.
The first case corresponds to the situation where
$\text{supp}(c) \subseteq \text{supp}(A)$, as
depicted in  Fig.\  \ref{fig:radon}.
In this case, for example, $\text{supp}(A)$
could correspond to the area occupied by a soft tissue
structure in which the SOS is
approximately the same as the background SOS
and $\text{supp}(c)$ would correspond to a
region within the tissue possessing a different SOS.
It can be verified readily that 
for every $\mathbf r\in \text{supp}(c)$, a collection 
of $\mathbf r_s \in \text{supp}(A)$ exists such that
values of the  data function
that specify the line integrals along all paths $L(\mathbf{r}_\beta^\prime , \mathbf{r}_\beta)$
that intersect $\mathbf r$ are accessible to the oracle. Stated otherwise,
in this case, all projection data that are needed to uniquely invert
the 2D Radon transform \cite{NattererBook} in Eq.\ (\ref{eq:tau}) are accessible and therefore
$s(\mathbf r)$ or, equivalently, $c(\mathbf r)$ can be exactly determined
in a mathematical sense.

\begin{figure}[h]
\centering
  \subfigure[]{\resizebox{1.5in}{!}{\includegraphics{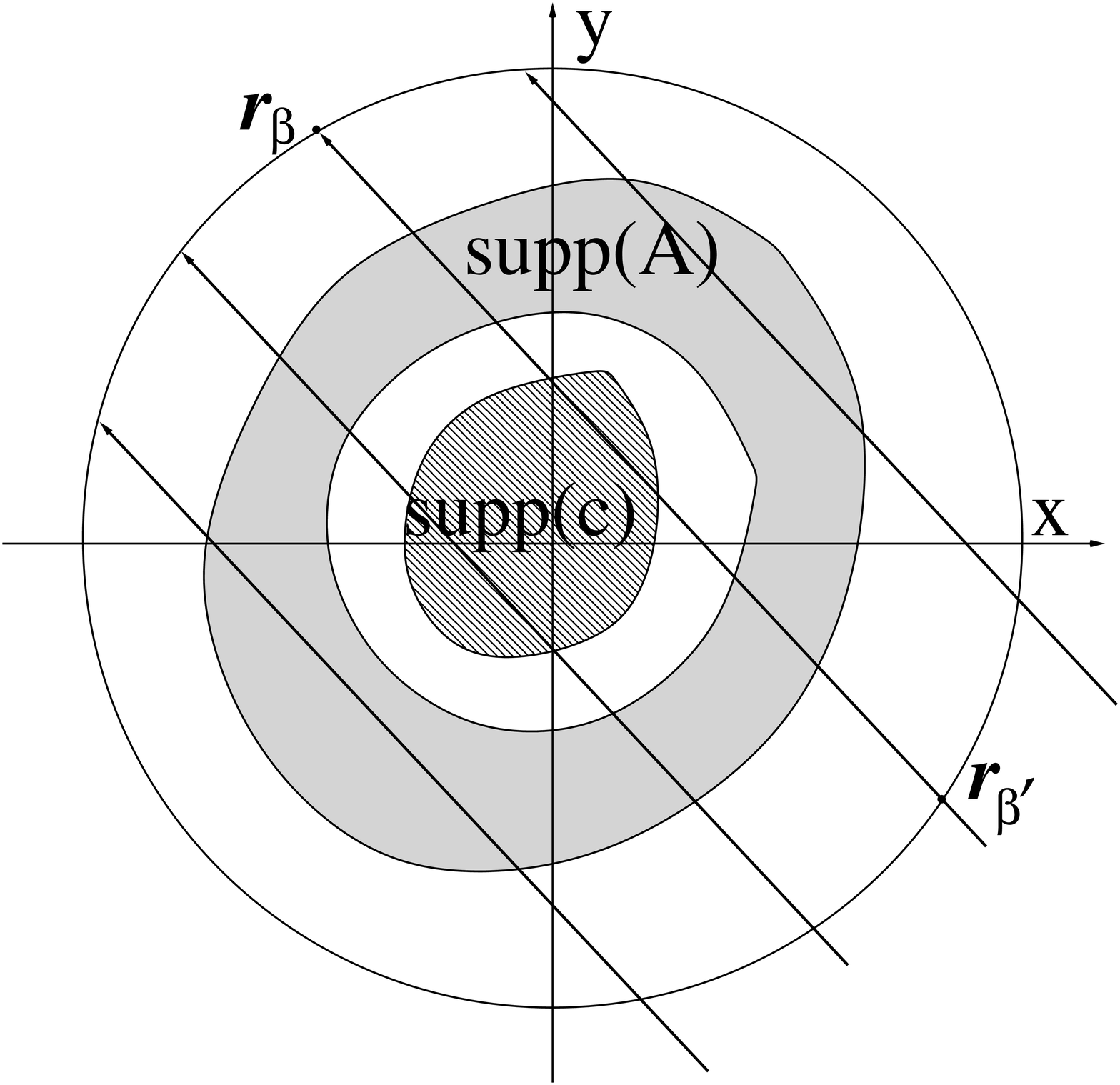}}}
  \subfigure[]{\resizebox{1.5in}{!}{\includegraphics{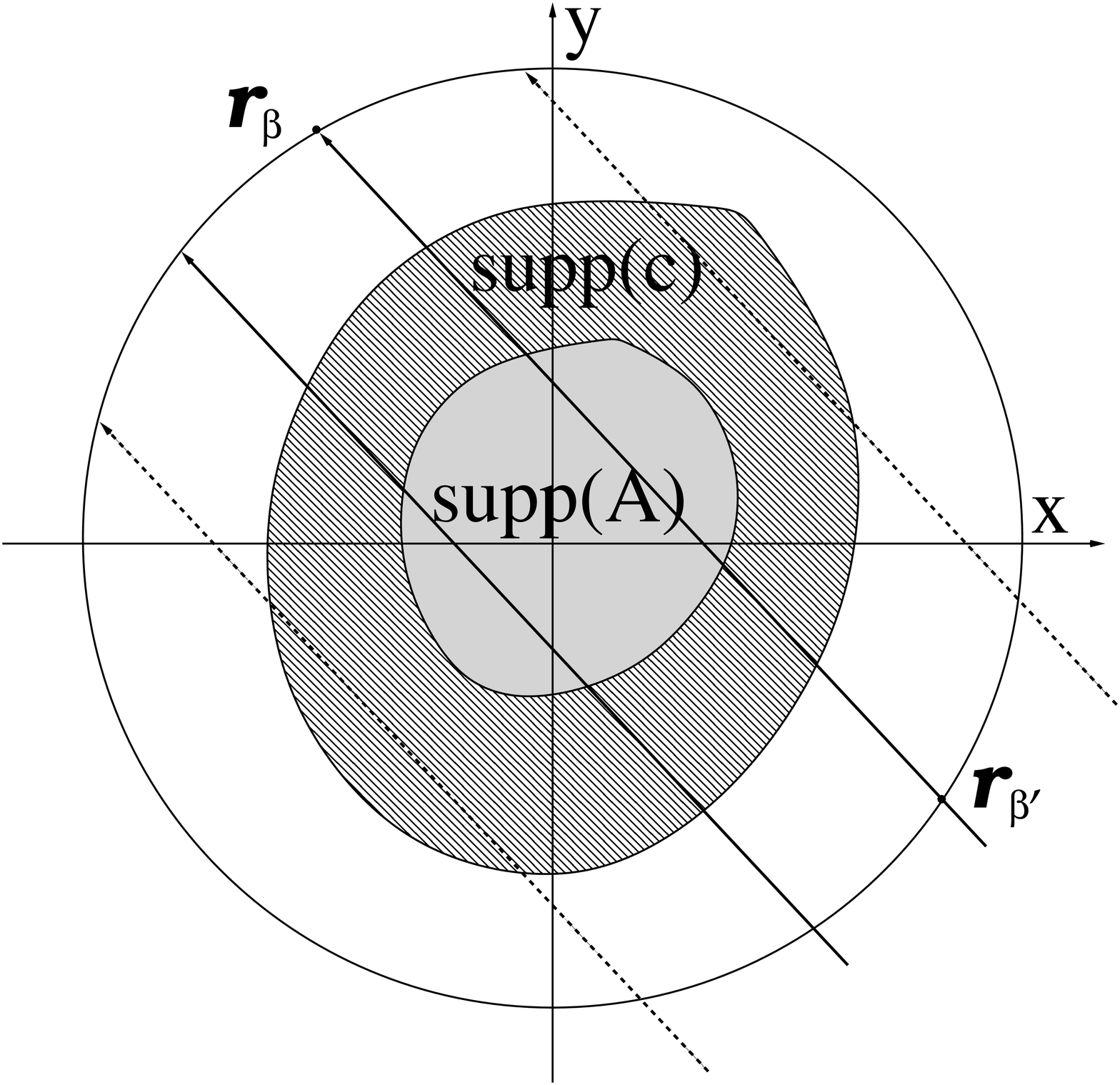}}}\\
\caption{\label{fig:Agb}
Two cases addressed in the heuristic analysis  of Sub-Problem \#2 described
in Sec.\ \ref{sect:supportTheory}:
 (a) Case where $\text{supp}(c)$ is enclosed by
$\text{supp}(A)$; and (b) Case where $\text{supp}(c)$ is
not enclosed by $\text{supp}(A)$.
In (b), certain line integral data for the slowness 
distribution in a subset of
$\text{supp}(c)$ are not measured, as indicated by the dashed
lines through the regions covered by lines.
 %covered by vertical lines
%and denoted as $C_{\alpha}$, are not measured
%by transducers.
%\rd{[Chao: Can you please update this figure?  All of the text and symbols
%are too small.  Also, we don't need $\mathbf r_0$ 
% or $\mathbf r_1$ and
%need to add $\mathbf r_\beta^\prime$.]}
}
\end{figure}

 In fact, the requirement
$\text{supp}(c) \subseteq \text{supp}(A)$
can be relaxed to $\text{supp}(c)$ being enclosed
by $\text{supp}(A)$, as shown in Fig. \ref{fig:Agb}(a).
This is because we only require
% the $t_f (\beta, \mathbf{r})$
$\mathbf r_s \in \partial \, \text{supp}(A)$, where
$\partial \, \text{supp}(A)$ denotes the boundary of $\text{supp}(A)$,
to determine the complete set of projection data as described above.

The second case corresponds to the situation where
 $\text{supp}(c)$ is not enclosed by $\text{supp}(A)$, as shown in
 Fig.\ \ref{fig:Agb}(b).
In this case, it can be verified  that
the complete data function cannot be determined, indicating that
the values of all line integrals through $s(\mathbf r)$ are
not accessible as depicted in Fig. \ref{fig:Agb}(b).
% for each projection angle $\alpha$,
%the projections of a subset of $\text{supp}(c)$ in this direction
%(indicated by $\rm{C}_{\alpha}$) are not measured
%because $A(\mathbf r)=0$ in that subset.
 That example is analogous to the interior problem in X-ray CT,
which possesses no unique solution \cite{NattererBook}.

These observations are summarized as follows.

\noindent \emph{\bf Support conjecture:}:  When geometrical acoustics is valid,
Sub-Problem \#2 cannot be accurately solved when
$\text{supp}(c)$ is not enclosed by $\text{supp}(A)$.

%Therefore, it is concluded that $c(\mathbf r)$ cannot be accurately reconstructed, in general, 
%if $\text{supp}(c)$ is not enclosed by $\text{supp}(A)$.

%In summary, through heuristic arguments and mathematical properties of
%the 2D Radon transform, a support condition has been identified.
%This condition suggests that, when geometrical acoustics is valid,
%Sub-Problem \#2 cannot be accurately solved when
%$\text{supp}(c)$ is not enclosed by $\text{supp}(A)$.

\section{Computer-Simulation Studies}
\label{sect:studies}

A general description of the computer-simulation methodologies
is described below.   
Although the optimization approach
to JR described above is based on the 3D
wave equation, 3D JR remains computationally intensive.
 For computational convenience, the 2D formulation
is investigated in this work.
The specific numerical experiments designed to investigate 
the numerical properties of the JR problem are described
subsequently in Sec.\ \ref{sect:investigations}.

%The primary reason for this is that 3D JR is computationally
%intensive, which would prohibit the studies described below.
%\footnote{3D simulations are not conducted in this
%study due to the intensive computational burden,  
%which is resulting from the line search used in 
%nonlinear optimization algorithms.}

%\section{Descriptions of numerical studies}
%\label{sect:describe}

\subsection{Simulation of idealized PA measurement data}
Numerical phantoms  were
utilized to represent $A(\mathbf r)$ and $c(\mathbf r)$, which were
described by Eqs.\ (\ref{eq:A}) and (\ref{eq:c}).
These phantoms, denoted by $\mathbf A$ and $\mathbf c$, contained  $512 \times 512$ pixels with
a pitch of $0.25$ mm.
({i.e.,} $N=512^2$ in Eqs. (\ref{eq:A}) and (\ref{eq:c}).)
For a given  $\mathbf A$ and $\mathbf c$, the lossless PA wave equation
was solved by use of the MATLAB k-Wave toolbox
\cite{treeby2010k} to produce simulated PA signals at
each transducer location.  Each PA signal
contained 6000 temporal samples recorded with a time step
$\Delta t = 50$ ns.
%The same procedure was repeated
%for noisy data, where 3\% (with respect to maximum
%value of noiseless data) additive white Gaussian
%noise (AWGN) was added to the simulated PA data.
The measurement geometry consisted
of 800 transducers that were
evenly distributed on the perimeter of a square with a side length of 100 mm.
%This measurement surface  enclosed the object.
The object was contained within this region.
Note that, unless stated otherwise, the generation of the simulated PA measurement
data in this way avoided an `inverse crime', due to a different
choice of discretization parameters from those employed by
the reconstruction algorithm as described below.

\subsection{Simulation of non-idealized PA measurement data}

To investigate the properties of JR under more realistic conditions,
additional PA data sets were simulated that 
considered certain physical factors.
Measurement noise was modeled by adding 3\% (with respect to the maximum
value of the noiseless data) white Gaussian
noise (AWGN) to the simulated PA data.
  Additional factors considered included acoustic attenuation,
the effect of the electrical and spatial impulse responses of the
transducers.  Specific details
are provided in Sec. \ref{sect:JR_real}.

\subsection{Implementation details for image reconstruction}
The implementations of the two image reconstruction sub-problems  in Algorithm 1 are described below.
The function `$\rm{F_c}$' that computes the solution of \eqref{eq:cost_c}
was implemented based on the MATLAB k-Wave toolbox 
\cite{treeby2010k}. Specifically, the 
wave equation \eqref{eq:we} and the adjoint 
wave equation \eqref{eq:adj_we} were solved numerically by use of the 
k-space pseudospectral method. The computed PA 
wavefield and the adjoint wavefield were employed 
to compute the gradient of the objective function 
in \eqref{eq:cost_c} {(see Appendix)}. 
The gradient was subsequently
utilized by the limited-memory BFGS (L-BFGS) 
algorithm to solve \eqref{eq:cost_c}
\cite{Poblano,DennisBook,NocedalBook}. 
The implementation of the function `$\rm{F_A}$' 
that solves \eqref{eq:cost_A} can be found in 
\cite{HuangchaoTMIiter}. In this study, 
a total variation (TV) penalty was adopted.
%to regularize the reconstrution of $\mathbf A$ 
%and $\mathbf c$.
 The `Dist' function measured 
the difference in terms of root mean squared 
error (RMSE), and the convergence tolerances
$\epsilon_A$ and $\epsilon_c$ were empirically 
chosen to have a value of $10^{-2}$ throughout
the studies. In all studies, the initial estimates
of $\mathbf A$ and $\mathbf c$ were set to be
$\mathbf{A}^{(0)}=0$ and $\mathbf{c}^{(0)}=1480$ m/s,
which is the background SOS.
Both $\mathbf A$ and $\mathbf c$ were reconstructed
on a uniform grid of $256 \times 256$ pixels with 
a pitch of 0.5 mm.

All simulations were computed in the MATLAB environment
on a workstation that contained dual hexa-core 
Intel(R) Xeon(R) E5645 CPUs and an NVIDIA Tesla C2075 
graphics processing unit (GPU). 
The GPU was equipped with 448 1.15 GHz CUDA cores and 
5 GB global memory. The Jacket toolbox \cite{Jacket} 
was employed to accelerate the computation of 
\eqref{eq:we} and \eqref{eq:adj_we} on the GPU.

\section{Numerical investigations and analyses}
\label{sect:investigations}

\subsection{ Topology of the JR cost function in Eq.\ (\ref{eq:cost}) }
\label{sect:cf}

The effectiveness of the optimization-based approach to
JR depends on the topology of the cost function that is minimized in Eq.\ (\ref{eq:cost}).
In these studies, only the data fidelity term was considered ($\lambda_1=\lambda_2=0$).
If the cost function is not convex or quasi-convex, a global minimum
(i.e., an accurate JR solution) may not be returned by the optimization
algorithm when the initial estimate is not close enough to the global minimizer.
Moreover, when the cost function is not strictly convex, there is no
guarantee that the  solution is unique.
In this section,  a low-dimensional stylized example is considered to yield
 insights into the general characteristics of the cost function topology.

\begin{figure}[ht]
\centering
  \subfigure[]{\resizebox{1.5in}{!}{\includegraphics{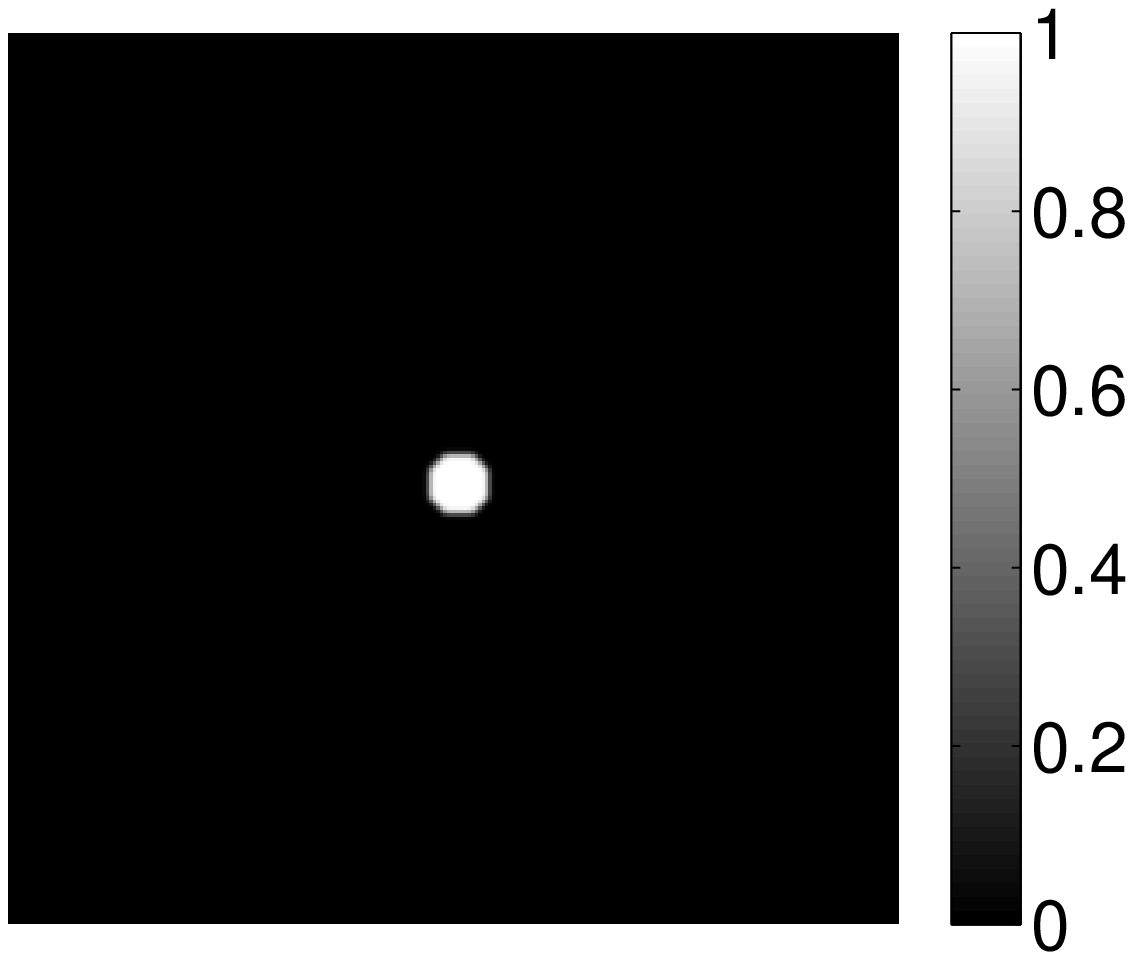}}}
  \subfigure[]{\resizebox{1.5in}{!}{\includegraphics{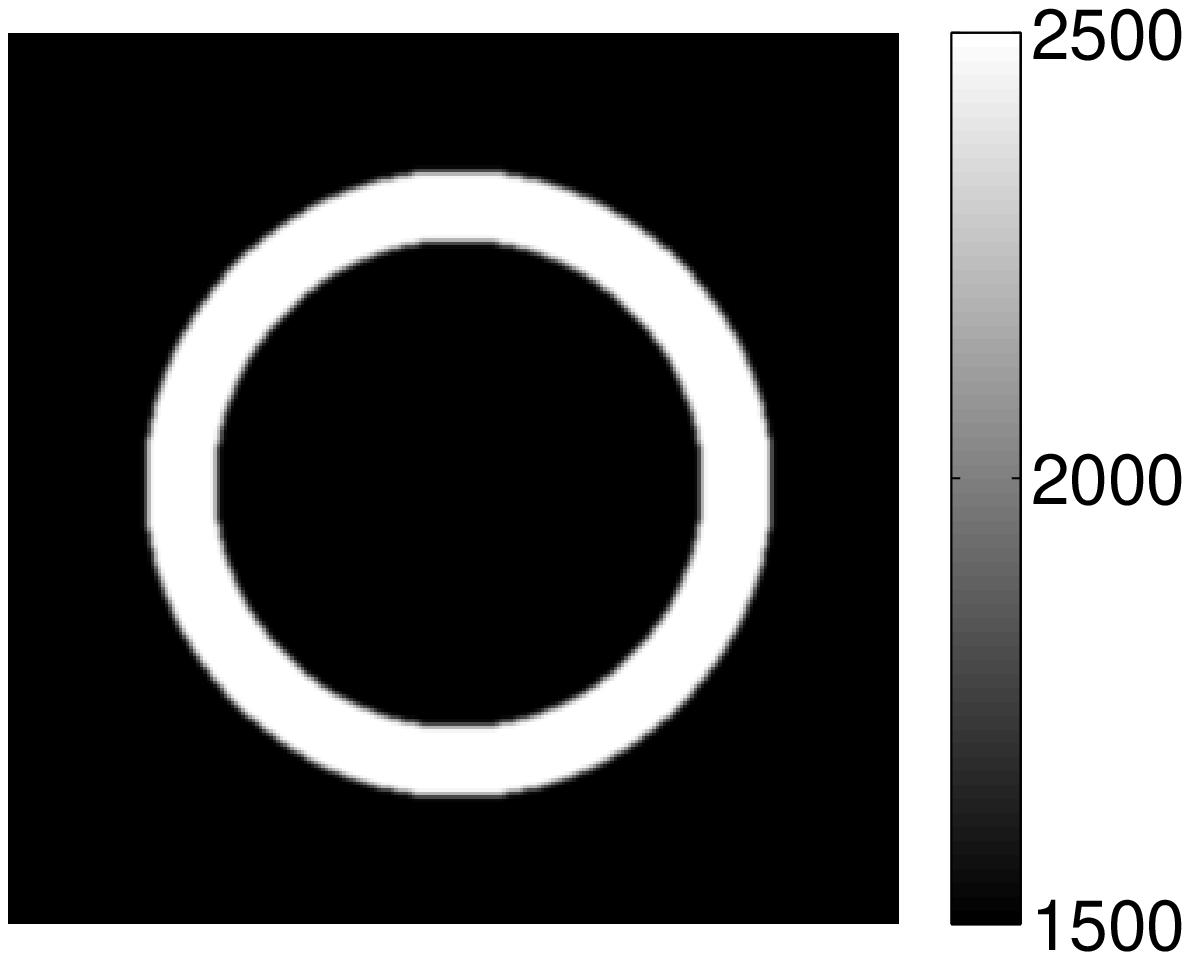}}}
\caption{\label{fig:phantoms_cf}
Subfigures (a) and (b) display the numerical phantoms that specified
$\mathbf A$ and $\mathbf c$ that were employed to investigate
the JR cost function topology in Sec.\ \ref{sect:cf}.
}
\end{figure}

The numerical phantoms  shown in Fig. \ref{fig:phantoms_cf} were
employed to represent $\mathbf A$ and $\mathbf c$.
To establish a low-dimensional representation of these phantoms,
a discretization scheme that differed from Eqs.\ (\ref{eq:A})
and (\ref{eq:c}) was employed (in this study only).  Namely, 
$\mathbf A$ was described as $\mathbf A=[A_s, A_b]$ where $A_s$ 
represented the value within the uniform disk and  $A_b$
is the known (assumed to be zero) background value.
Similarly, $\mathbf c$ was described as $\mathbf c=[c_s, c_b]$ where $c_s$ 
represented the value of the uniform  annulus and $c_b$ is
the fixed background value outside of the annulus.  
%The value of $\| \mathbf{H}(\mathbf c) \mathbf{A} - \hat{\mathbf p}\|^2$ was plotted as a function of the scale factors 
% $A_s$ and $c_s$. To accomplish this, 
For each $(A_s, c_s)$ pair,
 simulated PA data $ \hat{\mathbf p}$ were computed.
Subsequently, the value of  $\| \mathbf{H}(\mathbf c) \mathbf{A} - \hat{\mathbf p}\|^2$ 
was computed
and plotted as a function of the scale factors of
 $A_s$ and $c_s$.
%This procedure was repeated for the case  of uniform SOS distribution ($c_s=c_b$).
%In this case, the constant value describing the uniform SOS distribution was varied.

% with the same disk phantom of 
%$\mathbf A$ and constant $\mathbf c$. Figure \ref{fig:cf} shows
%the cost functions, which were computed with regularization parameters
%$\lambda_1 = \lambda_2 = 0$.

\begin{figure}[h]
\centering
  \subfigure[]{\resizebox{1.5in}{!}{\includegraphics{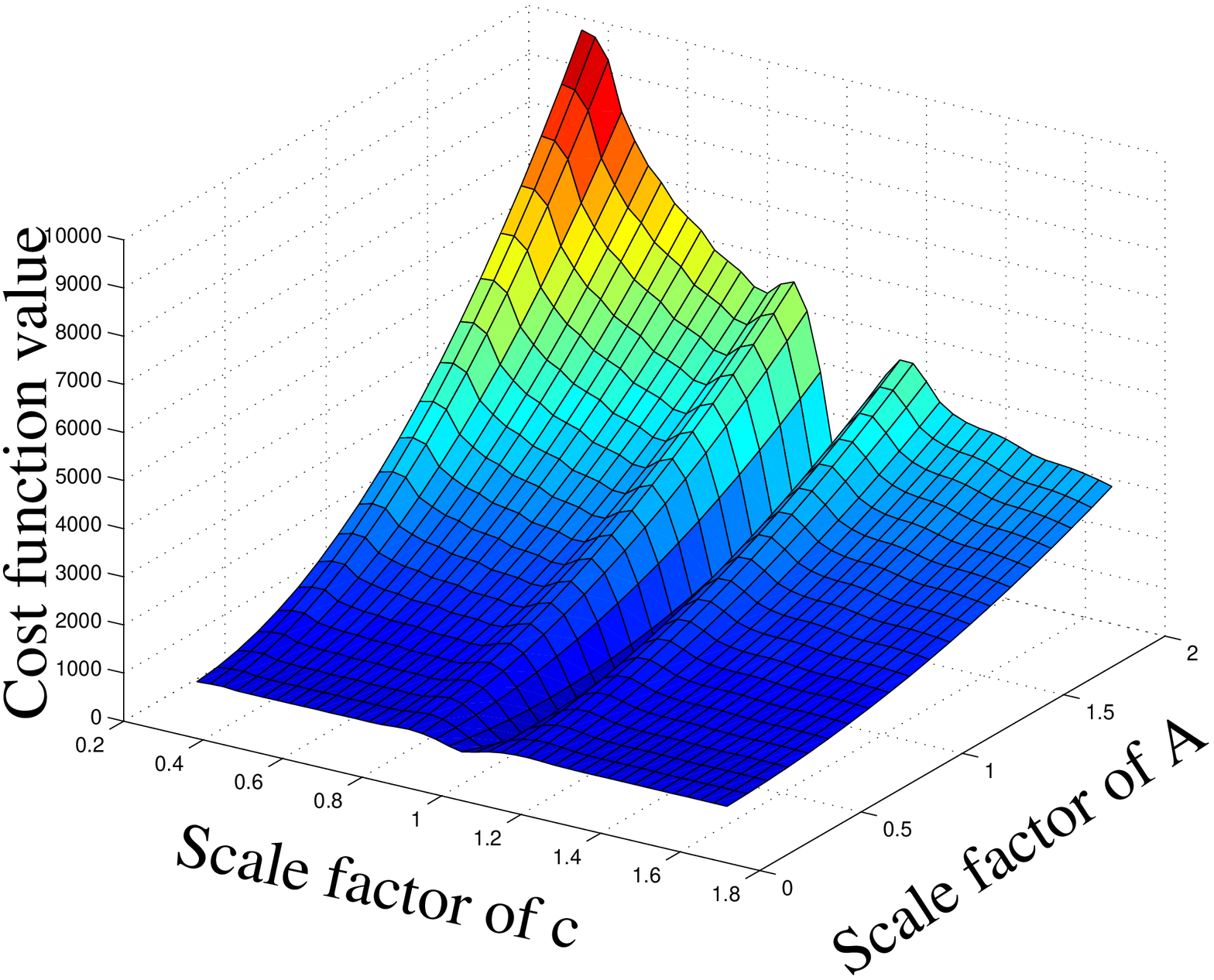}}}
  \subfigure[]{\resizebox{1.5in}{!}{\includegraphics{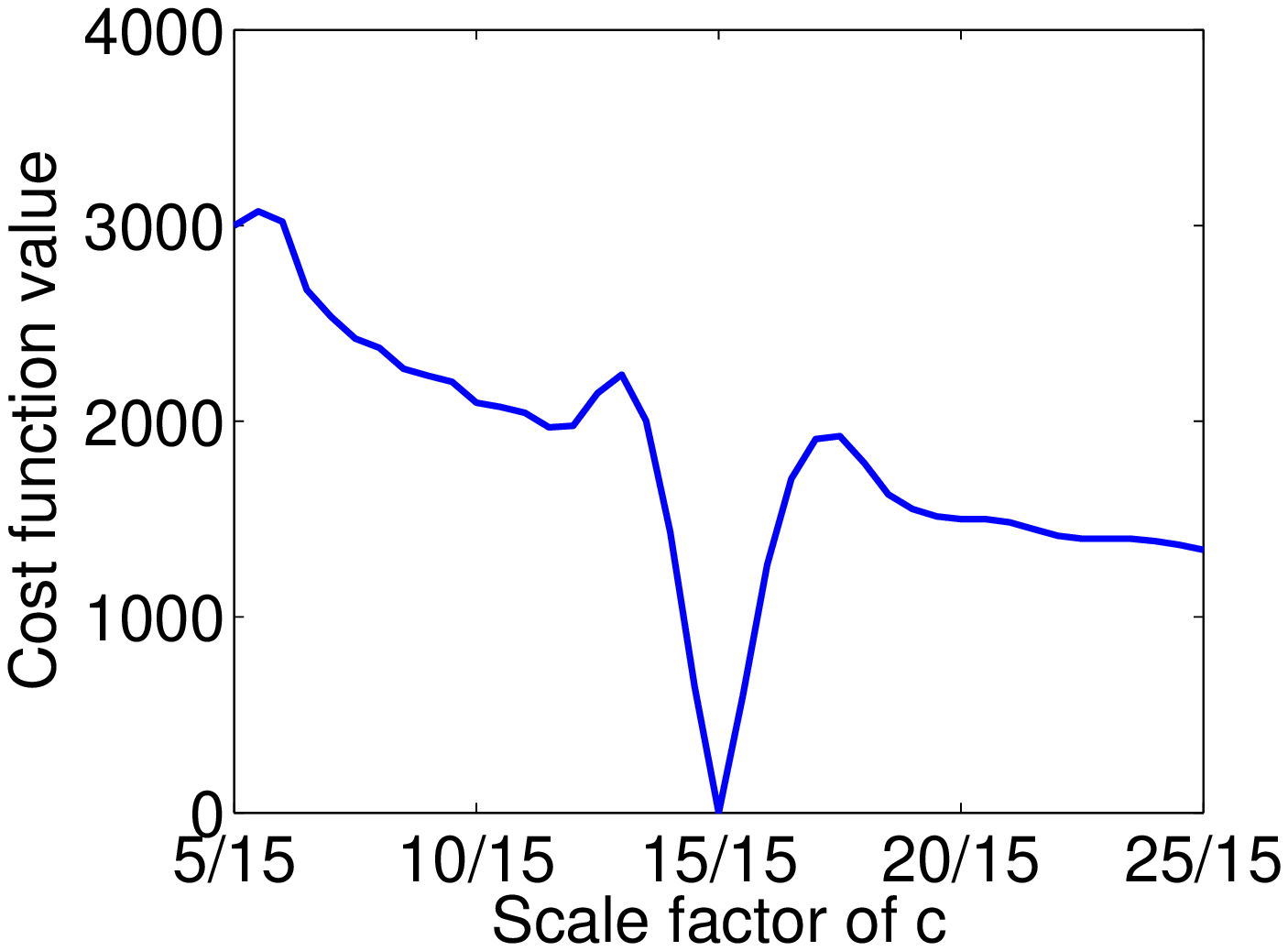}}}\\
%  \subfigure[]{\resizebox{2.7in}{!}{\includegraphics{cf_r9_const.eps}}}
%  \subfigure[]{\resizebox{2.7in}{!}{\includegraphics{cf_r9_const_pfl.eps}}}
\caption{\label{fig:cf}
%Cost functions (without regularizations) correspond to 
%annulus  $\mathbf c$ (first row) and constant  $\mathbf c$ 
%(second row).
Subfigure (a) displays a surface plot of the data fidelity term in the JR
cost function
%term for the cases where the SOS phantom was specified by an annulus
%and a constant value, respectively.
and the corresponding profile corresponding to $A_s=1$ is shown
in subfigure (b). Details are provided in Sec.\ \ref{sect:cf}.
}
\end{figure}

%tom rows of Fig.\  \ref{fig:cf} display 
%the data for cases of the annulus
%and uniform SOS distributions, respectively.
Figure  \ref{fig:cf}(a) displays a surface plot of the
 data fidelity term  $\| \mathbf{H}(\mathbf c) \mathbf{A} - \hat{\mathbf p}\|^2$
as a function of the scale factors of $A_s$ and $c_s$, while profiles
corresponding to $A_s=1$ are shown in subfigure (b).
 The results show
that the data fidelity term is not convex with respect to 
 $\mathbf c_s$. These results suggest the cost function may also be non-convex with respect to
more general and higher-dimensional SOS distributions $\mathbf c$.

\begin{figure}[h]
\centering
  \subfigure[]{\resizebox{1.5in}{!}{\includegraphics{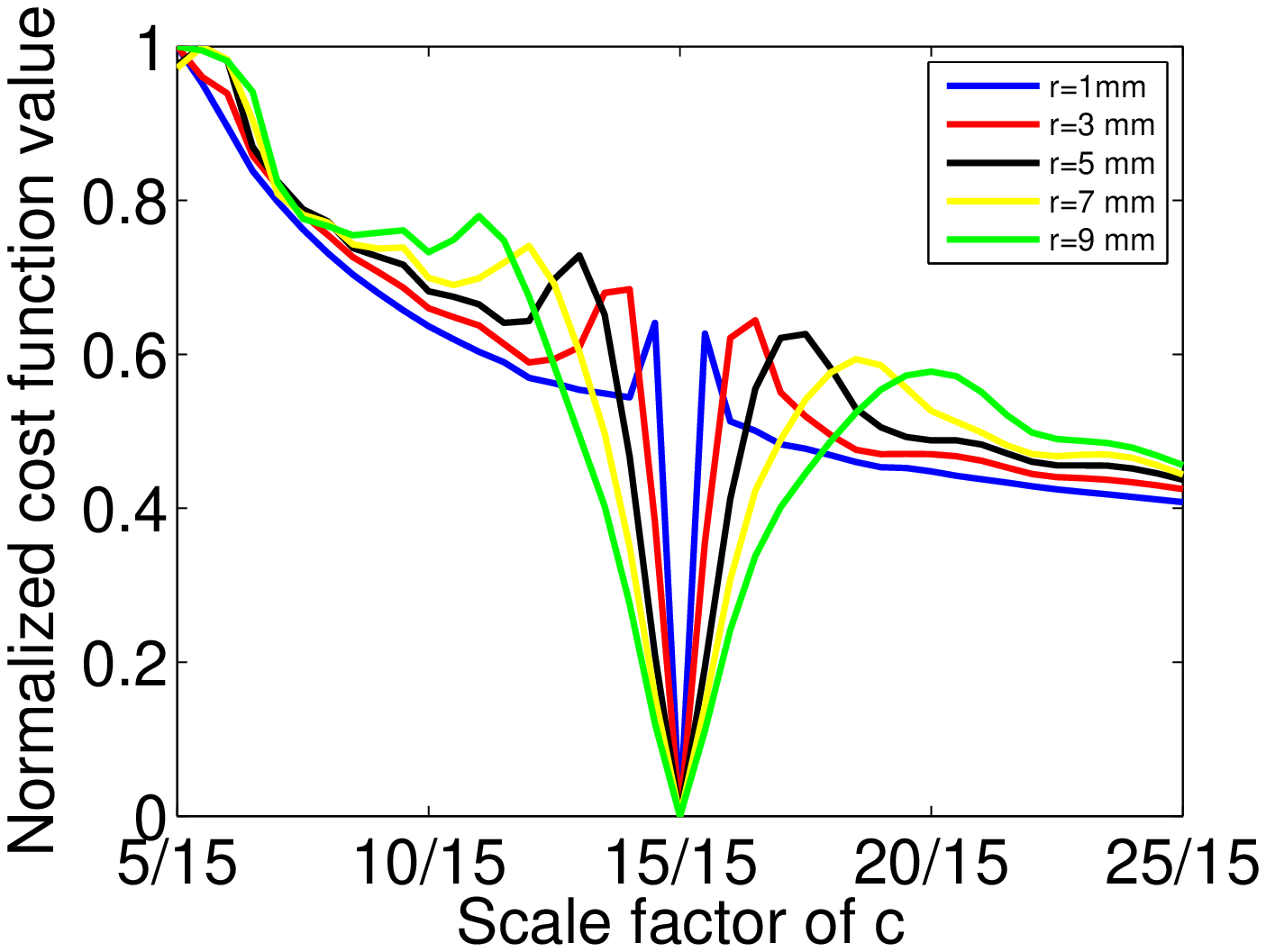}}}
  \subfigure[]{\resizebox{1.5in}{!}{\includegraphics{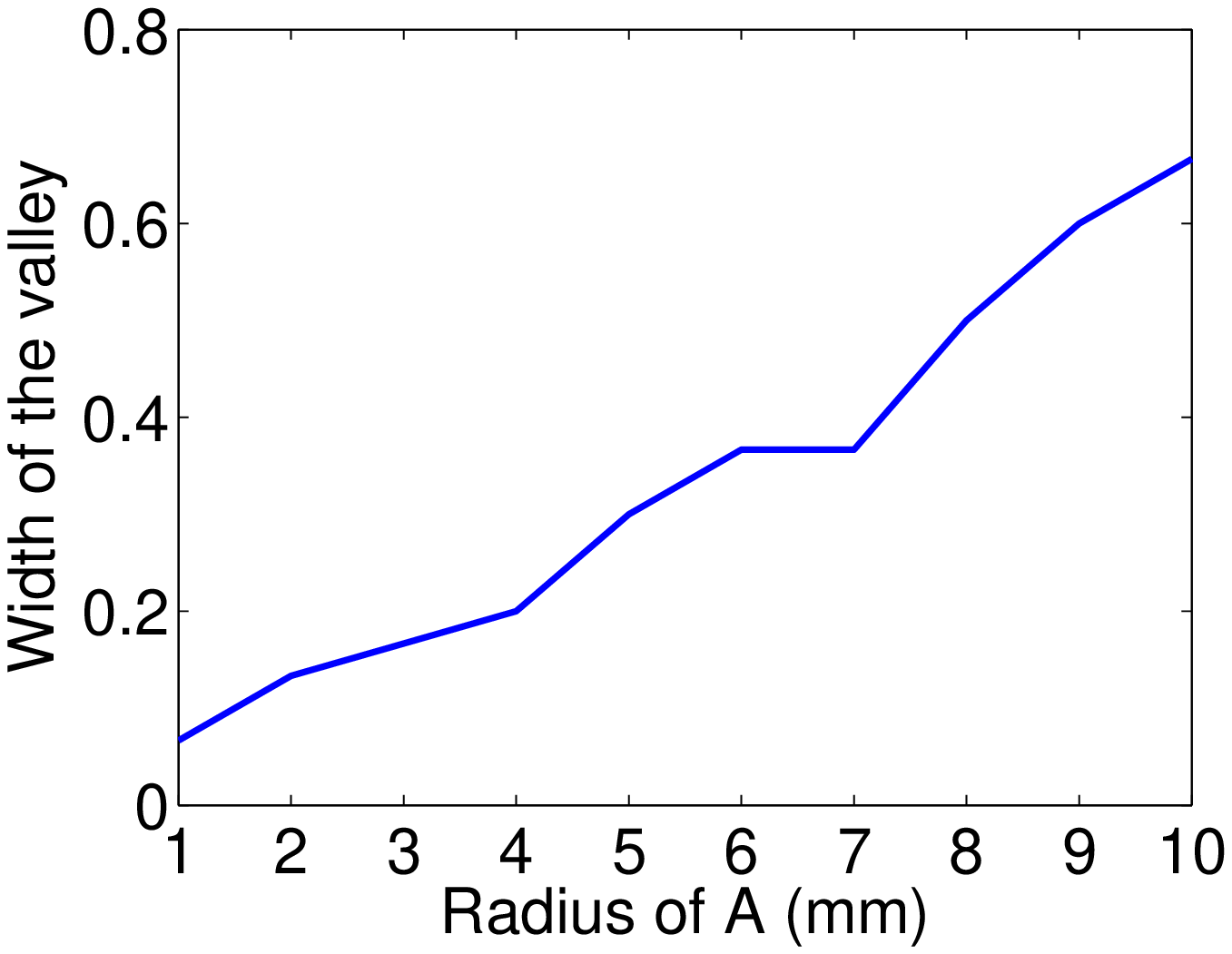}}}\\
%  \subfigure[]{\resizebox{2.7in}{!}{\includegraphics{cf_const_pfl.eps}}}
%  \subfigure[]{\resizebox{2.7in}{!}{\includegraphics{valley_const.eps}}}
\caption{\label{fig:valley}
Profiles of the normalized data fidelity term     
corresponding to $A_s=1$ for different radii of the disk phantom that specified $\mathbf A$  are displayed in  subfigure (a).
Subfigure (b) displays a plot of the width of the valleys in (a)  versus the disk radius.
}
\end{figure}

To investigate how the spatial structure of $\mathbf A$ can influence
the topology of the cost function, the above procedure was 
repeated when the radius of the disk in $\mathbf A$ was varied.
%and the heterogeneous SOS phantom was utilized.
Profiles of the normalized data fidelity term 
corresponding to $A_s=1$  are displayed in  Fig.\ \ref{fig:valley}(a).
% the left column of Fig.\ \ref{fig:valley},
%where subfigures (a) and (c) correspond to the annulus 
%and constant $\mathbf c$, respectively.
These data reveal that 
the `valley' containing the global minimum of the cost function widens 
as the radius of the disk heterogeneity in $\mathbf A$ increases. This can be seen more clearly in 
the plots of the width of the valley versus the disk radius
that are shown in Fig.\ \ref{fig:valley}(b).
 This observation 
suggests that the quality of the initial estimate for $\mathbf c$ 
can be relaxed  when the radius of the support of $\mathbf A(\mathbf r)$ increases.
%hich is consistent with the heuristic support condition described below.

In summary, these results establish that, for a fixed $\mathbf A$, the JR cost function is generally
non-convex with respect to $\mathbf c$.  Accordingly, Sub-Problem \#2 is generally
a non-convex problem.  Sub-Problem \#1 is convex if the penalty is convex.
  The overall non-convexity of the JR problem indicates that accurate initial
estimates of $\mathbf c$ will generally be necessary to avoid local minima that represent
inaccurate solutions.  Additionally, the non-convexity of the problem
suggests that uniqueness of the solution is not guaranteed.
%\bl{[I'm not sure how the non-convexity of the problem
%is related to the uniqueness of the solution]}

%\subsection{Conditions for accurate reconstruction of $\mathbf c$ given $\mathbf A$ (sub-problem (\ref{eq:cost_c}))}
\subsection{Numerical investigations of Sub-Problem \#2}
\label{sect:conditions}

As described above, Sub-Problem \#2 corresponds to a non-convex optimization problem that
can be difficult to solve in practice. Accordingly, errors that arise when solving
Sub-Problem \#2 can accumulate in Algorithm 1 and hinder the ability to perform accurate JR.
Below, numerical experiments are reported that reveal insights into some mathematical
properties of  $A(\mathbf r)$ and $c(\mathbf r)$ that influence the ability to accurately
solve Sub-Problem \#2.

\subsubsection{Effect of spatial supports of $A(\mathbf r)$ and $c(\mathbf r)$}
\label{sect:support}
%Numerical phantoms representing `adequate' 
%$\mathbf A$ (left column of Fig. \ref{fig:goodA})
%and `defective' $\mathbf A$ (left column of 
%Fig. \ref{fig:defectiveA}) were chosen

\begin{figure}[h]
  \centering
    \resizebox{1.5in}{!}{
      \includegraphics{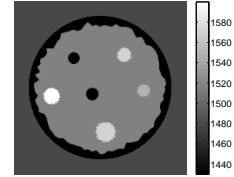}
  }
  \caption{\label{fig:breastSOS}
The numerical SOS phantom described in Sec.\ \ref{sect:conditions}
that was utilized in the investigations of Sub-Problem \#2.
}
\end{figure}

\begin{figure}[h]
\centering
  \subfigure[]{\resizebox{1.1in}{!}{\includegraphics{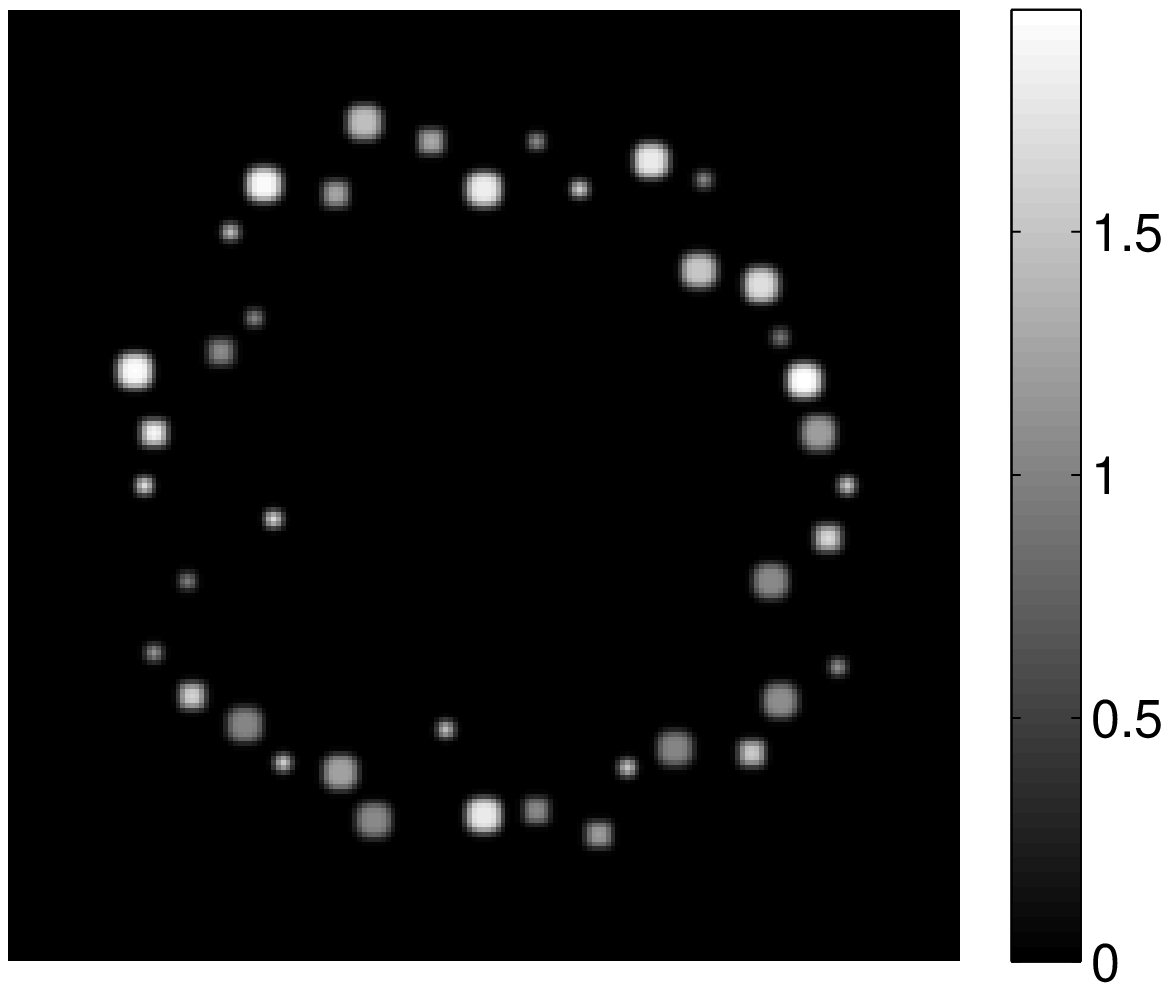}}}
  \subfigure[]{\resizebox{1.1in}{!}{\includegraphics{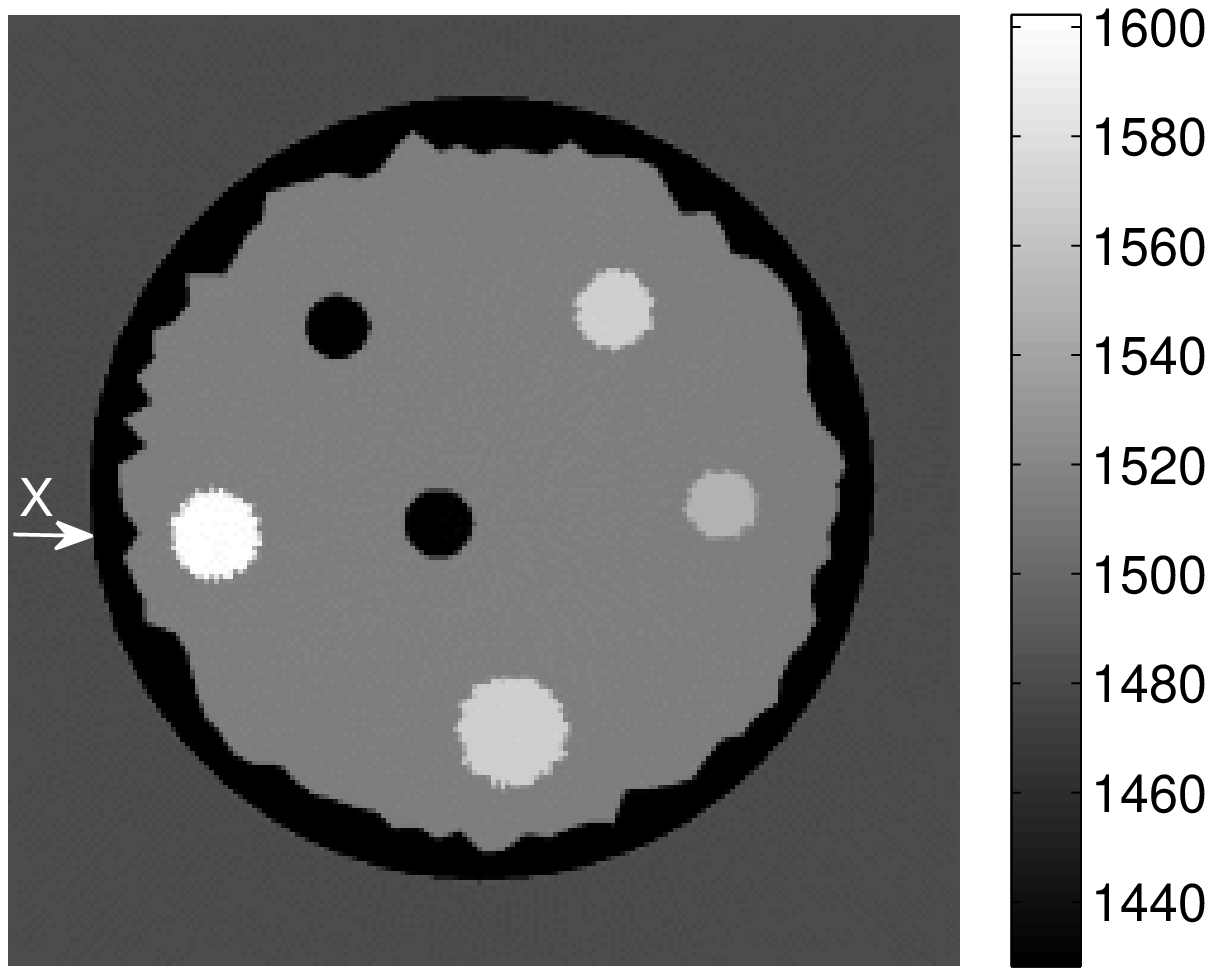}}}
  \subfigure[]{\resizebox{1.1in}{!}{\includegraphics{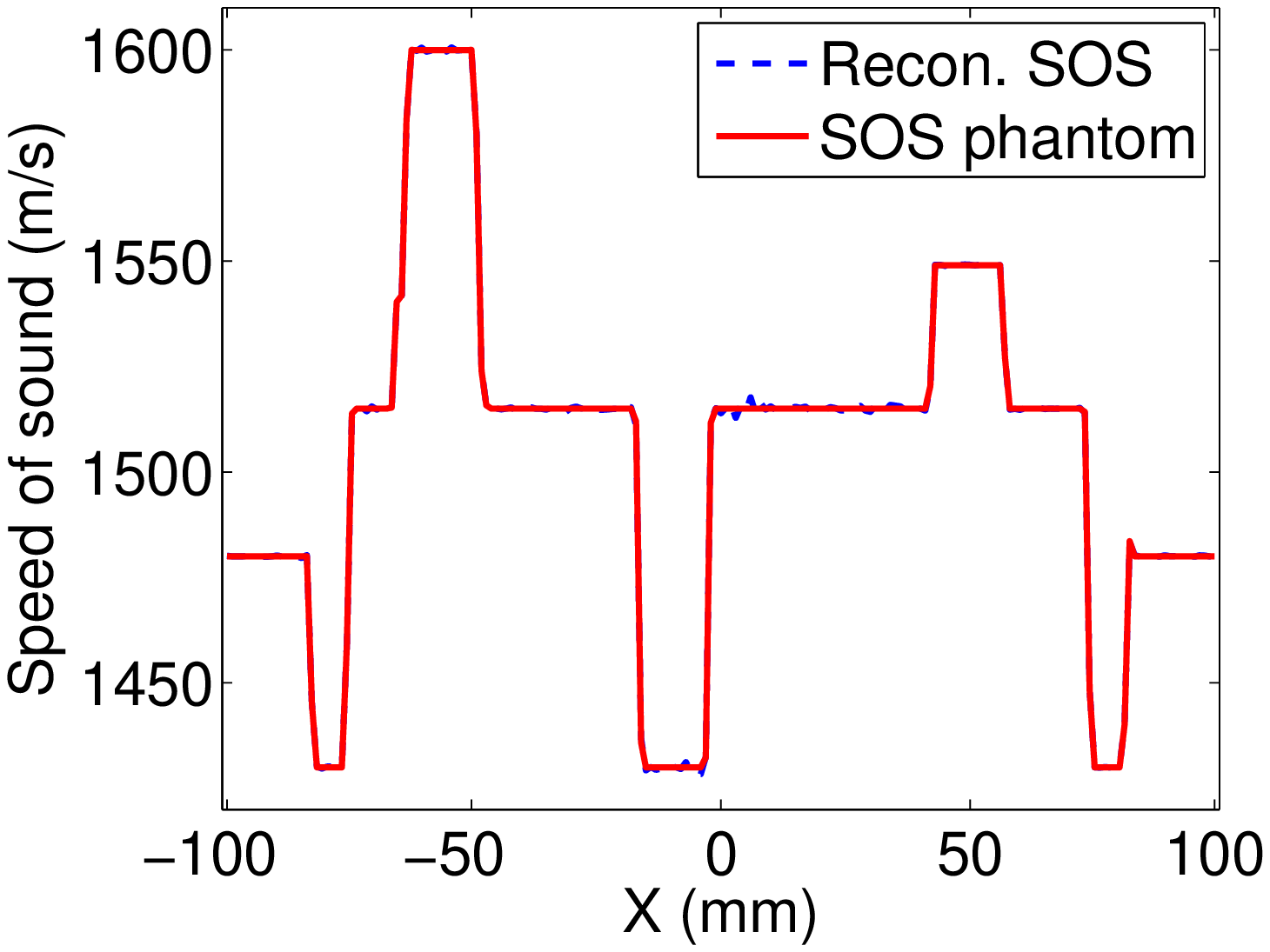}}}\\
\vskip -0.1cm
  \subfigure[]{\resizebox{1.1in}{!}{\includegraphics{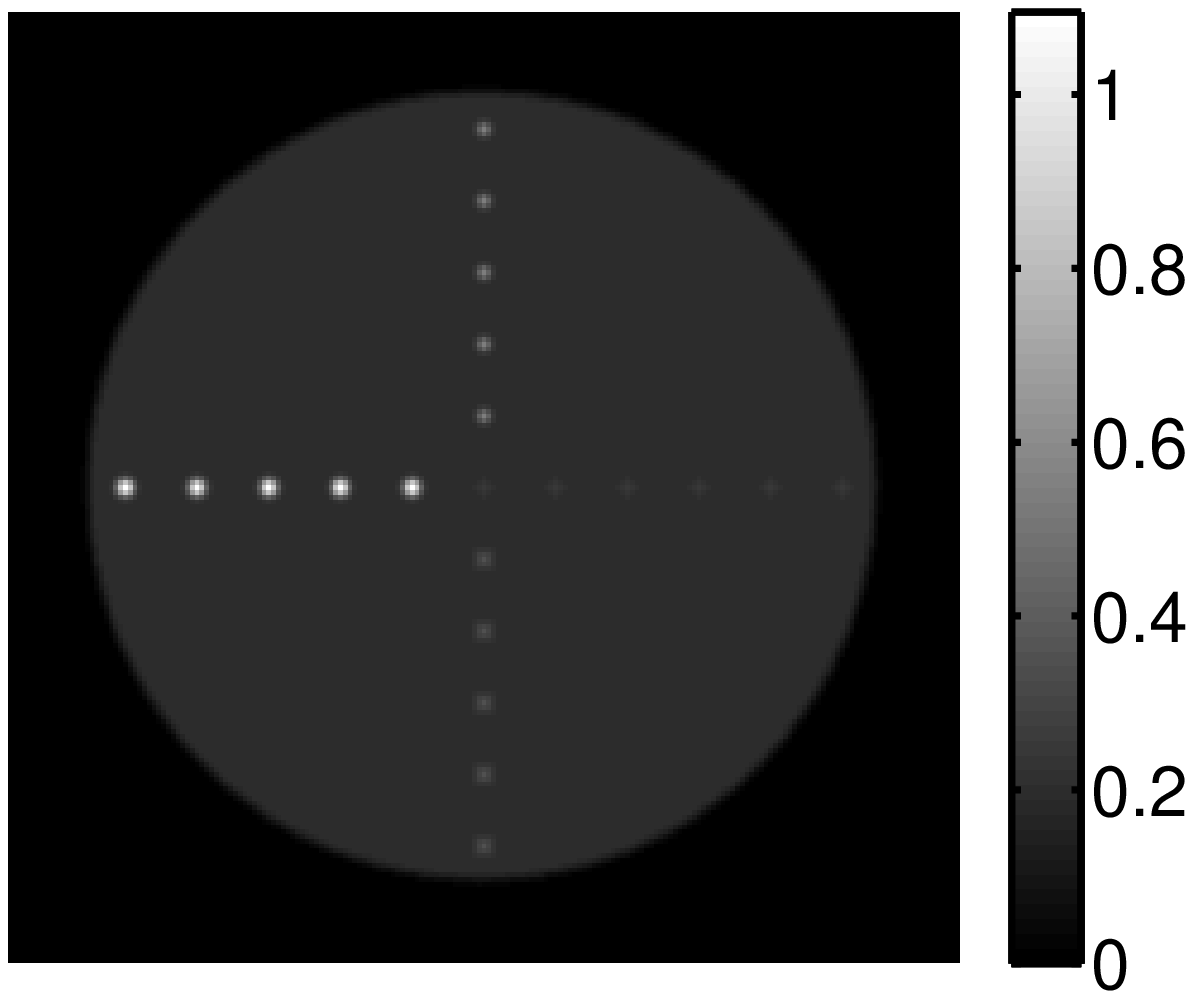}}}
  \subfigure[]{\resizebox{1.1in}{!}{\includegraphics{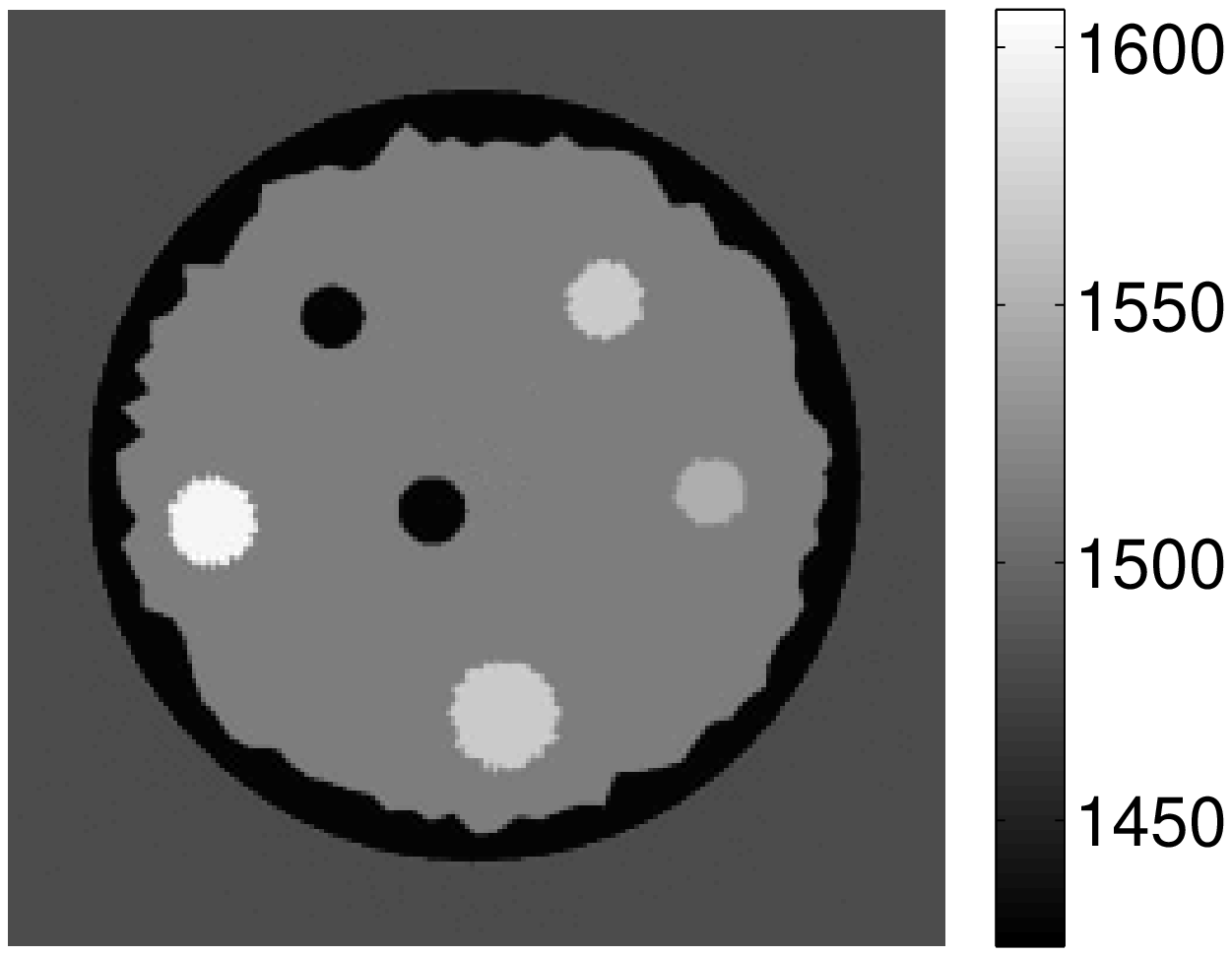}}}
  \subfigure[]{\resizebox{1.1in}{!}{\includegraphics{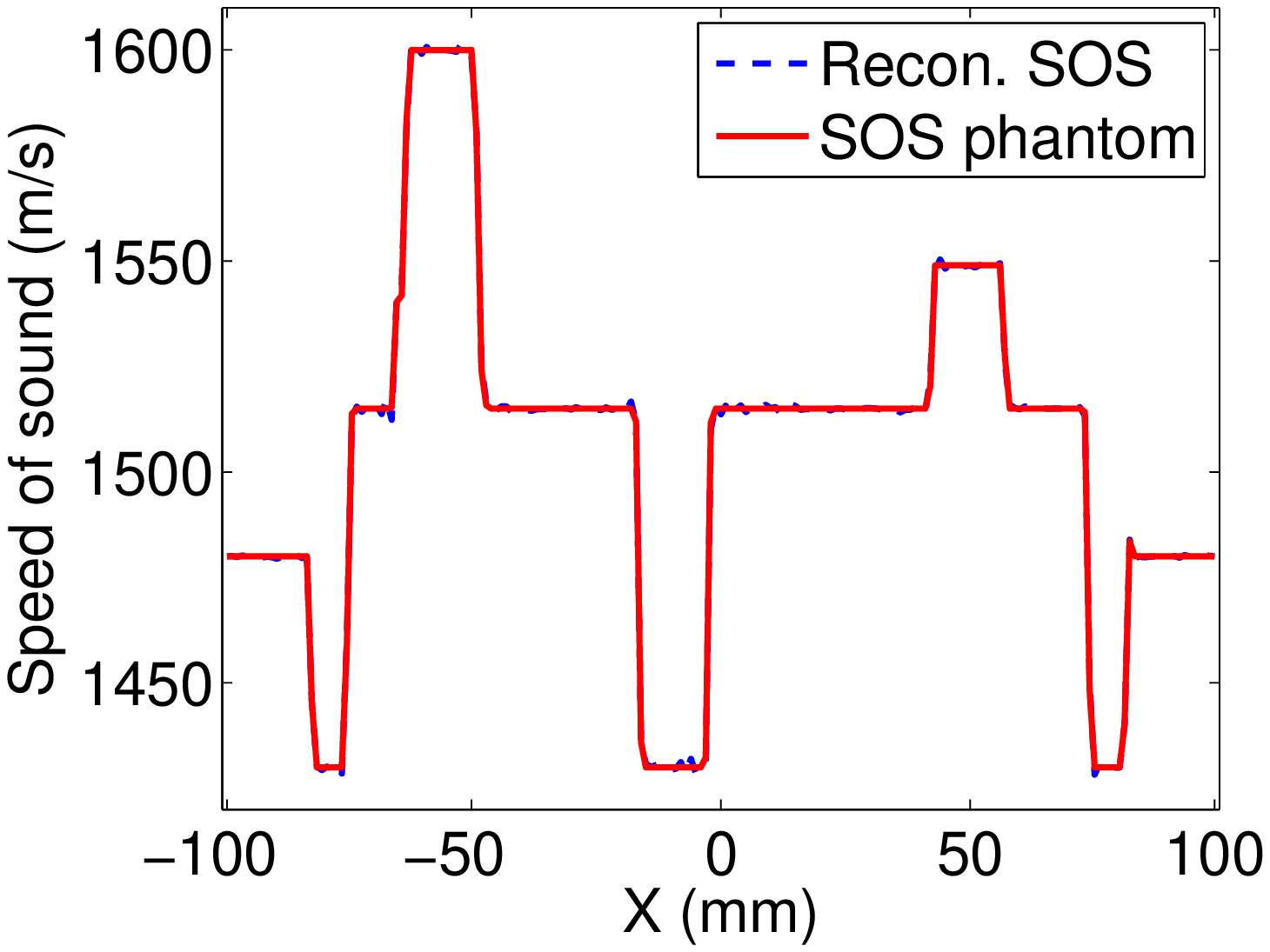}}}
\caption{\label{fig:goodA}
Numerical investigations of Sub-Problem \#2 - Spatial support effects:
Two different phantoms $\mathbf A$ that satisfy the support conjecture are shown
in subfigures (a) and (d).  The estimates of $\mathbf c$
obtained by solving Sub-Problem \#2 by use of noiseless data and the
corresponding $\mathbf A$
are shown in subfigures (b) and (e). Image profiles through the estimates
of $\mathbf c$ are shown in subfigures (c) and (f).
}
\end{figure}

Studies were conducted to investigate
the extent to which the support conjecture, provided in Sec.\ \ref{sect:supportTheory},
influences the ability to accurately solve Sub-Problem \#2, and hence perform JR,
 by use of perfect measurements.
The numerical phantom employed to represent
the SOS distribution, $\mathbf c$, is shown in Fig.\ \ref{fig:breastSOS}.
The SOS values were representative of human breast tissues.
We first considered two choices for $\mathbf A$, shown in
Figs.\   \ref{fig:goodA}(a) and  \ref{fig:goodA}(d),
which were designed to satisfy the support conjecture 
($\text{supp}(c)$ {is enclosed by} $\text{supp}(A)$).
For each choice of $\mathbf A$, a corresponding estimate of $\mathbf c$ was estimated
by solving Sub-Problem \#2 from noiseless simulated PACT measurements.
The value of the regularization parameter $\lambda_c$ in Eq.\ (\ref{eq:cost_c})
was zero.
The reconstructed estimates of $\mathbf c$ corresponding to the two
choices of $\mathbf A$ are shown in  Figs.\ \ref{fig:goodA}(b)
and \ref{fig:goodA}(c). The corresponding profiles extracted
from the central rows of the reconstructed estimates are displayed in Figs.\
 \ref{fig:goodA}(c) and  \ref{fig:goodA}(f), respectively.
These results  reveal
that it is possible to reconstruct accurate estimates of
$\mathbf c$ from perfect PACT measurements via Sub-Problem \#2 when the specified $\mathbf A$
satisfies the support conjecture. The effects of noise and other measurement errors are
addressed later.

\begin{figure}[h]
\centering
  \subfigure[]{\resizebox{1.1in}{!}{\includegraphics{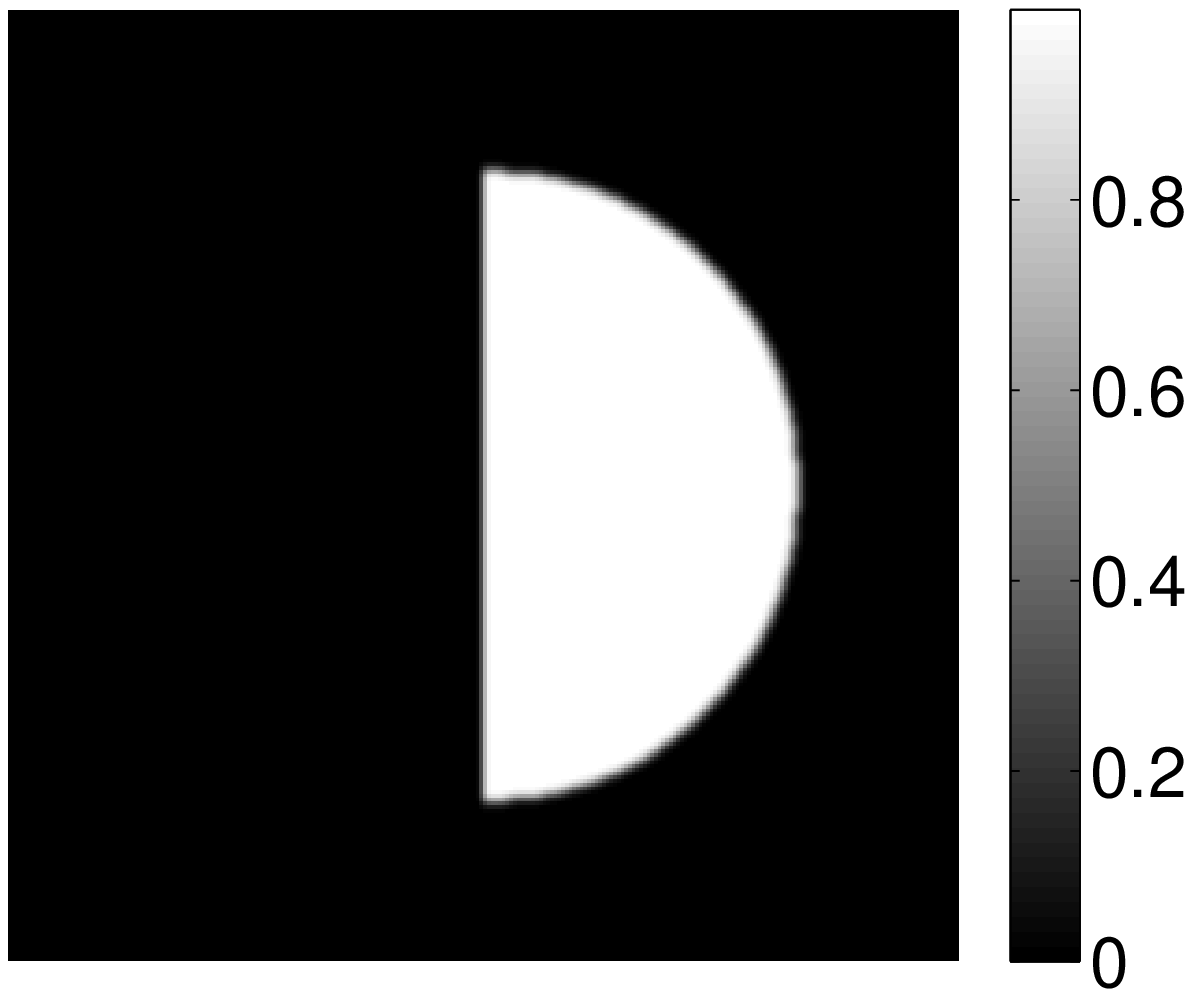}}}
  \subfigure[]{\resizebox{1.1in}{!}{\includegraphics{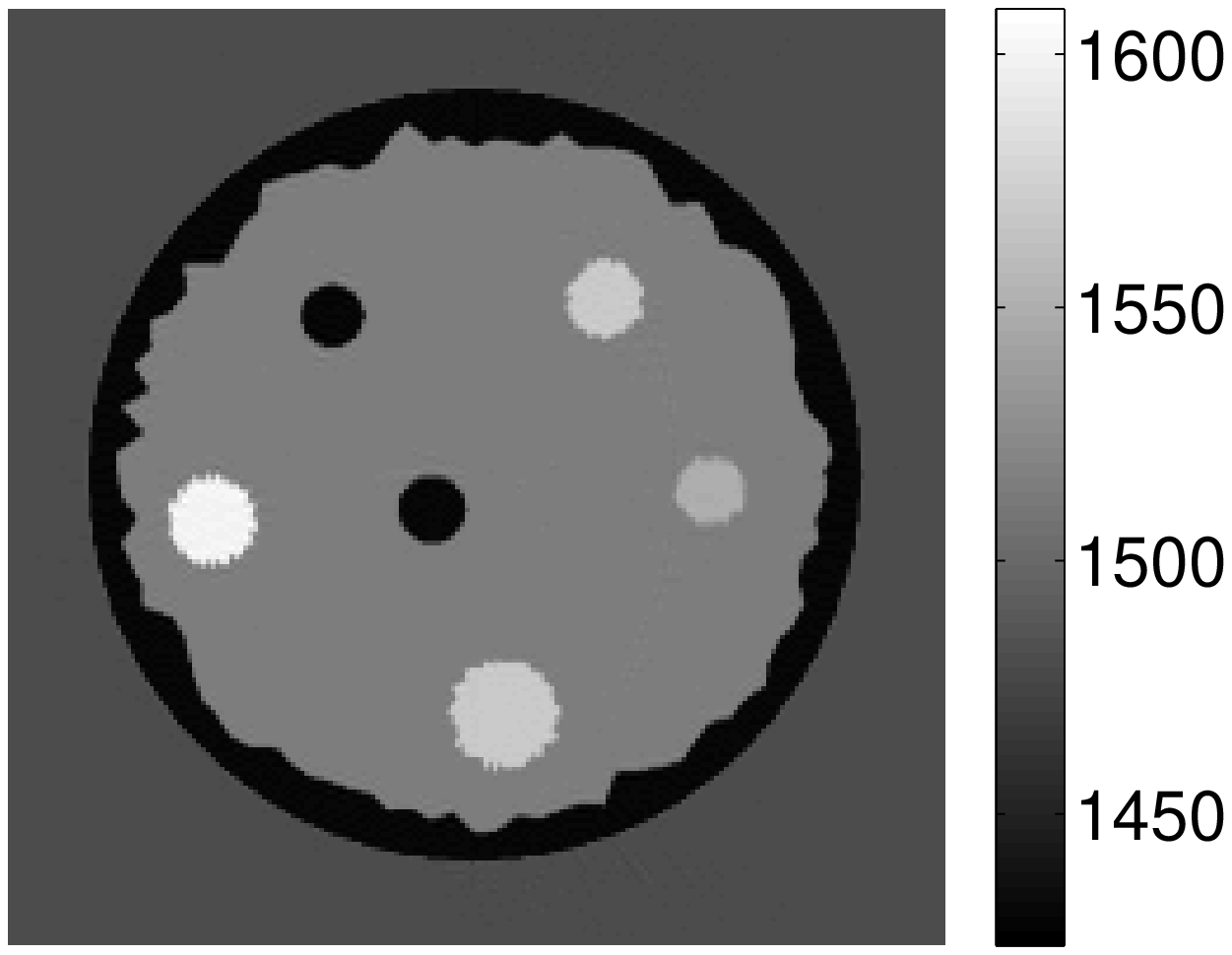}}}
  \subfigure[]{\resizebox{1.1in}{!}{\includegraphics{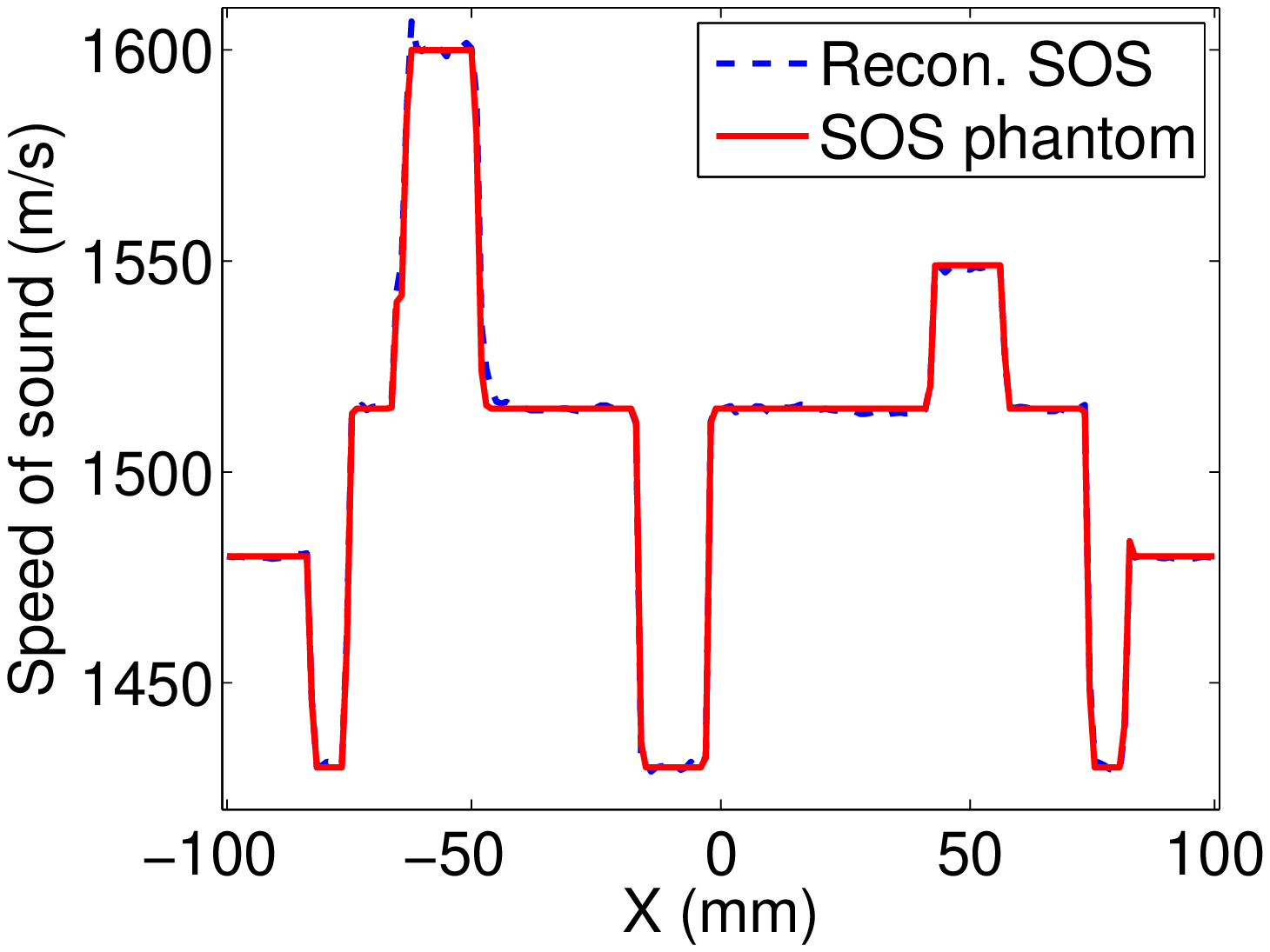}}}\\
\vskip -0.1cm
  \subfigure[]{\resizebox{1.1in}{!}{\includegraphics{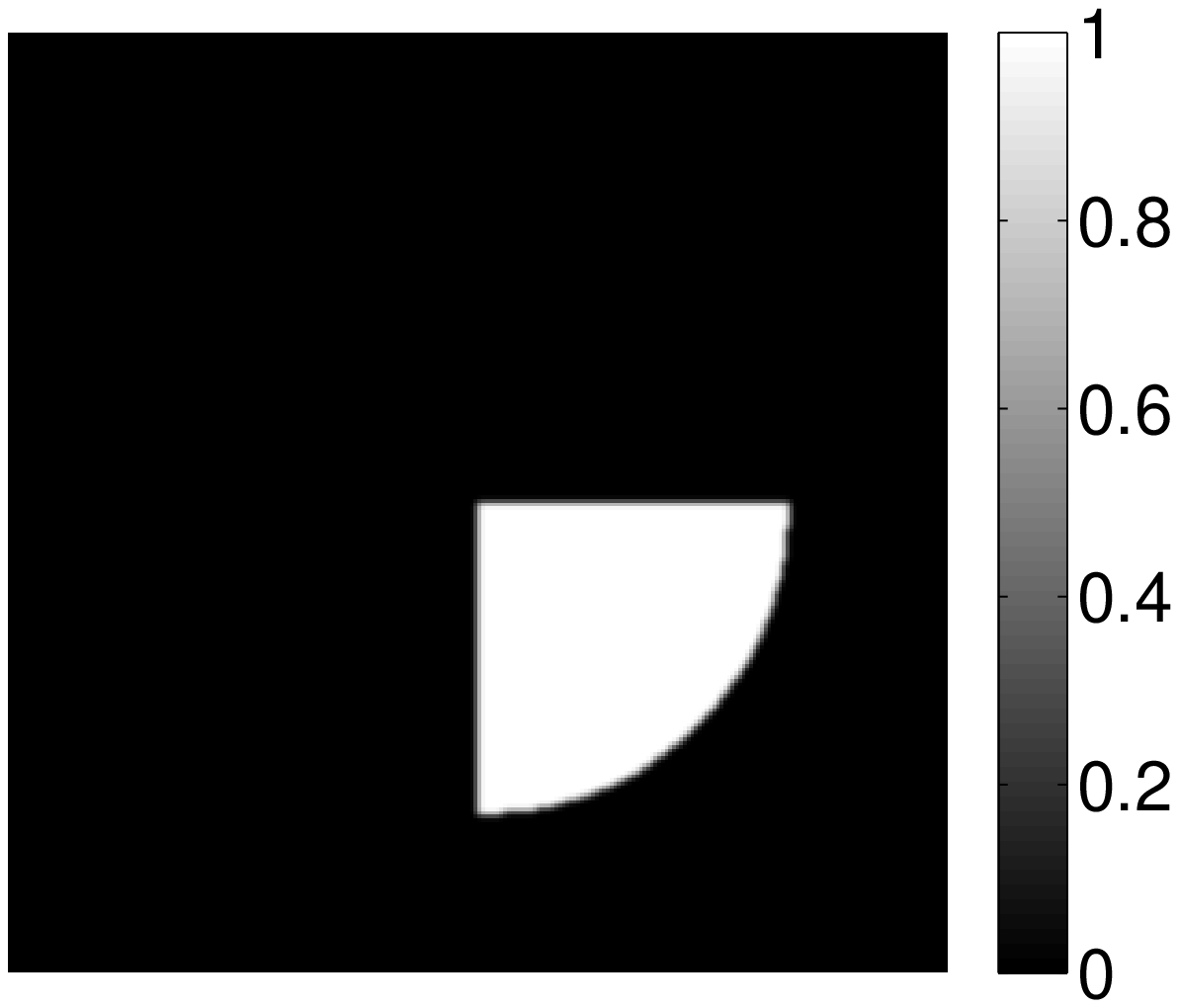}}}
  \subfigure[]{\resizebox{1.1in}{!}{\includegraphics{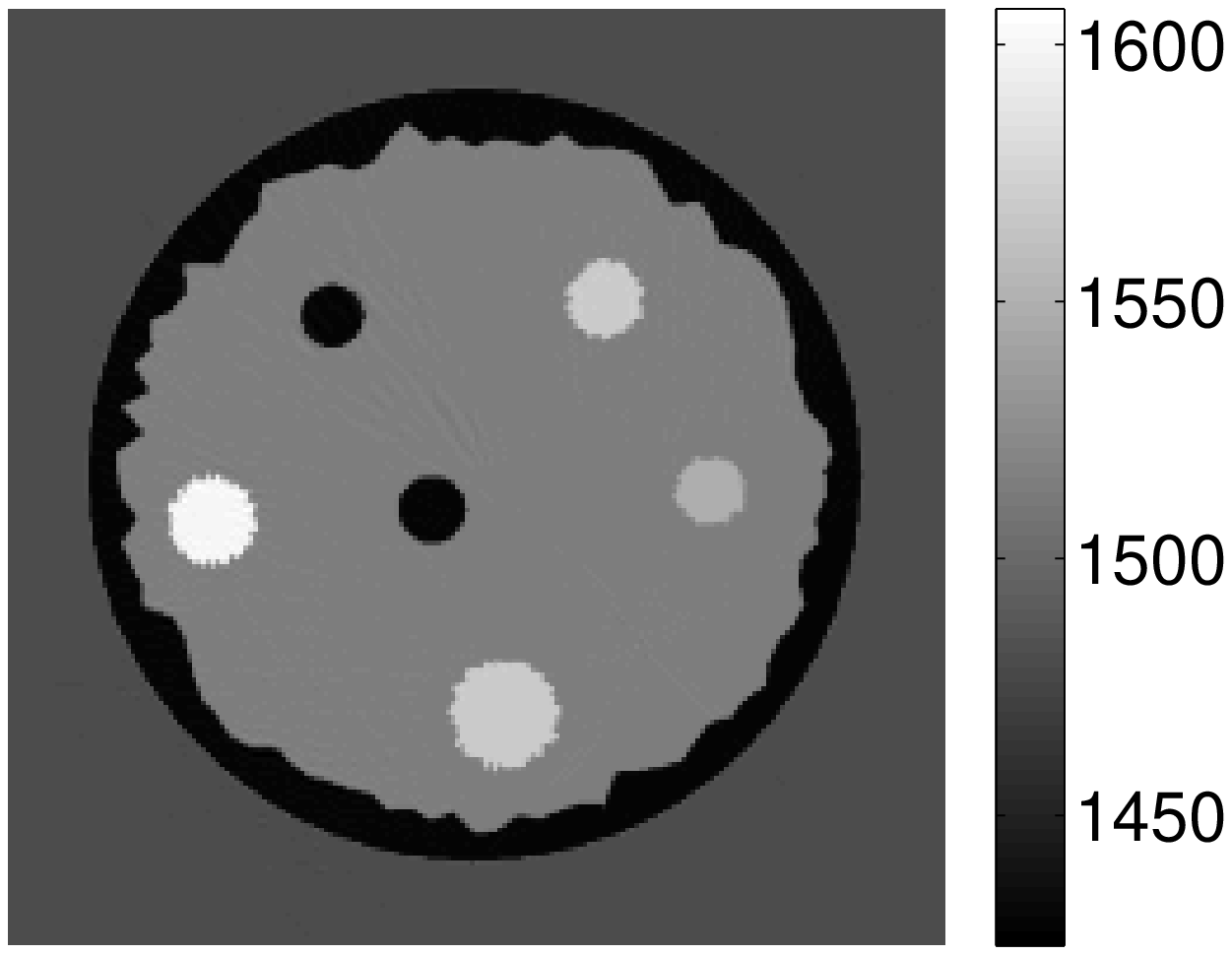}}}
  \subfigure[]{\resizebox{1.1in}{!}{\includegraphics{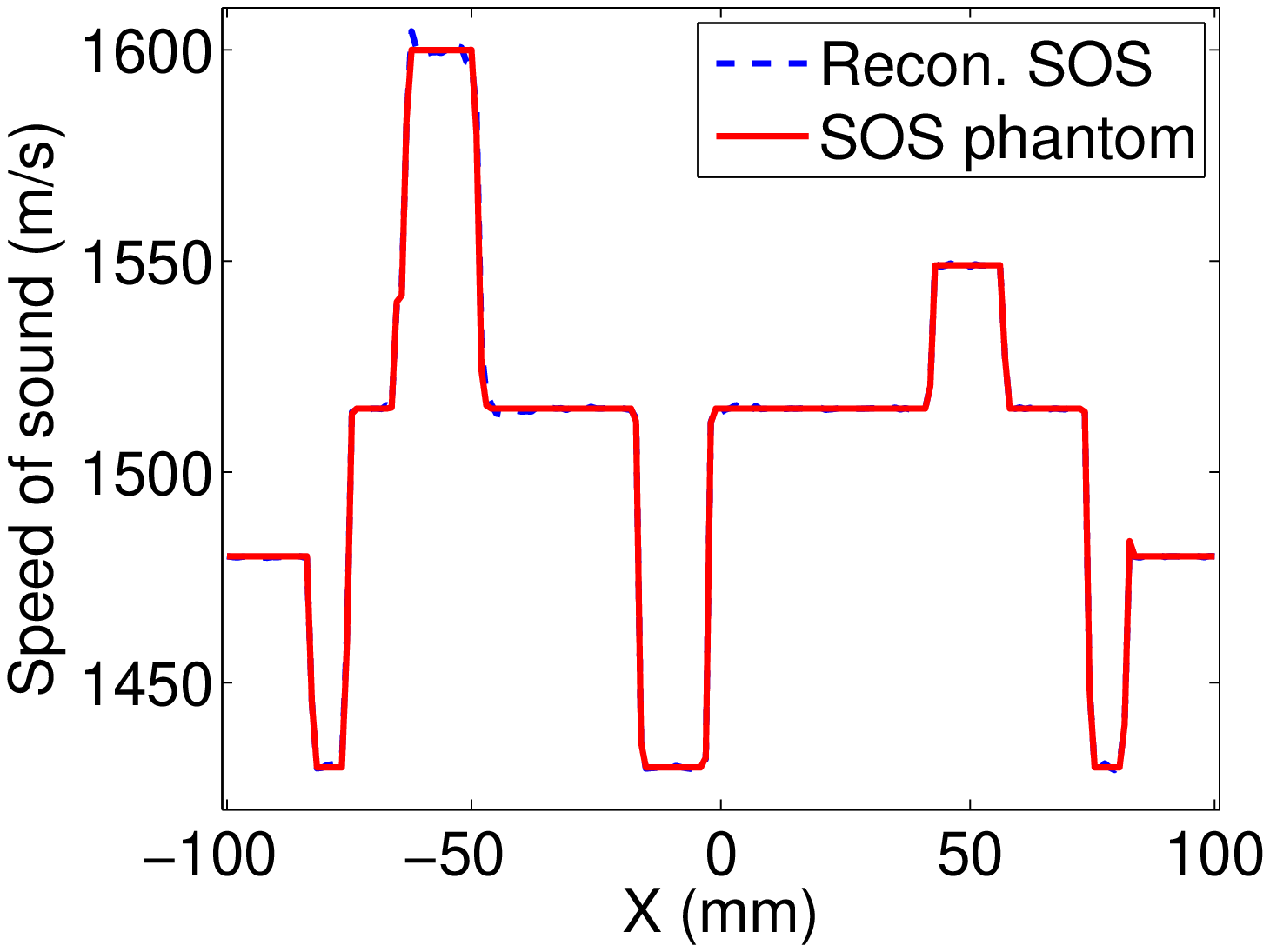}}}\\
\vskip -0.1cm
  \subfigure[]{\resizebox{1.1in}{!}{\includegraphics{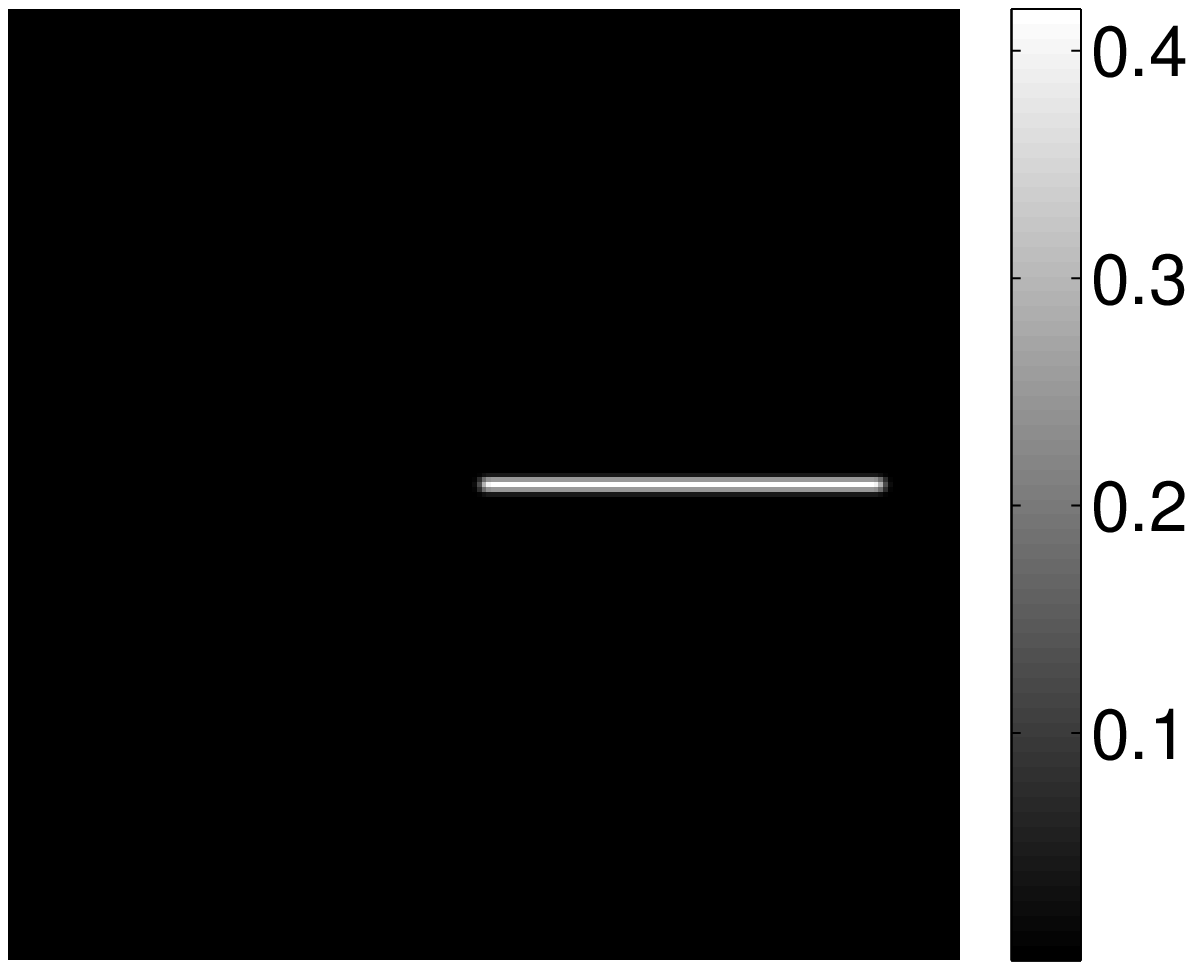}}}
  \subfigure[]{\resizebox{1.1in}{!}{\includegraphics{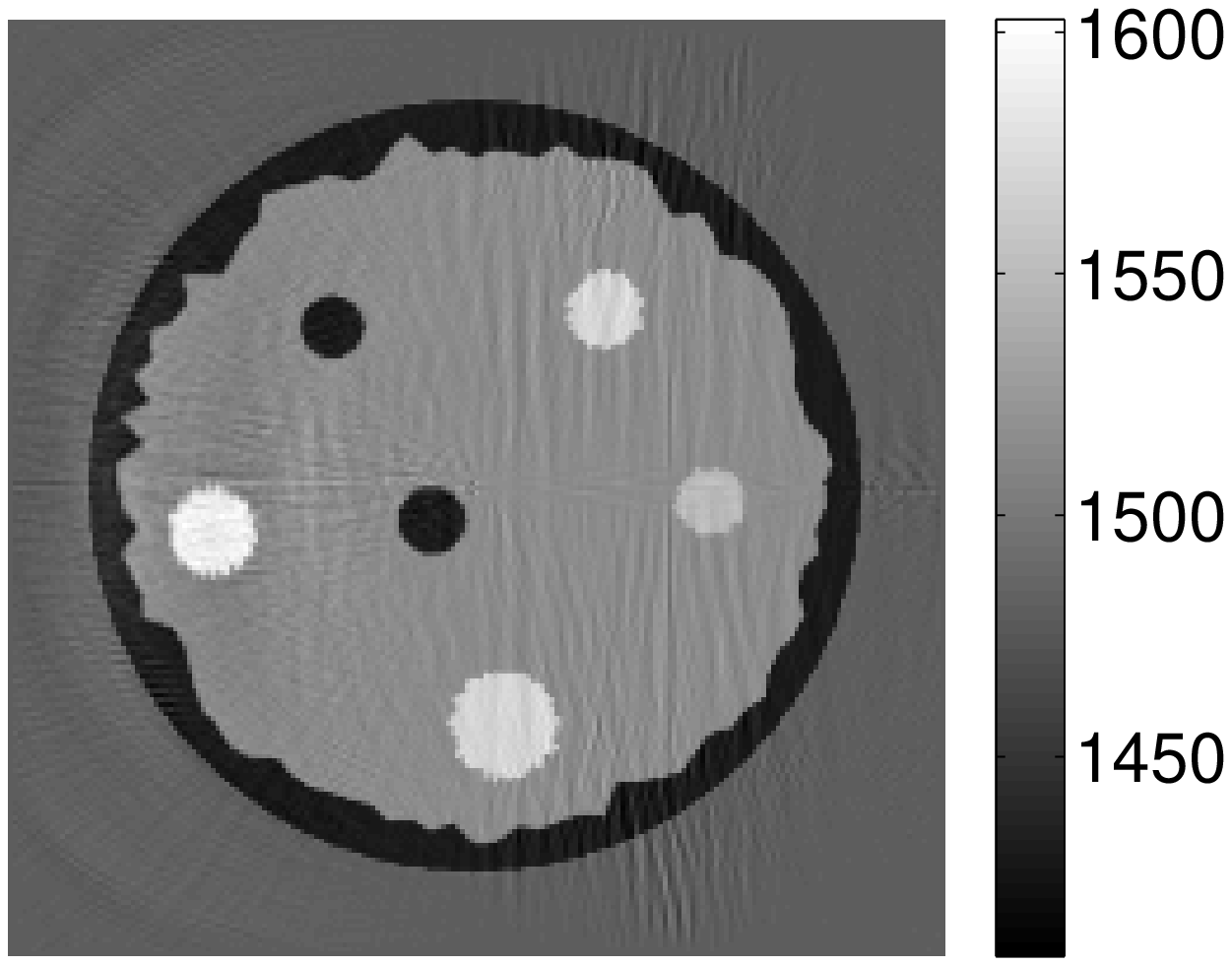}}}
  \subfigure[]{\resizebox{1.1in}{!}{\includegraphics{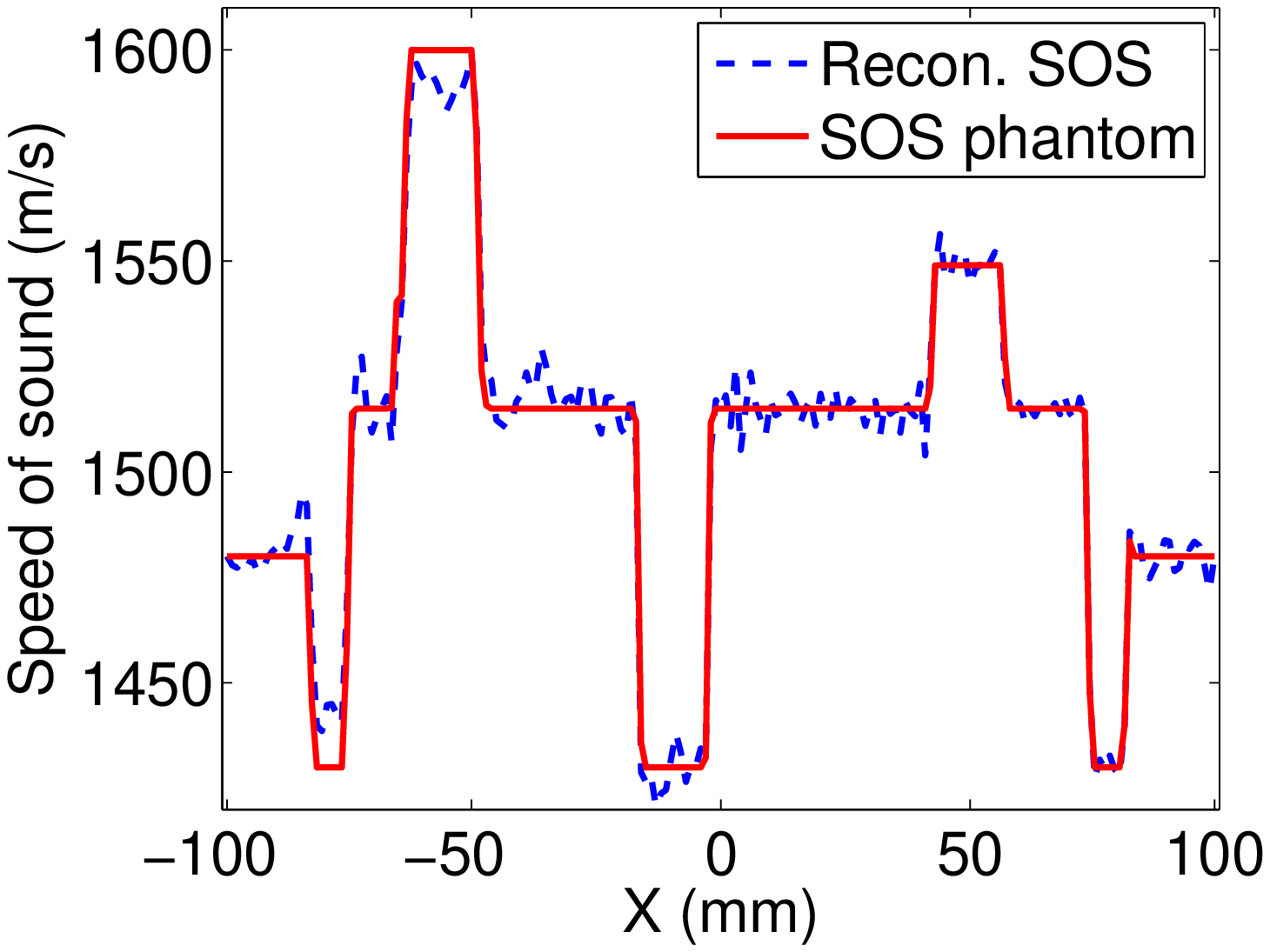}}}\\
\vskip -0.1cm
  \subfigure[]{\resizebox{1.1in}{!}{\includegraphics{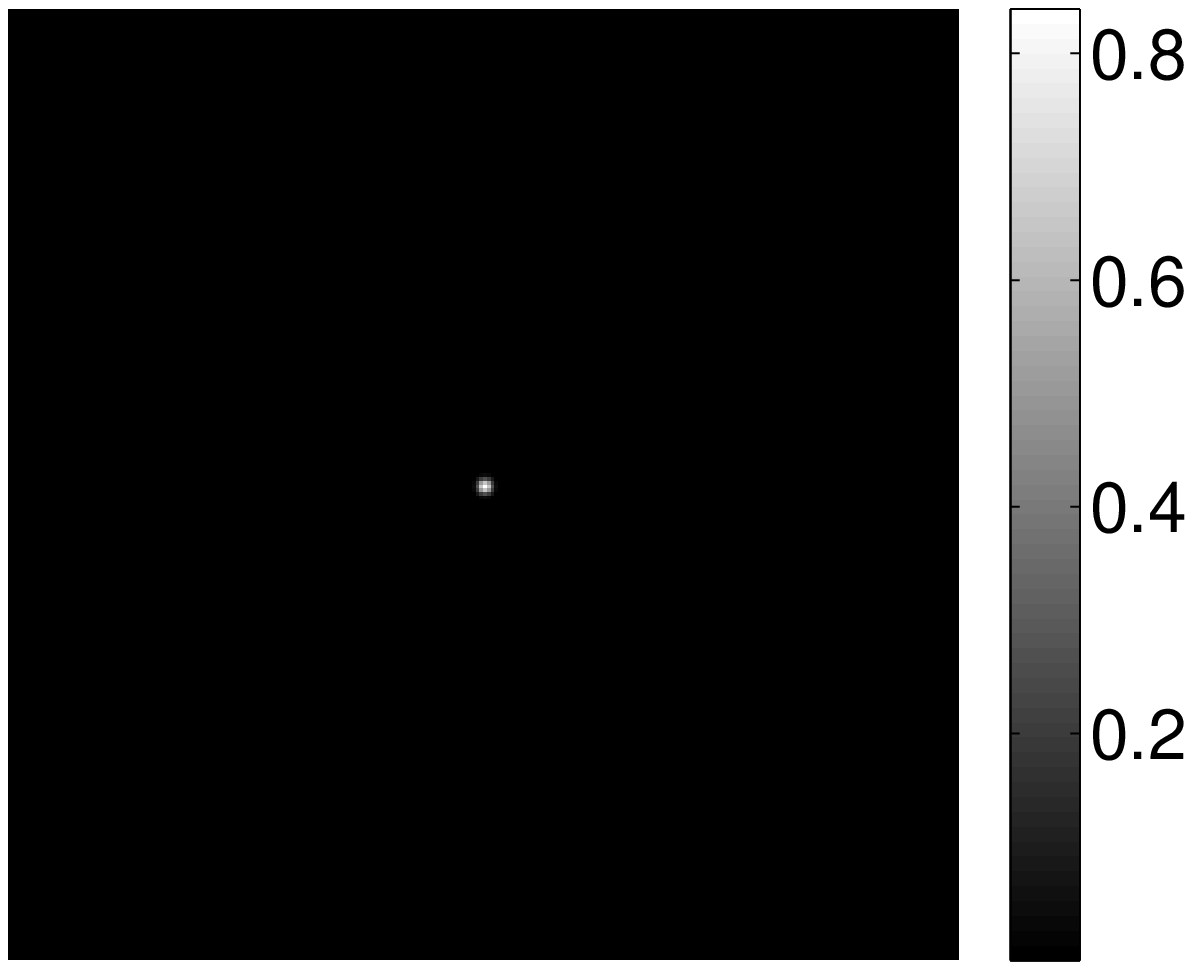}}}
  \subfigure[]{\resizebox{1.1in}{!}{\includegraphics{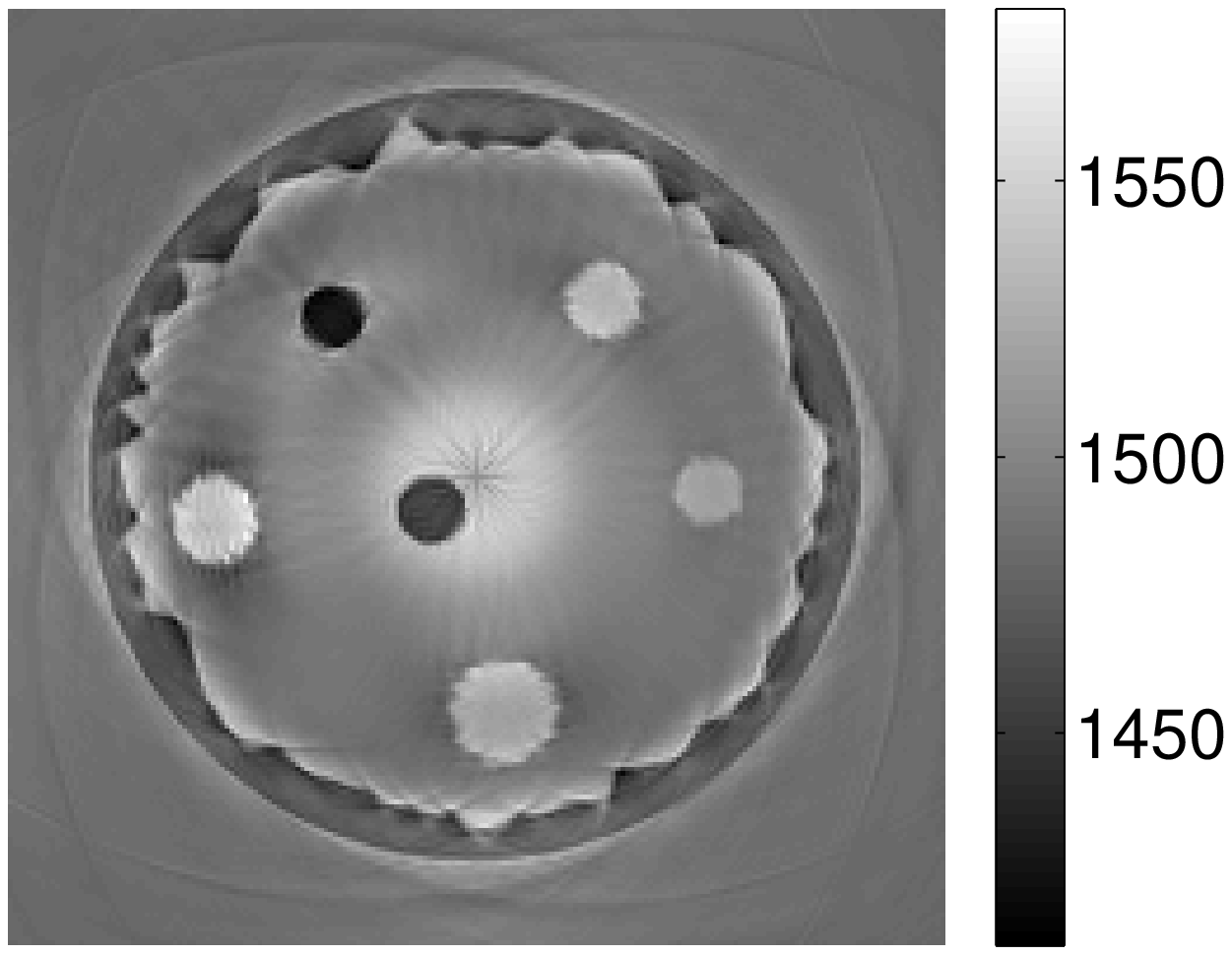}}}
  \subfigure[]{\resizebox{1.1in}{!}{\includegraphics{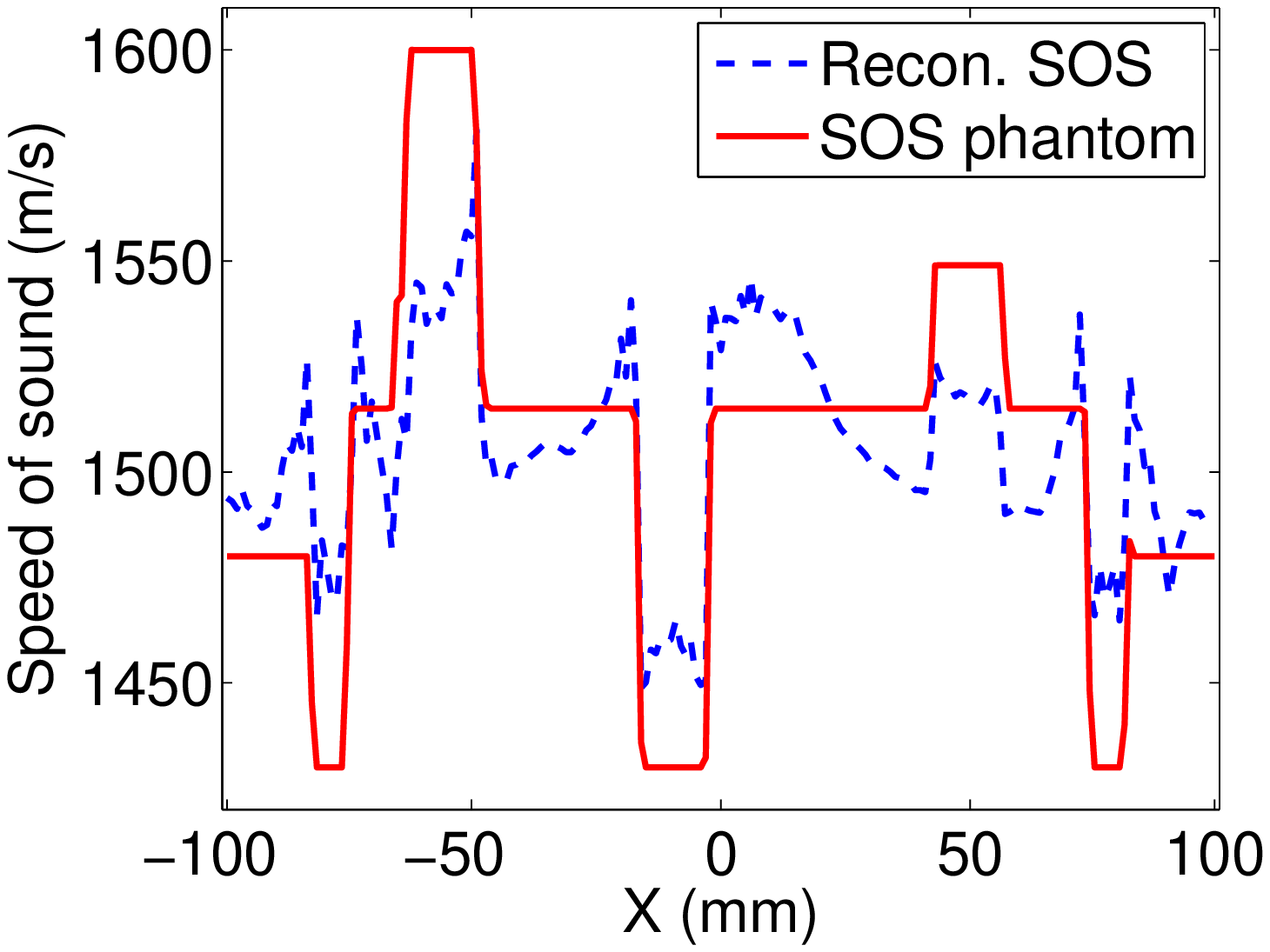}}}
\caption{\label{fig:defectiveA}
Numerical investigations of Sub-Problem \#2 - Spatial support effects:
Four different  phantoms $\mathbf A$ that violated the support conjecture are shown
in subfigures (a), (d), (g), and (j).  The  estimates of $\mathbf c$
obtained by solving Sub-Problem \#2 by use of noiseless data and the
corresponding $\mathbf A$
are shown in subfigures (b), (e), (h), and (k). Image profiles through the estimates
of $\mathbf c$ are shown in subfigures (c), (f), (i), and (l).
}
\end{figure}

We next considered the four
choices for $\mathbf A$, shown in the
left column of Fig.\ \ref{fig:defectiveA} (Figs.\ \ref{fig:defectiveA}(a), (d), (g), and (j)),
which, roughly speaking, were designed to violate the support conjecture to different extents.
%In order to exclude the effect of the 
%spatial spectrum of $\mathbf A$, both `adequate' 
%and `defective' $\mathbf A$ employed here have
%sharp boundaries so that the k-space condition
%discussed in Section \ref{sect:k-space} 
%is satisfied.
As above, corresponding unregularized estimates of $\mathbf c$ were estimated
by solving Sub-Problem \#2 from noiseless simulated PACT measurements.
The reconstructed estimates of $\mathbf c$ corresponding to the 
different choices of $\mathbf A$ are shown in the middle column of
  Fig.\ \ref{fig:defectiveA}.
% (subfigures (b), (e), (h), and (k)), and \ref{fig:goodA}(c).
 The corresponding profiles extracted 
from the central rows of the reconstructed estimates are displayed in the
right column.
% (Figs.\ \ref{fig:defectiveA}(c), (f), (i), and (l)). 
%The RMSE between the SOS phantom 
%in Fig. \ref{fig:phantoms}(c) and the reconstructed 
%$\mathbf c$ were also computed. The RMSE of images 
%in Fig. \ref{fig:rdnA}(a) and (c) and Fig. 
%\ref{fig:goodA}(a) and (c) are 0.27, 1.66, 0.17, 
%and 1.66, respectively. 
%The results in Fig. \ref{fig:defectiveA} 
Figures \ref{fig:defectiveA}(b) and (e)
reveal that, despite that the violation of the support conjecture,
accurate estimates of $\mathbf c$ could still be reconstructed.
%This suggests, in general, that the support condition is not a necessary condition
%for accurate reconstruction of $\mathbf c$ for an exactly known
%$\mathbf A$ and perfect measurement data.
Figures \ref{fig:defectiveA}(h) and (k)
demonstrate that the accuracy of the reconstructed
estimate of $\mathbf c$ degrades as the size of the support of $A(\mathbf r)$
is reduced.

In summary, these observations reveal that,
in general, the support conjecture does not need to be satisfied in order to achieve
accurate reconstruction of $\mathbf c$ for an exactly known
$\mathbf A$ and perfect measurement data. 
This may be explained by the
fact that the support conjecture was established by use of geometrical
acoustics that represents a simplification of the complicated 
acoustic wave propagation in both our simulations or in practice. 
However, in the cases where the support conjecture was satisfied,
$\mathbf c$ was accurately estimated.
% accurate estimates of $\mathbf c$ were obtained.   
Moreover, the simulation
results indicate that the extent to which the supports of $A(\mathbf r)$
and $c(\mathbf r)$ overlap does affect the ability to accurately solve
Sub-Problem \#2, and hence the JR problem.

\subsubsection{Effect of relative spatial bandwidths of $A(\mathbf r)$ and $c(\mathbf r)$}   
\label{sect:k-space}

%The condition on $\text{supp}(A)$ is not the only factor affecting the reconstruction of $\mathbf c$.
 %In this 
%section, we use a series of computer
%simulations to show that the spatial
%spectrum of $\mathbf A$ will also affect
%the accuracy of the reconstructed $\mathbf c$.

\begin{figure}[h]
\centering
  \subfigure[]{\resizebox{1.5in}{!}{\includegraphics{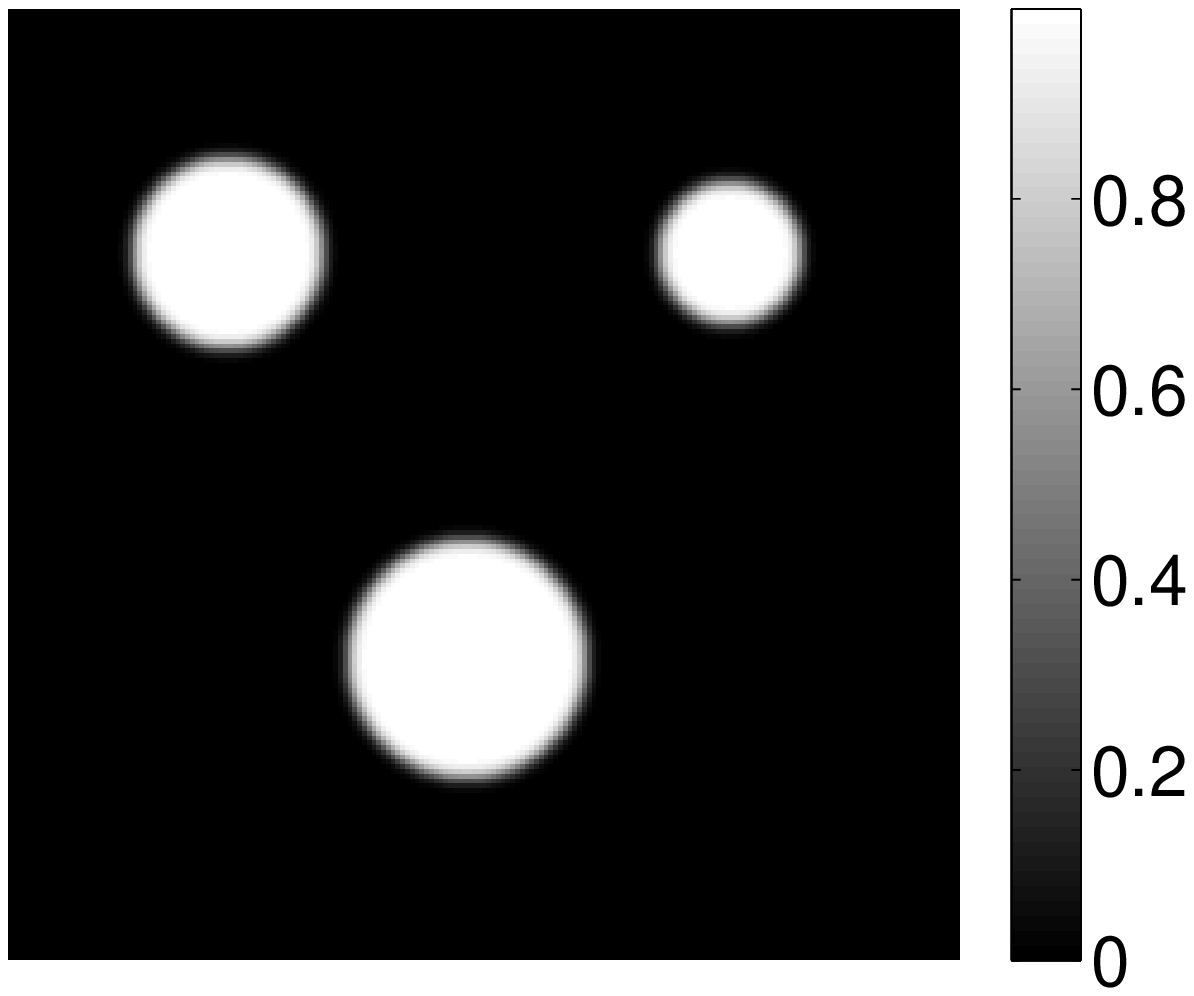}}}
  \subfigure[]{\resizebox{1.5in}{!}{\includegraphics{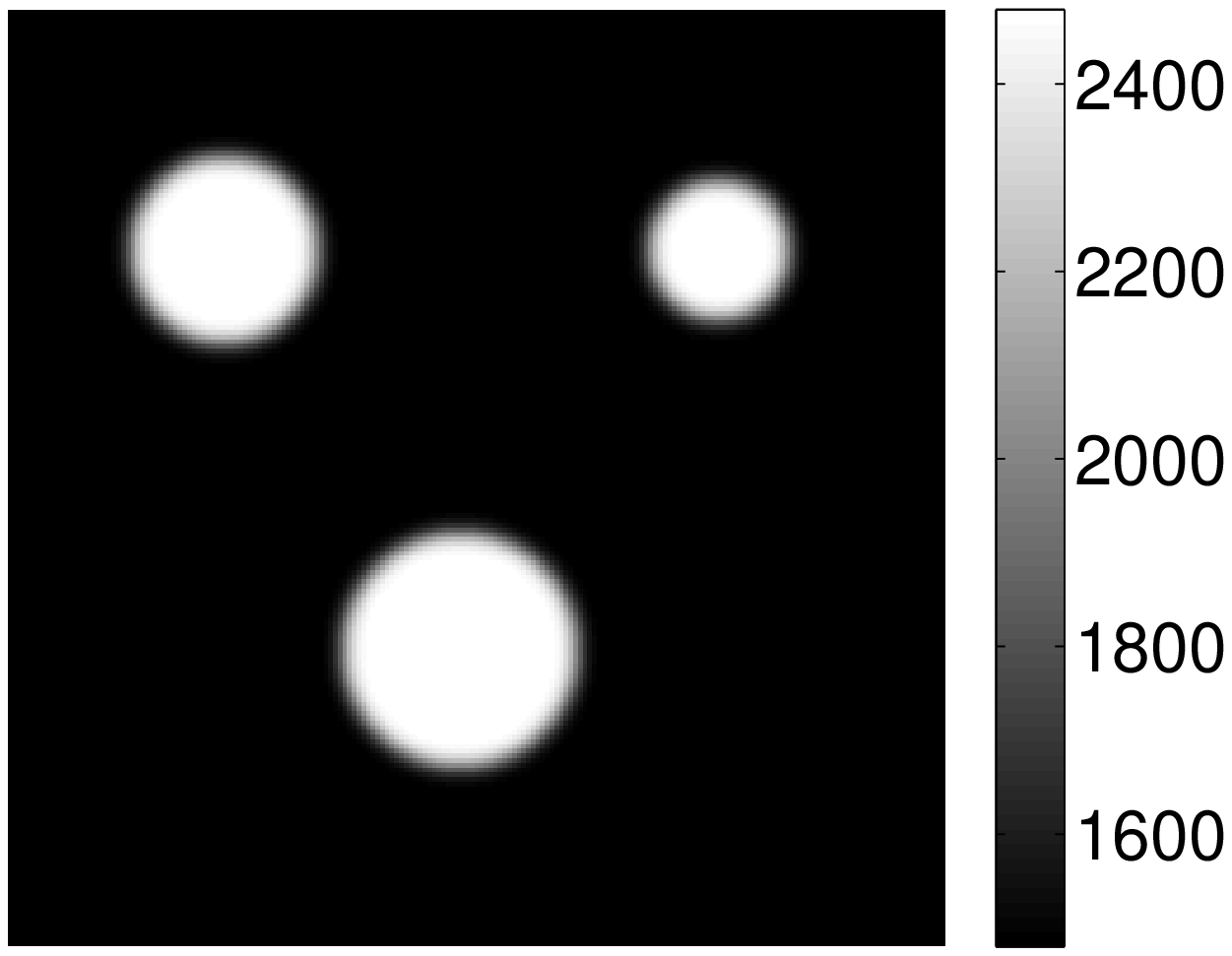}}}
\caption{\label{fig:disk_phantoms}
The numerical phantoms representing (a) $\mathbf A$ and (b) $\mathbf c$
that are described in Sec.\ \ref{sect:k-space}.
}
\end{figure}

Studies were conducted to investigate
the extent to which the relative spatial bandwidths of $A(\mathbf r)$ and $c(\mathbf r)$
influence the ability to accurately solve Sub-Problem \#2 by use of perfect measurements.
Figure \ref{fig:disk_phantoms} shows
the numerical phantoms for $\mathbf A$
and $\mathbf c$.
To exclude the effects related to the supports of $A(\mathbf r)$ and $c(\mathbf r)$,
the phantoms were designed to satisfy the support conjecture.
% related to the support condition discussed above,
%the phantom of $\mathbf A$ was chosen 
%such that $\text{supp}(c) \subseteq 
%\text{supp}(A)$, i.e. the support condition 
%was satisfied.
 The spatial structures of the original
phantoms in Fig.\ \ref{fig:disk_phantoms} are identical, indicating
that their spatial frequency bandwidths are identical.
 The phantom depicting $\mathbf A$
was subsequently convolved with different Gaussian
kernels to generate additional phantoms that possessed
different spatial frequency bandwidths.
The full width at half maximum (FWHM) of the blurring kernel 
was employed as a summary measure of the relative
spatial bandwidths of the smoothed phantoms.
Sub-Problem \#2 was subsequently solved with $\lambda_c=0$
 by use of both perfect
and noisy simulated measurement data for each of the smoothed versions of
$\mathbf A$.

\begin{figure*}[ht]
\centering
  \subfigure[]{\resizebox{1.5in}{!}{\includegraphics{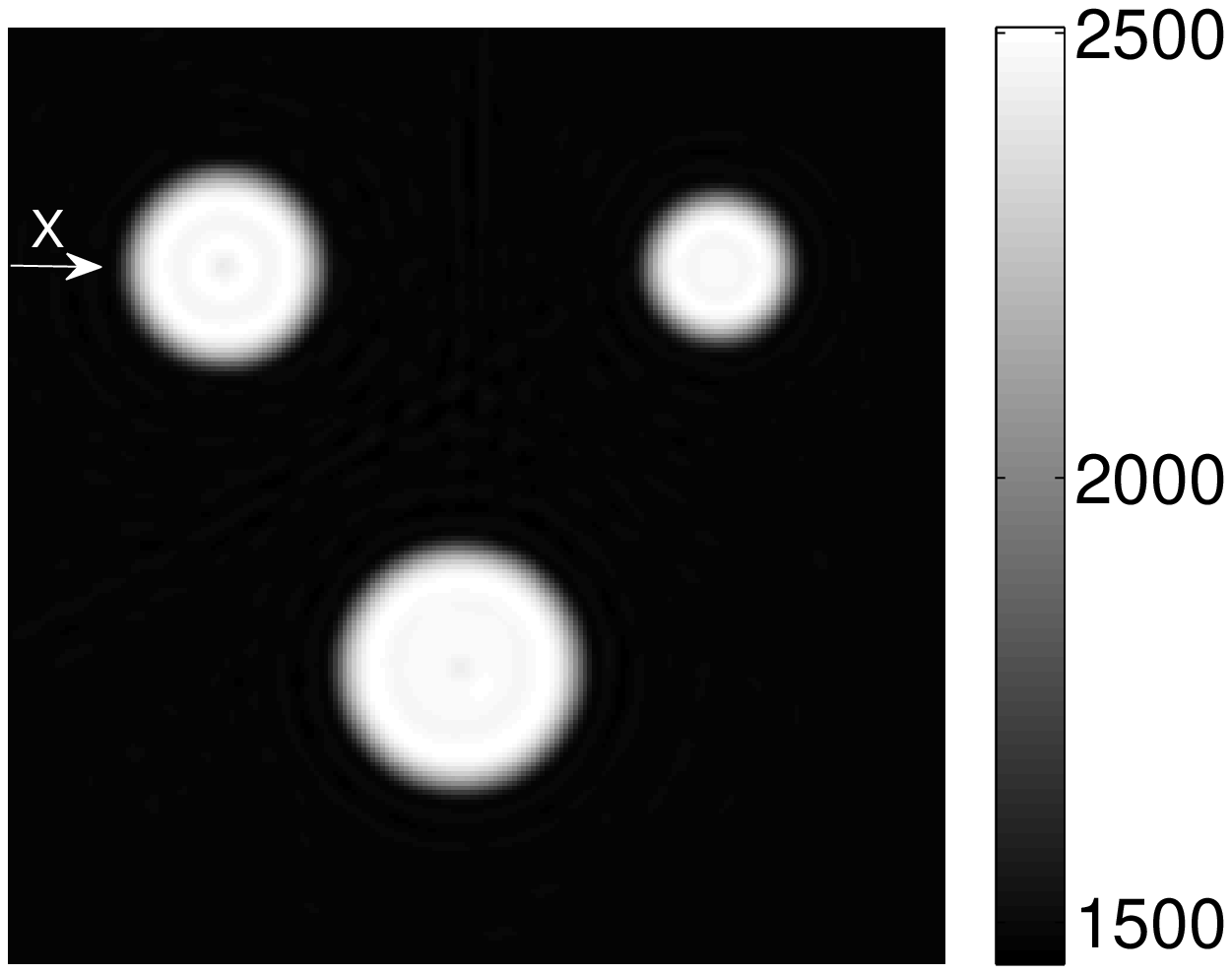}}}
  \subfigure[]{\resizebox{1.5in}{!}{\includegraphics{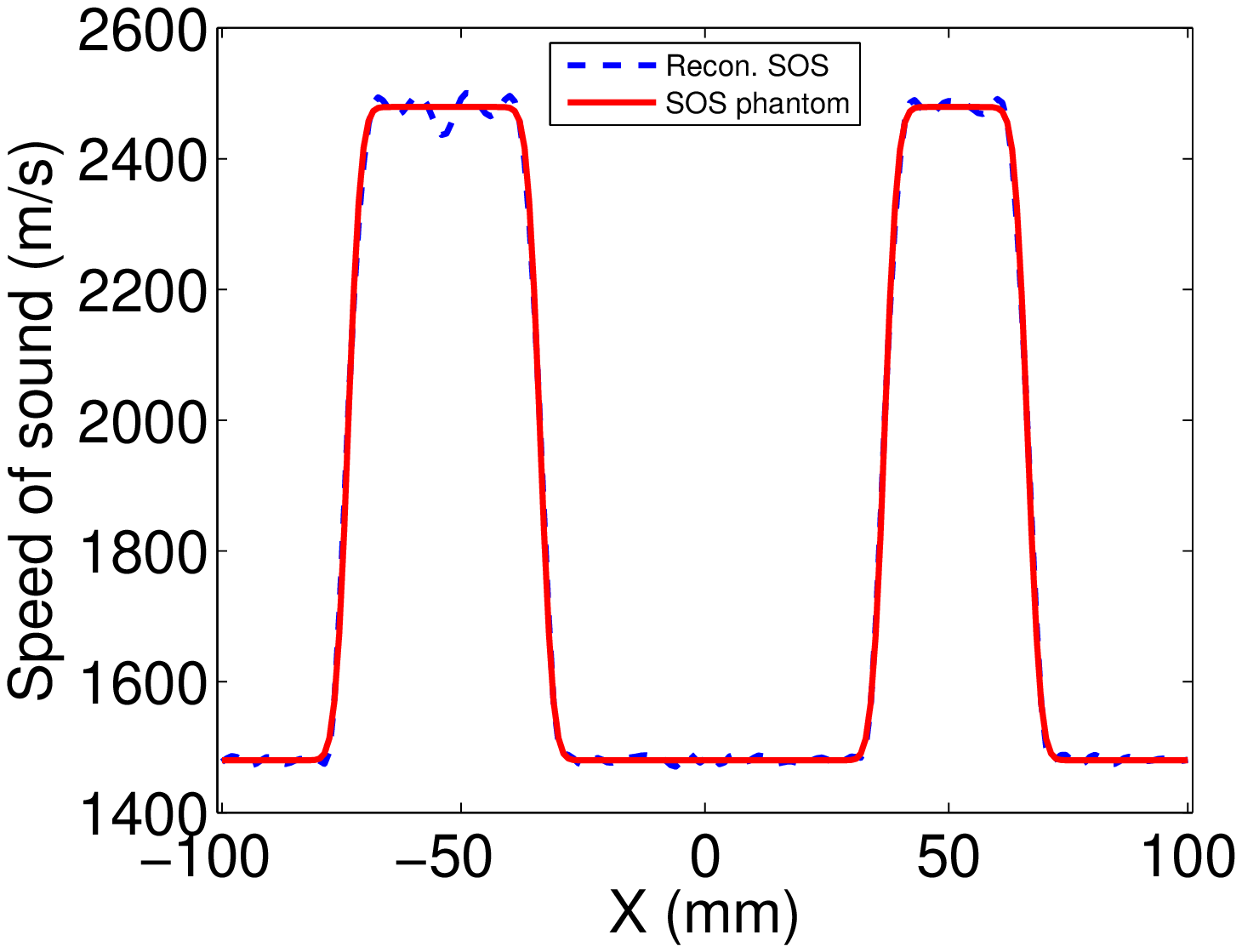}}} 
%  \subfigure[]{\resizebox{1.5in}{!}{\includegraphics{diskSOS_Ac025n.eps}}}
  \subfigure[]{\resizebox{1.5in}{!}{\includegraphics{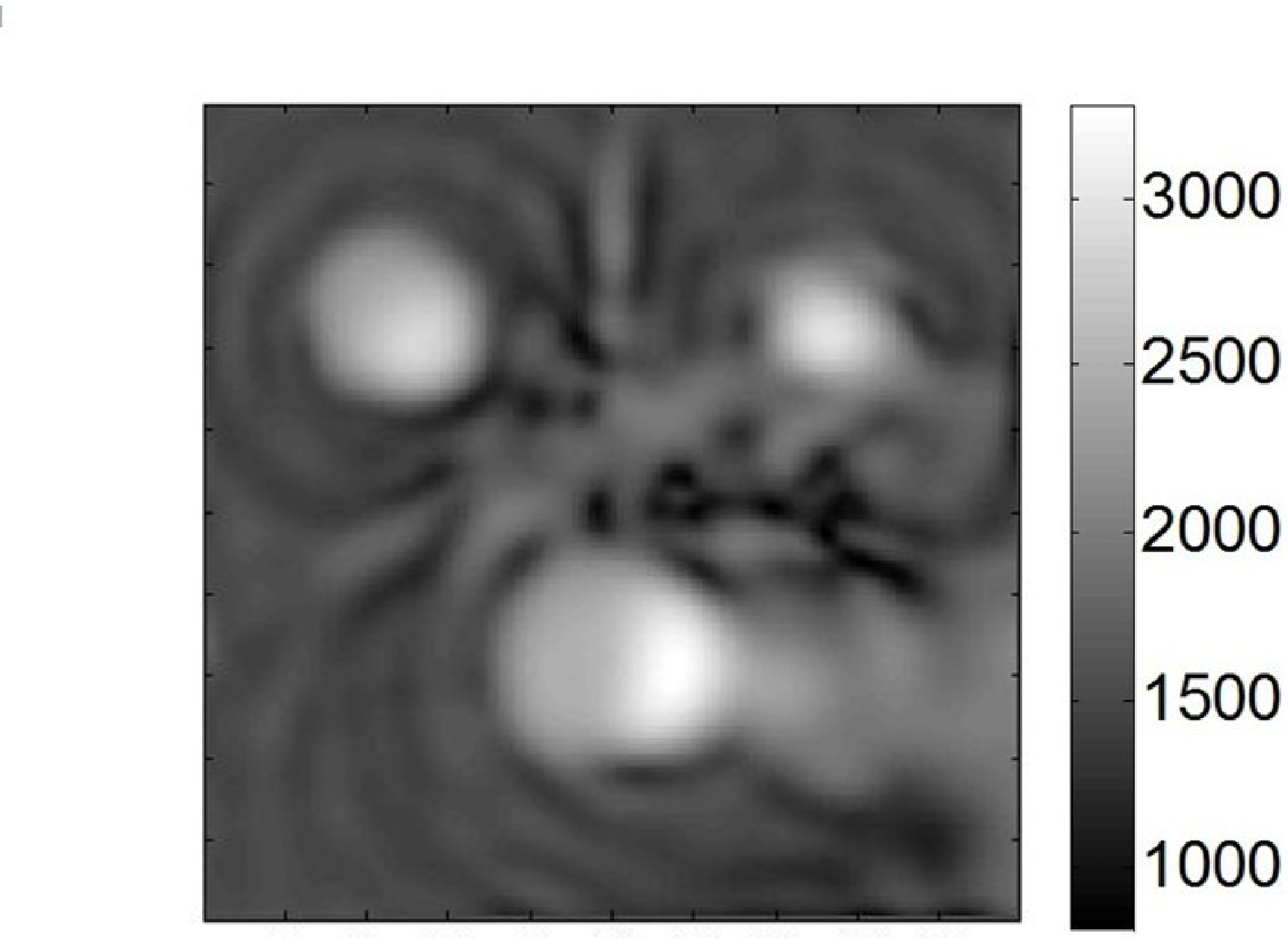}}}
  \subfigure[]{\resizebox{1.5in}{!}{\includegraphics{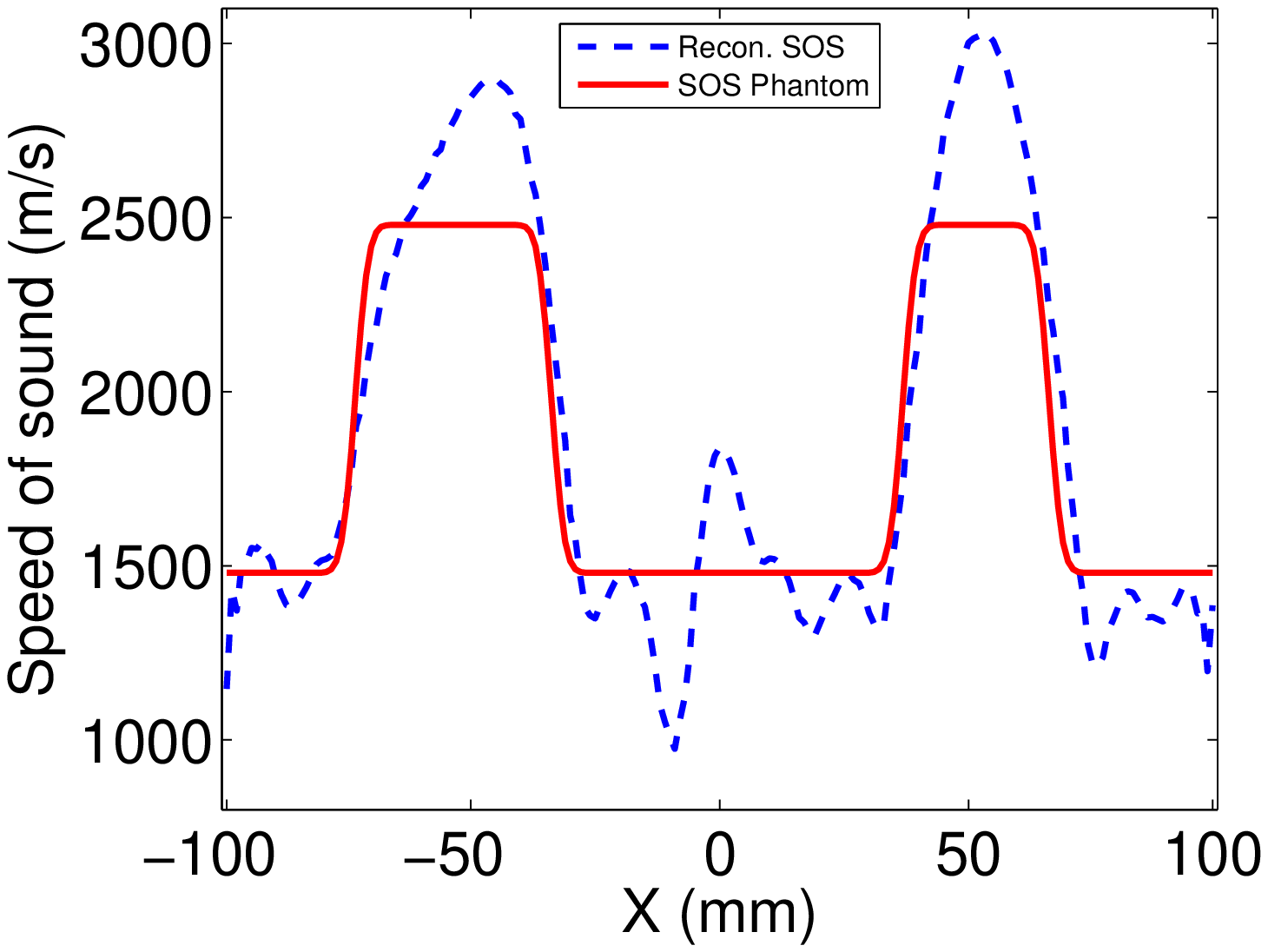}}}\\
\vskip -0.1cm
  \subfigure[]{\resizebox{1.5in}{!}{\includegraphics{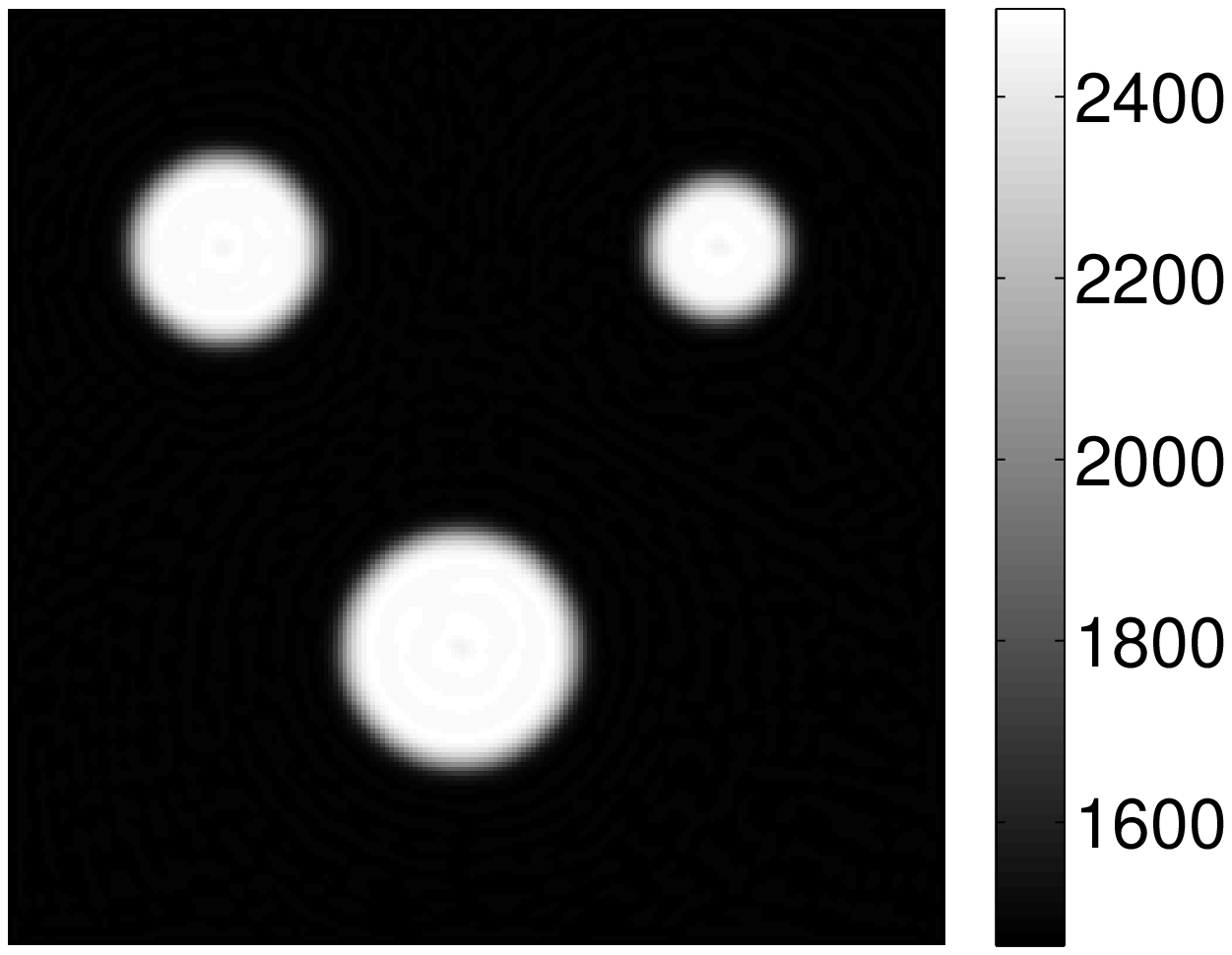}}}
  \subfigure[]{\resizebox{1.5in}{!}{\includegraphics{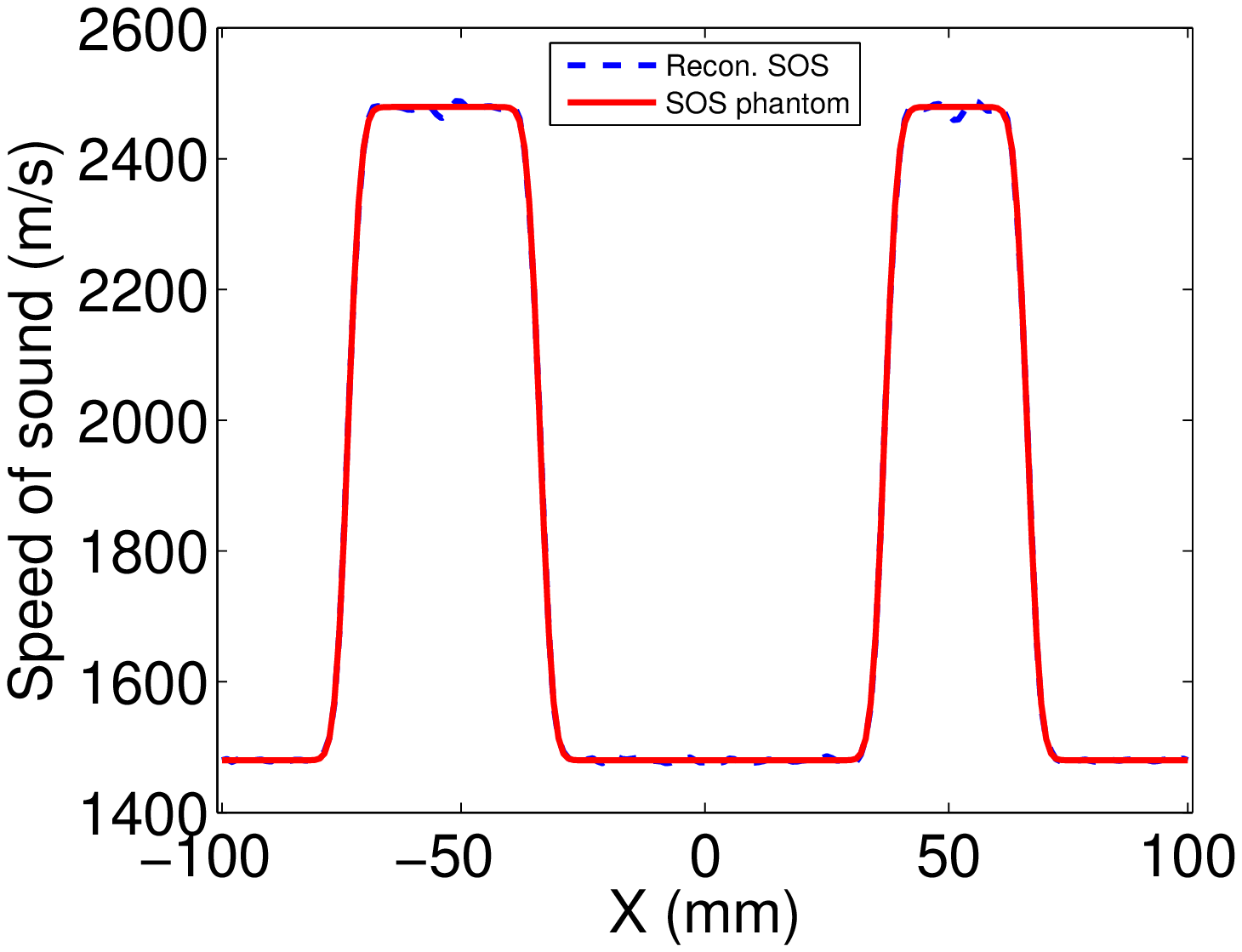}}} 
  \subfigure[]{\resizebox{1.5in}{!}{\includegraphics{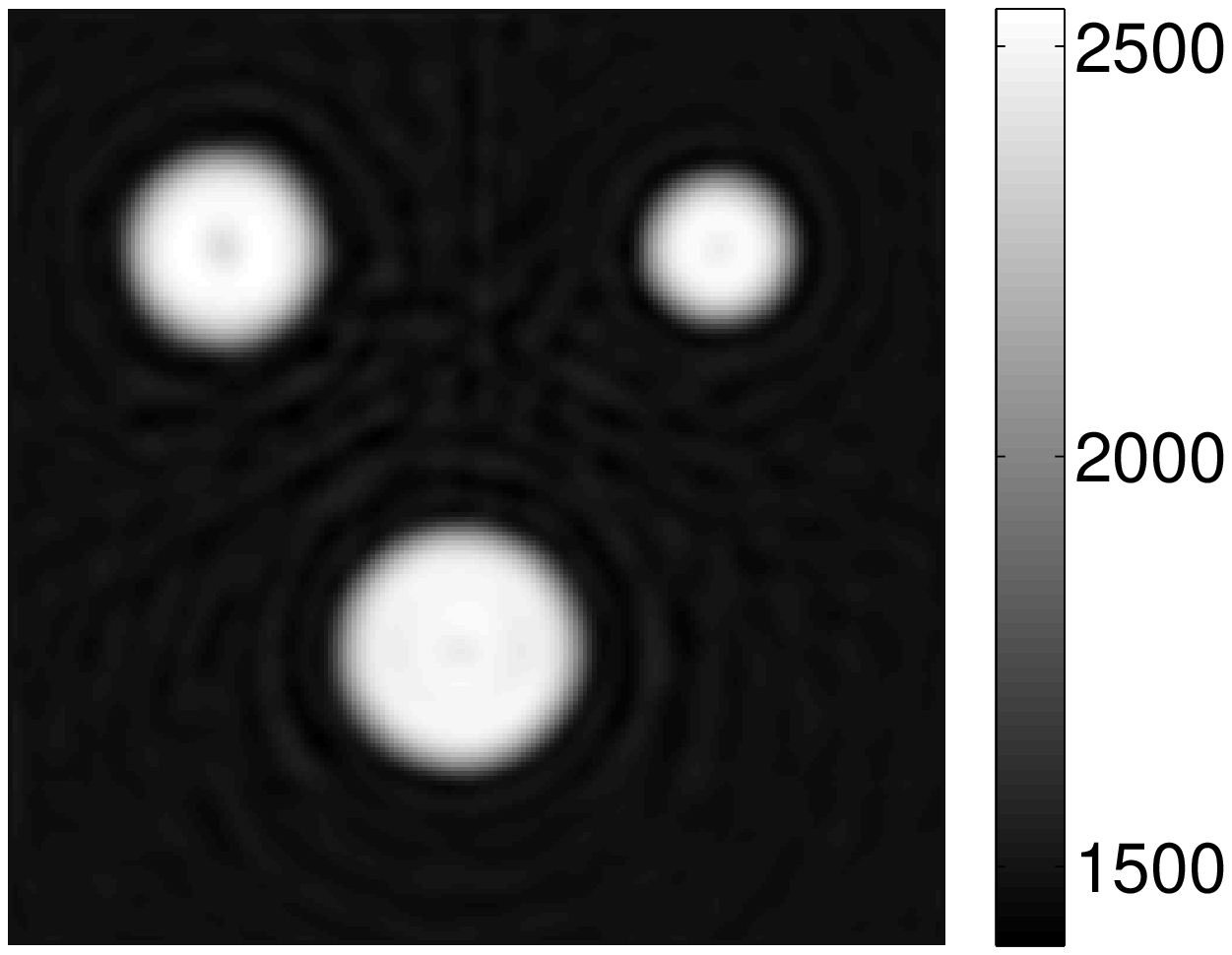}}}
  \subfigure[]{\resizebox{1.5in}{!}{\includegraphics{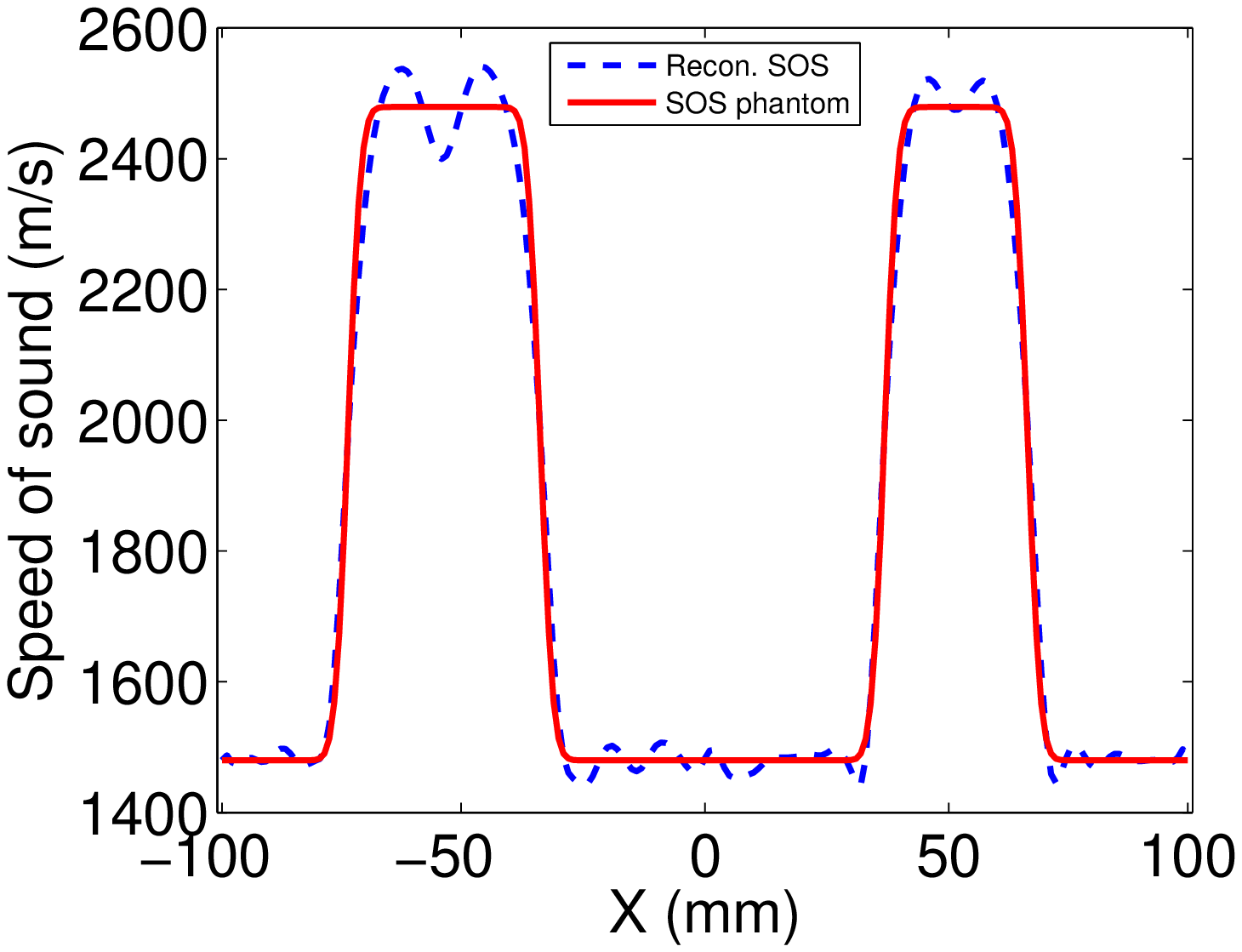}}}\\
\vskip -0.1cm
%  \subfigure[]{\resizebox{1.5in}{!}{\includegraphics{diskSOS_Ac063.eps}}}
%  \subfigure[]{\resizebox{1.5in}{!}{\includegraphics{diskSOS_Ac063_pfl.eps}}}
%  \subfigure[]{\resizebox{1.5in}{!}{\includegraphics{diskSOS_Ac063n.eps}}}
%  \subfigure[]{\resizebox{1.5in}{!}{\includegraphics{diskSOS_Ac063n_pfl.eps}}}\\
  \subfigure[]{\resizebox{1.5in}{!}{\includegraphics{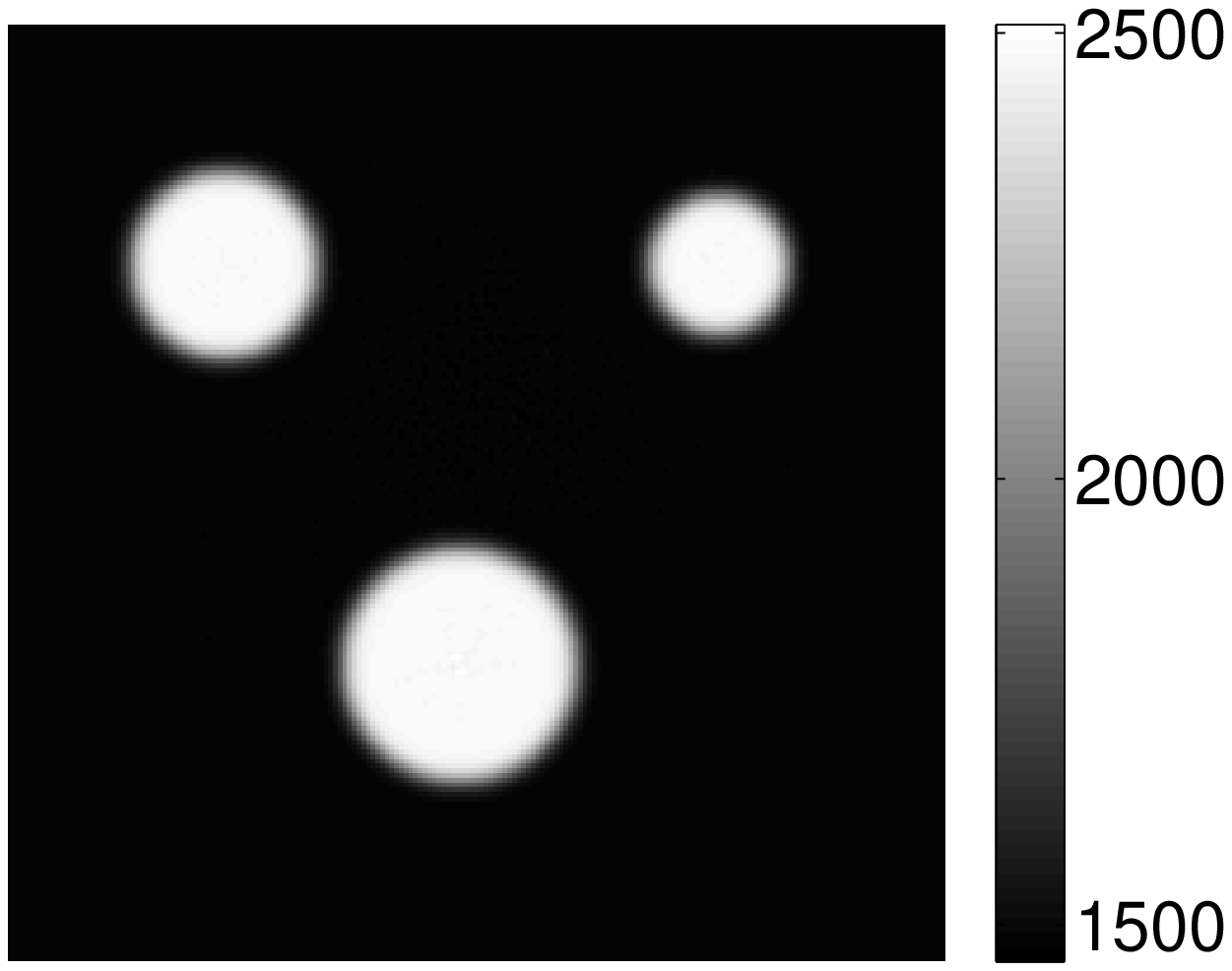}}}
  \subfigure[]{\resizebox{1.5in}{!}{\includegraphics{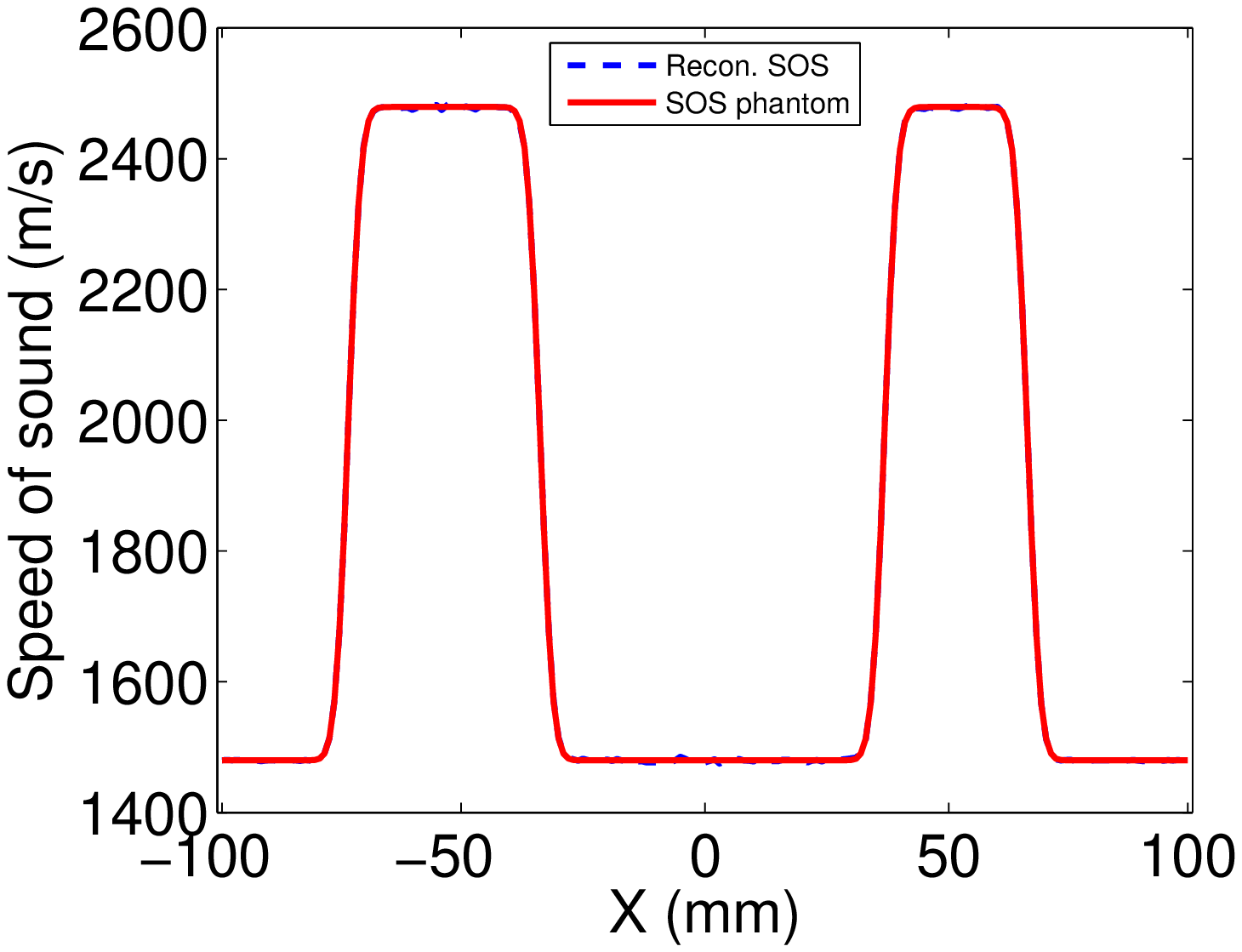}}} 
  \subfigure[]{\resizebox{1.5in}{!}{\includegraphics{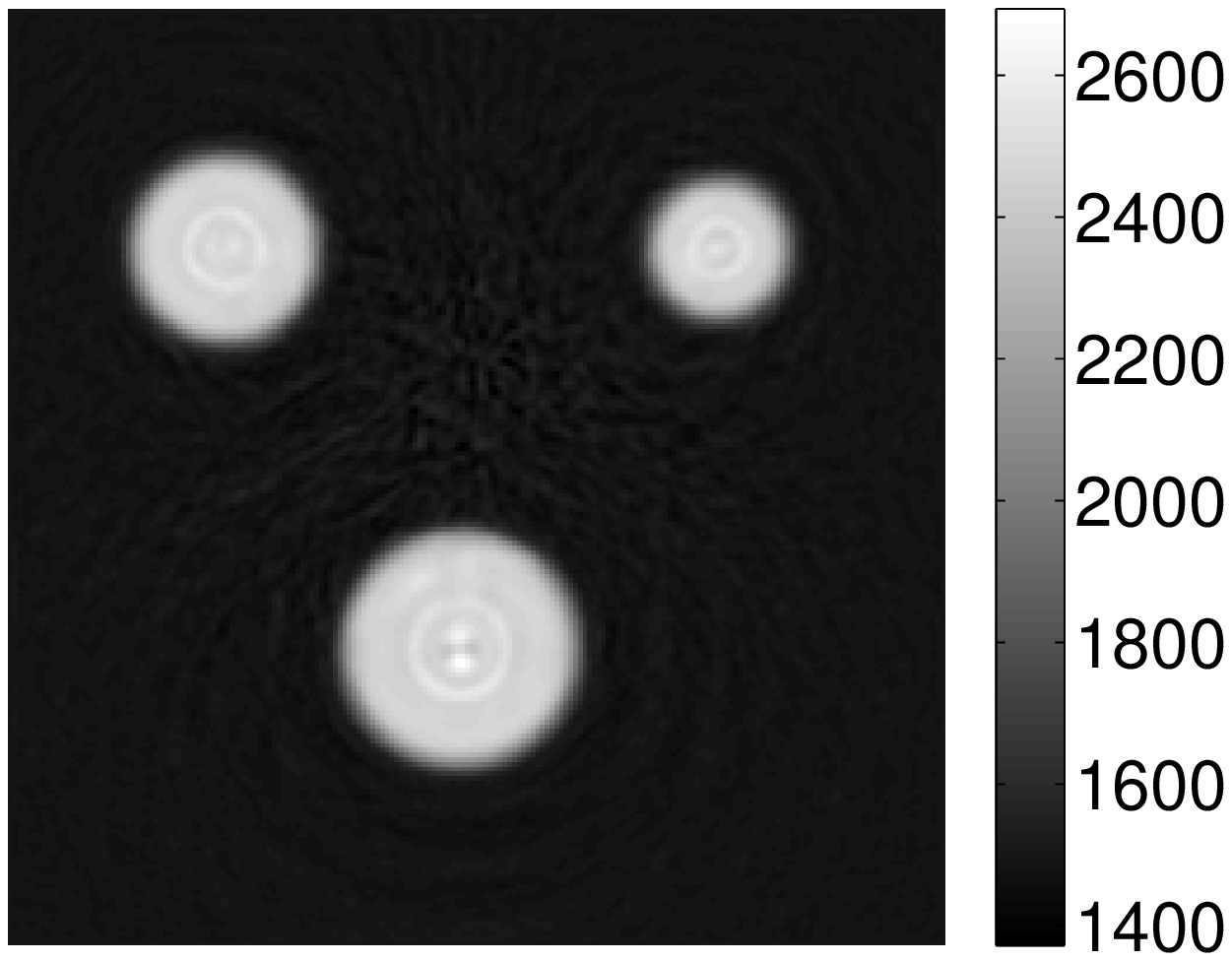}}}
  \subfigure[]{\resizebox{1.5in}{!}{\includegraphics{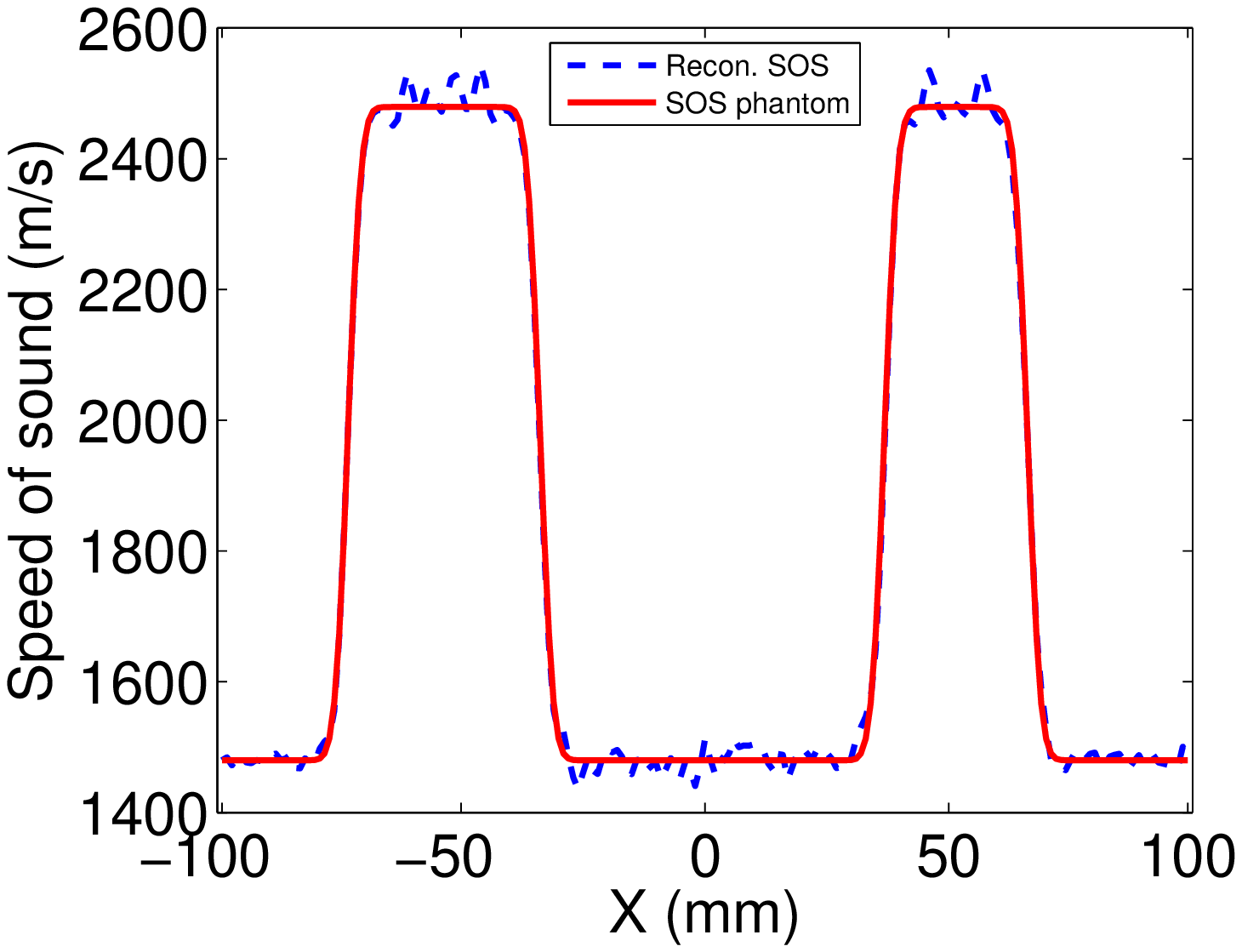}}}
%  \subfigure[]{\resizebox{1.5in}{!}{\includegraphics{diskSOS_Ac2.eps}}}
%  \subfigure[]{\resizebox{1.5in}{!}{\includegraphics{diskSOS_Ac2_pfl.eps}}}
%  \subfigure[]{\resizebox{1.5in}{!}{\includegraphics{diskSOS_Ac2n.eps}}}
%  \subfigure[]{\resizebox{1.5in}{!}{\includegraphics{diskSOS_Ac2n_pfl.eps}}}
\caption{\label{fig:k-space}
Numerical investigations of Sub-Problem \#2 - Spatial bandwidth effects:
As described in Sec.\ \ref{sect:k-space}, unregularized estimates of $\mathbf c$
were reconstructed from noiseless and noisy data and a series of $\mathbf A$
that had different spatial bandwidths relative to that of the sought-after
$\mathbf c$.
From the top to the bottom rows, the results
 correspond to relative bandwidths of $\mathbf A$ to
$\mathbf c$ of 0.25, 0.44, and 1.0, respectively.
Subfigures (a), (e), and (i) correspond
to images reconstructed from perfect measurement data and
the associated image profiles are displayed in
subfigures (b), (f), and (j).
Subfigures (c), (g), and (k) correspond
to images reconstructed from noisy measurement data and
the associated image profiles are displayed in
subfigures (d), (h), and (l).}
\end{figure*}

\begin{figure}[h]
\centering
  \subfigure[]{\resizebox{1.5in}{!}{\includegraphics{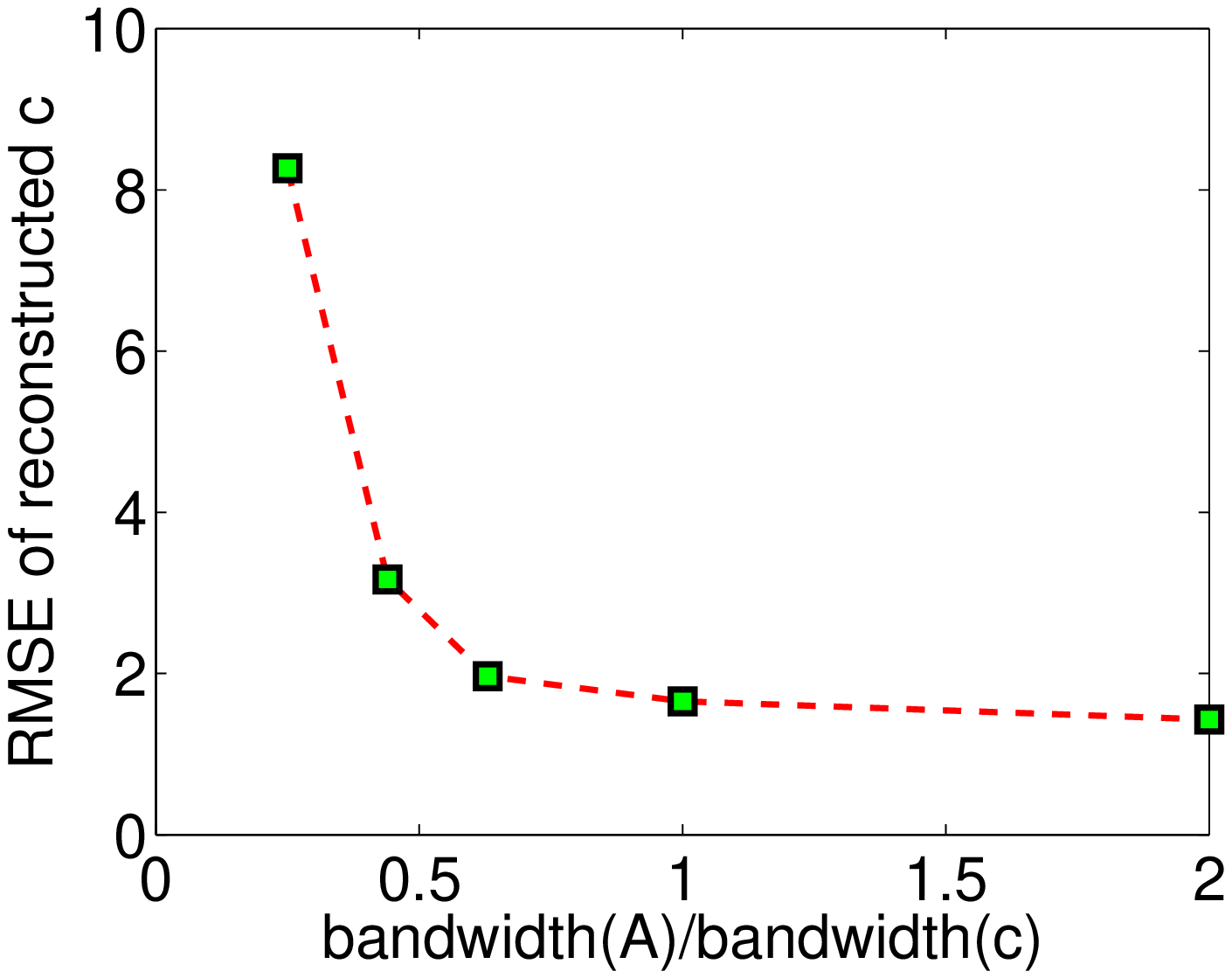}}}
  \subfigure[]{\resizebox{1.5in}{!}{\includegraphics{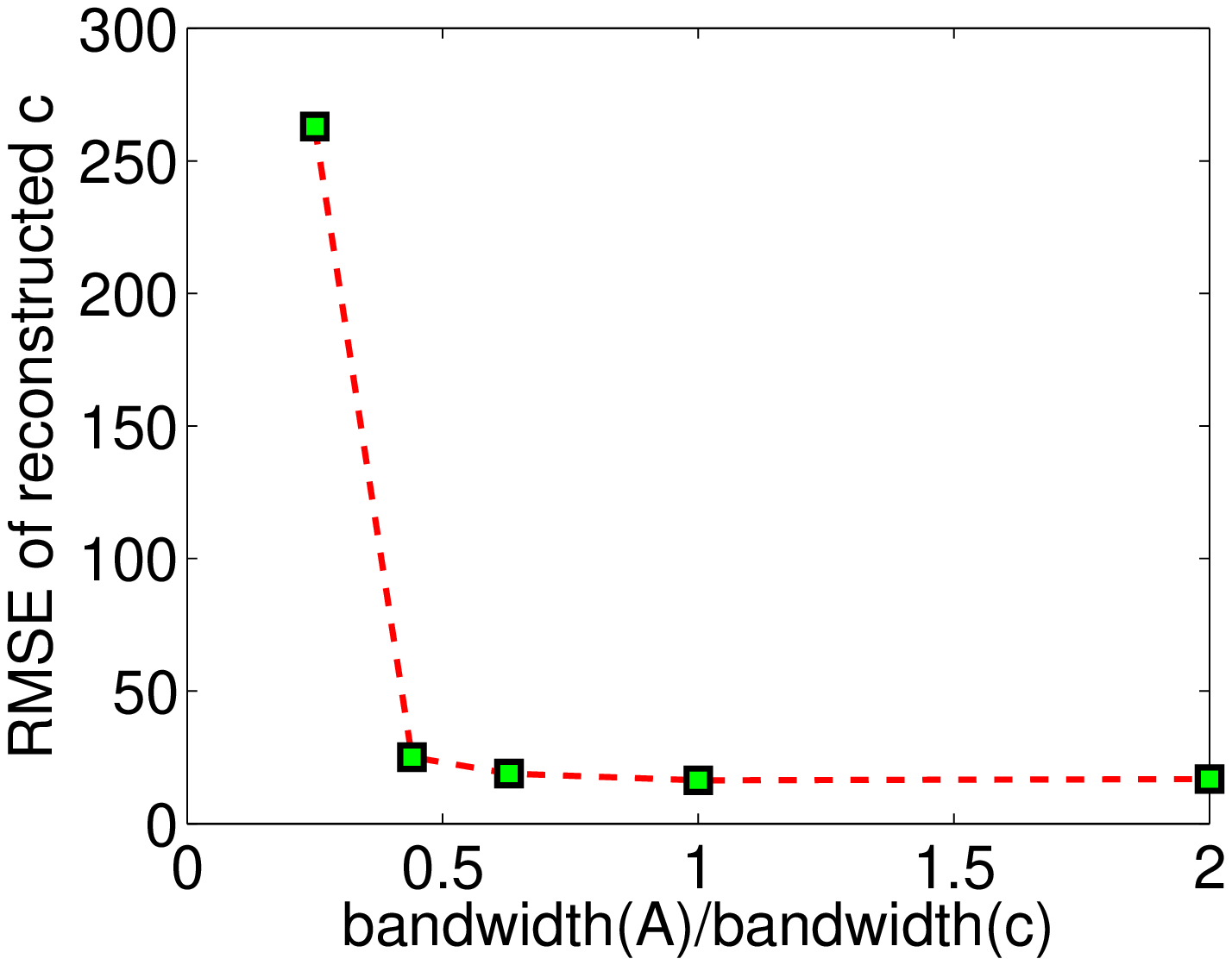}}}
\caption{\label{fig:k-space_p}
Numerical investigations of Sub-Problem \#2 - Spatial bandwidth effects:
Plots of RMSE of the estimated $\mathbf c$ as a function of the
bandwidth ratio of $\mathbf A$ to $\mathbf c$.
Subfigures (a) and (b) correspond to
the noiseless and noisy results, repectively.
}
\end{figure}

The reconstructed estimates of $\mathbf c$ 
 are shown in Fig. \ref{fig:k-space}. 
From  top to bottom, the results
 correspond to relative bandwidth ratios (of $\mathbf A$ to
$\mathbf c$) of 0.25, 0.44, and 1.0, respectively.
Figures  \ref{fig:k-space}(a), (e), and (i) correspond
to images reconstructed from perfect measurement data and
the associated image profiles are displayed in
Figs.\  \ref{fig:k-space}(b), (f), and (j).
Figures  \ref{fig:k-space}(c), (g), and (k) correspond
to images reconstructed from noisy measurement data and
the associated image profiles are displayed in
Figs.\  \ref{fig:k-space}(d), (h), and (l).
The RMSE of the reconstructed $\mathbf c$ 
with respect to the relative bandwidth ratio of 
$\mathbf A$ to $\mathbf c$ for both 
noiseless and noisy cases is  displayed
in Fig. \ref{fig:k-space_p}.

These results reveal
that the relative spatial frequency bandwidths of
$A(\mathbf r)$ and $c(\mathbf r)$
affect the numerical stability of Sub-Problem \#2 and, hence,
that of the JR problem.
The accuracy 
of the reconstructed $\mathbf c$ in the present of
measurement noise for a specified $\mathbf A$ was observed to be influenced
strongly by the relative spatial bandwidths of $\mathbf A$ and $\mathbf c$.
Specifically, in order to accurately 
reconstruct $\mathbf c$ in the presence of noise,
 the presented results suggest that the spatial
frequency  bandwidth of $A(\mathbf r)$ 
should be comparable to or larger than the spatial frequency bandwidth 
of $c(\mathbf r)$. We will refer to this as the \emph{k-space conjecture}
 hereafter.
%In practice, the smoothness may not be
%satisfied if the variation of $\mathbf A$
%is weaker than the variation of $\mathbf c$.
%Another senario where the smoothness
%condition could be violated is 
%An object function $A(\mathbf r)$ that satisfies both
%support condition and smoothness condition
%will be called `sufficient' in the remainder 
%of this paper, otherwise it will be called
%`deficient'. 

%\subsection{Relative numerical instability of the sub-problems in (\ref{eq:cost_A}) and (\ref{eq:cost_c})}
\subsubsection{Effect of perturbations of $A(\mathbf r)$}
\label{sect:instable}

\if 0
Below, we  investigate the relative numerical stability
of the problems of reconstructing $\mathbf A$ given
$\mathbf c$ (sub-problem  (\ref{eq:cost_A})) and the problem of 
reconstructing $\mathbf c$ given $\mathbf A$ (sub-problem (\ref{eq:cost_c})).
 The phantoms of $\mathbf A$
and $\mathbf c$ used here are the same
as the ones in Fig. \ref{fig:disk_phantoms}.
To investigate the numerical instability,
$\mathbf A$ (resp. $\mathbf c$) was 
perturbed by additive white Gaussian noise 
(AWGN) when reconstructing $\mathbf c$ 
(resp. $\mathbf A$). The perturbation is 
measured by the relative error, which is 
defined as the ratio of the $l_2$ norm 
of the AWGN to the $l_2$ norm of $\mathbf A$ 
(or $\mathbf c$). Figure \ref{fig:ic} 
shows the reconstructed $\mathbf A$ 
(first and second columns) and $\mathbf c$ 
(third and fourth columns) corresponding 
to the perturbed $\mathbf c$ and $\mathbf A$, 
respectively. The results from the top 
to the bottom row correspond to relative 
error of 0.2\%, 1.0\% and 5.0\%, 
respectively. These results are summarized 
in Fig. \ref{fig:ic_p} \rd{[Axis labels in that fig 
are confusing to me]} \bl{[fixed]}, which shows that, 
for a fixed relative error, the 
reconstructed estimate of $\mathbf c$ has larger 
RMSE than does the reconstructed  estimate of
$\mathbf A$.
This demonstrates that the problem of 
of reconstructing $\mathbf c$ given  $\mathbf A$ is more ill-conditioned
than  the problem 
of reconstructing $\mathbf A$ given  $\mathbf c$.
\fi

In the studies described above, perfect knowledge
of $\mathbf A$ was assumed.  Below, a numerical experiment is described
that provides insights into how small perturbations in the assumed $\mathbf A$
affects the accuracy of the reconstructed $\mathbf c$ obtained
 by solving  Sub-Problem \#2.

\begin{figure}[h]
\centering
  \subfigure[]{\resizebox{1.5in}{!}{\includegraphics{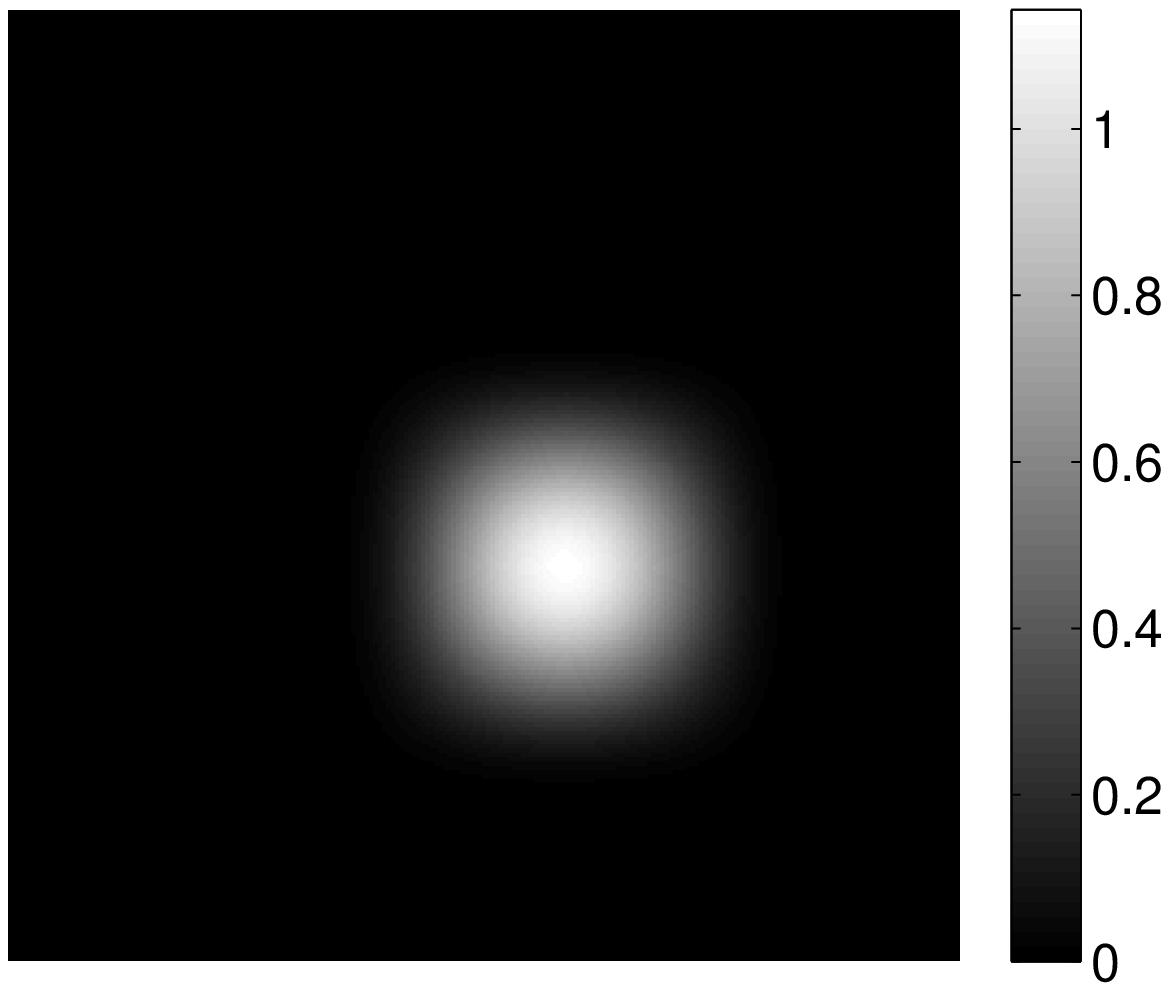}}}
  \subfigure[]{\resizebox{1.5in}{!}{\includegraphics{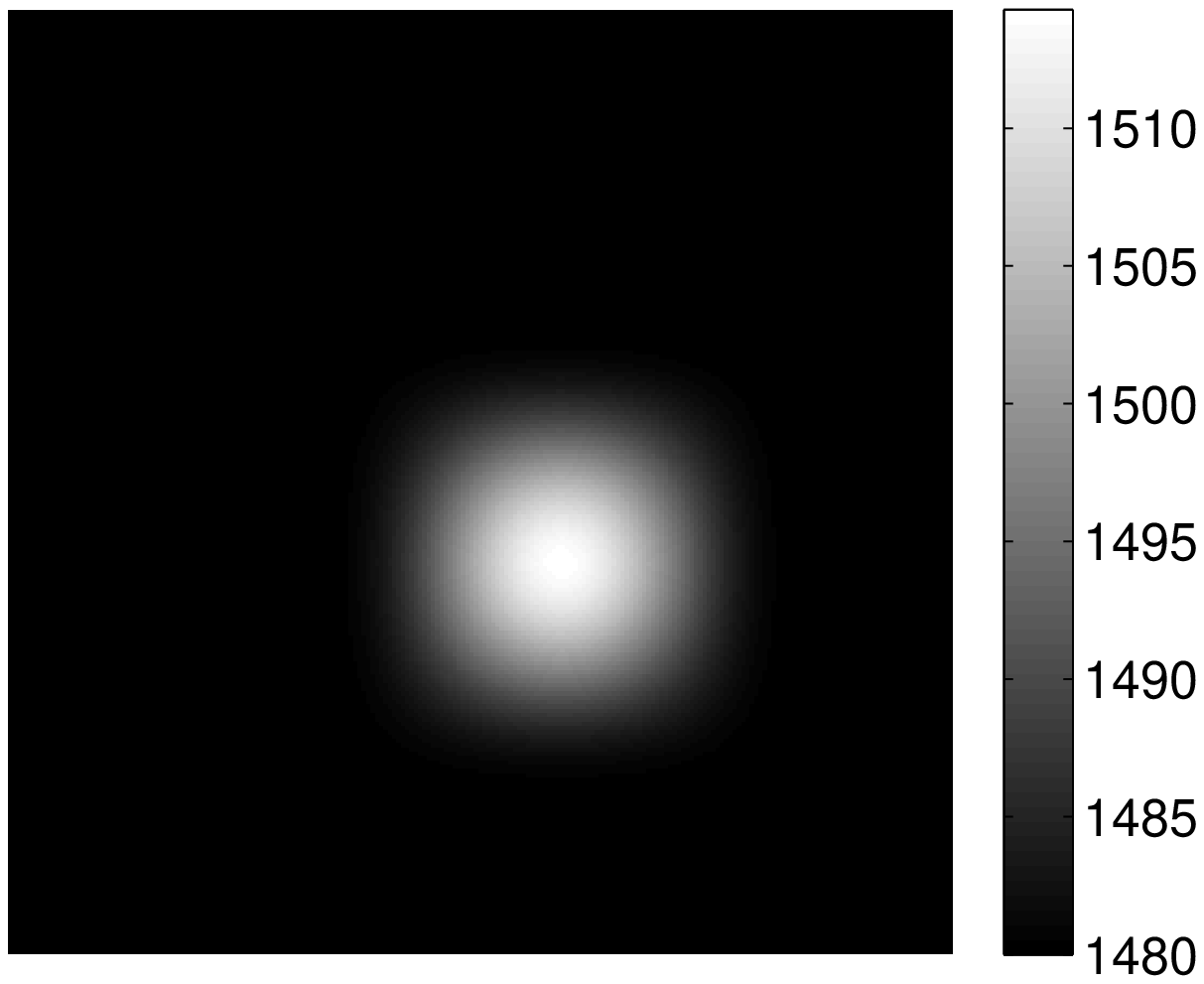}}}\\
\vskip -0.1cm
  \subfigure[]{\resizebox{1.5in}{!}{\includegraphics{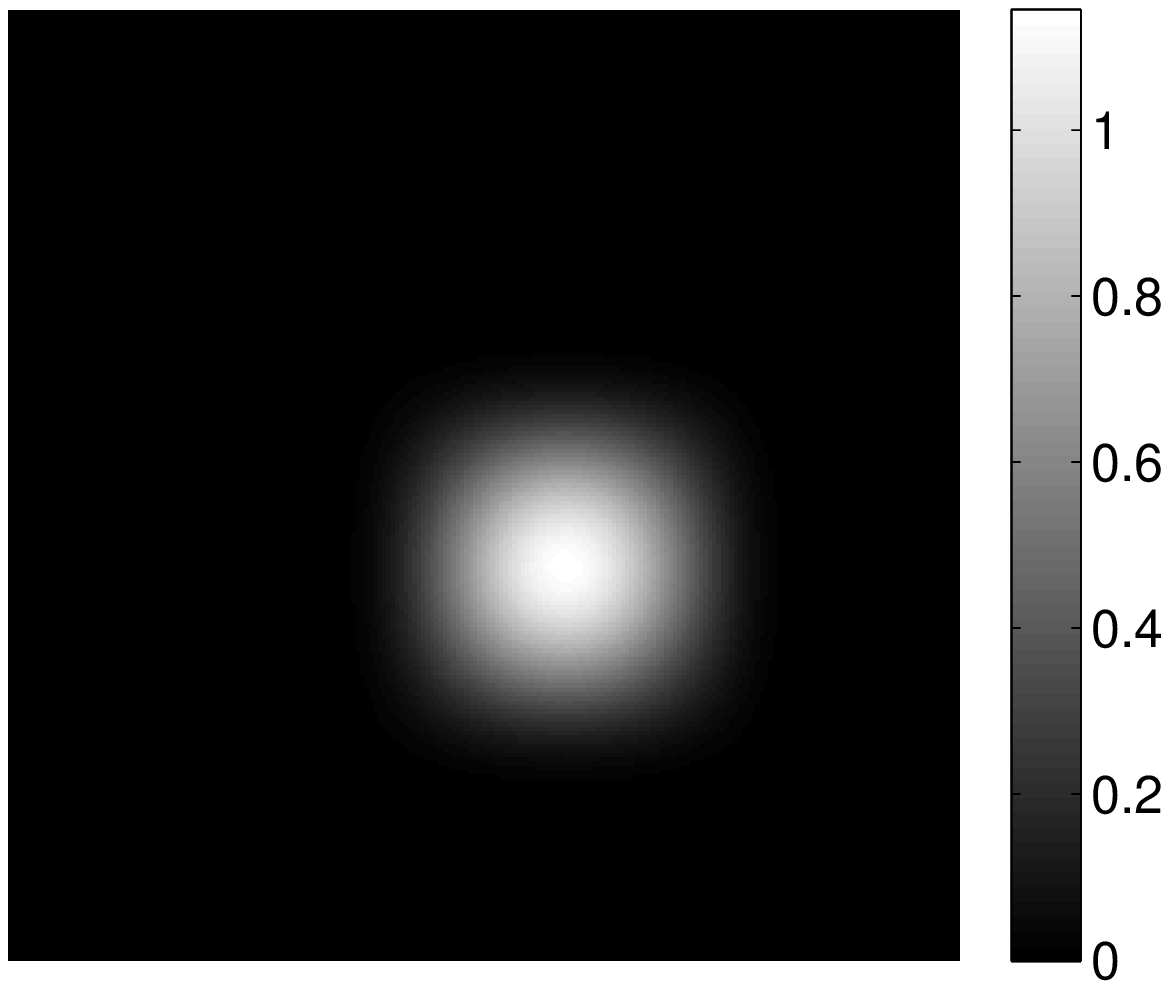}}}
  \subfigure[]{\resizebox{1.5in}{!}{\includegraphics{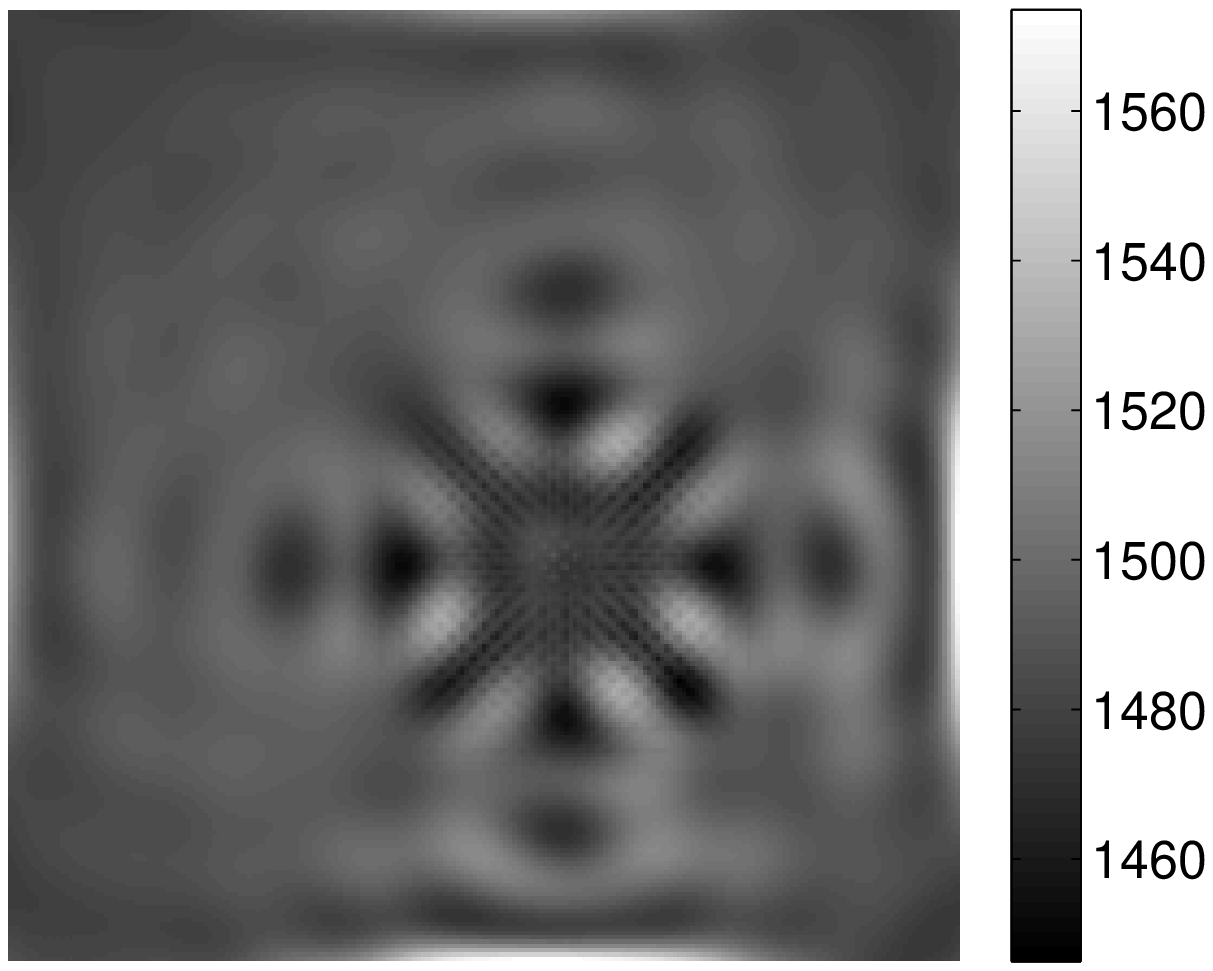}}}
\caption{\label{fig:ill-cond}
Numerical investigations of Sub-Problem \#2 - Effect of perturbation of $\mathbf A$:
Two numerical phantoms representing $\mathbf A$ are shown
in subfigures (a) and (c). As described in Sec.\ \ref{sect:instable},
these two phantoms are very similar, with a RMSE between them of only $0.004$.
Unregularized estimates of $\mathbf c$ reconstructed by use of noiseless
simulated measurement data and the $\mathbf A$ specified in (a) and (c)
are shown in subfigures (b) and (d), respectively.  Although the phantoms
depicting $\mathbf A$ are very similar, the reconstructed estimates
of $\mathbf c$ are not.
}
\end{figure}

 Figures \ref{fig:ill-cond}(a) and (c) display two 
similar numerical phantoms depicting $\mathbf A$. 
The RMSE between these phantoms is $0.004$.
% Simulated perfect PACT measurements were produced
Perfect PACT measurements were simulated
for each of the two  $\mathbf A$, for a given $\mathbf c$ (not shown).
 Figures \ref{fig:ill-cond}(b) 
and (d) display the reconstructed estimates of
$\mathbf c$ when the $\mathbf A$ specified in
 Fig. \ref{fig:ill-cond}(a) and (c) was assumed, respectively.
 These results demonstrate that the problem of reconstructing $\mathbf c$ for a given $\mathbf A$
is ill-conditioned in the sense that small changes in  $\mathbf A$
can produce significant changes in the reconstructed estimate
of $\mathbf c$.
%be completely different given similar 
%$\mathbf A$. Again, these results confirm
%the numerical instability of the problem 
%of reconstruction of $\mathbf c$, which 
This observation is consistent with the theoretical results in 
\cite{StefanovArxiv12}. 

\if 0
In fact, $\mathbf A$ and $\mathbf c$ 
shown in Fig. \ref{fig:ill-cond} were 
discovered when conducting JR. Specifically, 
the phantoms in the top row of Fig. 
\ref{fig:ill-cond} were employed to 
generate the simulated PA data, and the 
ones in the bottom row are the JR results.
The JR was performed without regularization, 
and the results were obtained when the 
objective function value was equal to 
zero (up to round-off error). Therefore, 
\fi

\begin{figure}[h]
\centering
  \subfigure[]{\resizebox{1.5in}{!}{\includegraphics{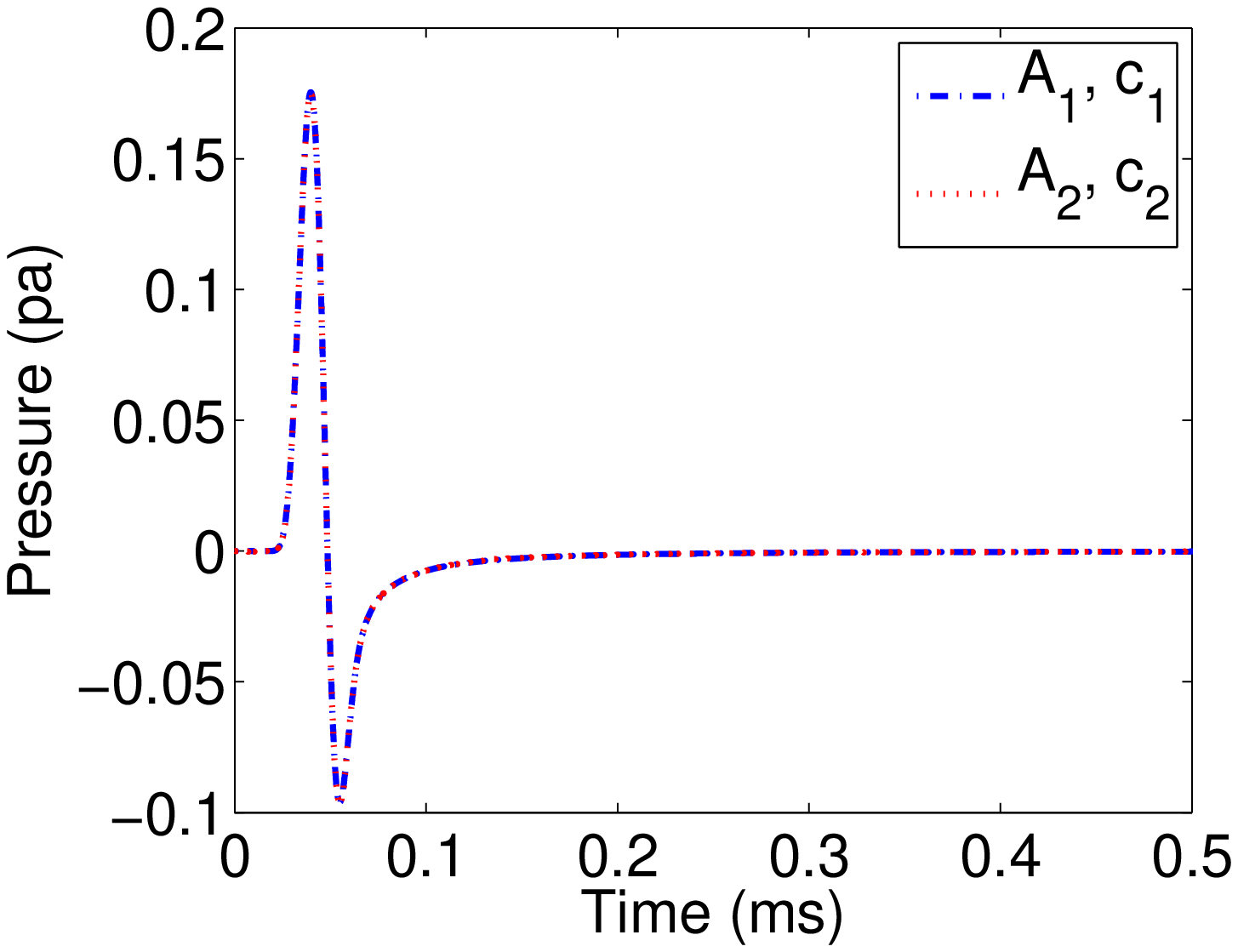}}}
  \subfigure[]{\resizebox{1.5in}{!}{\includegraphics{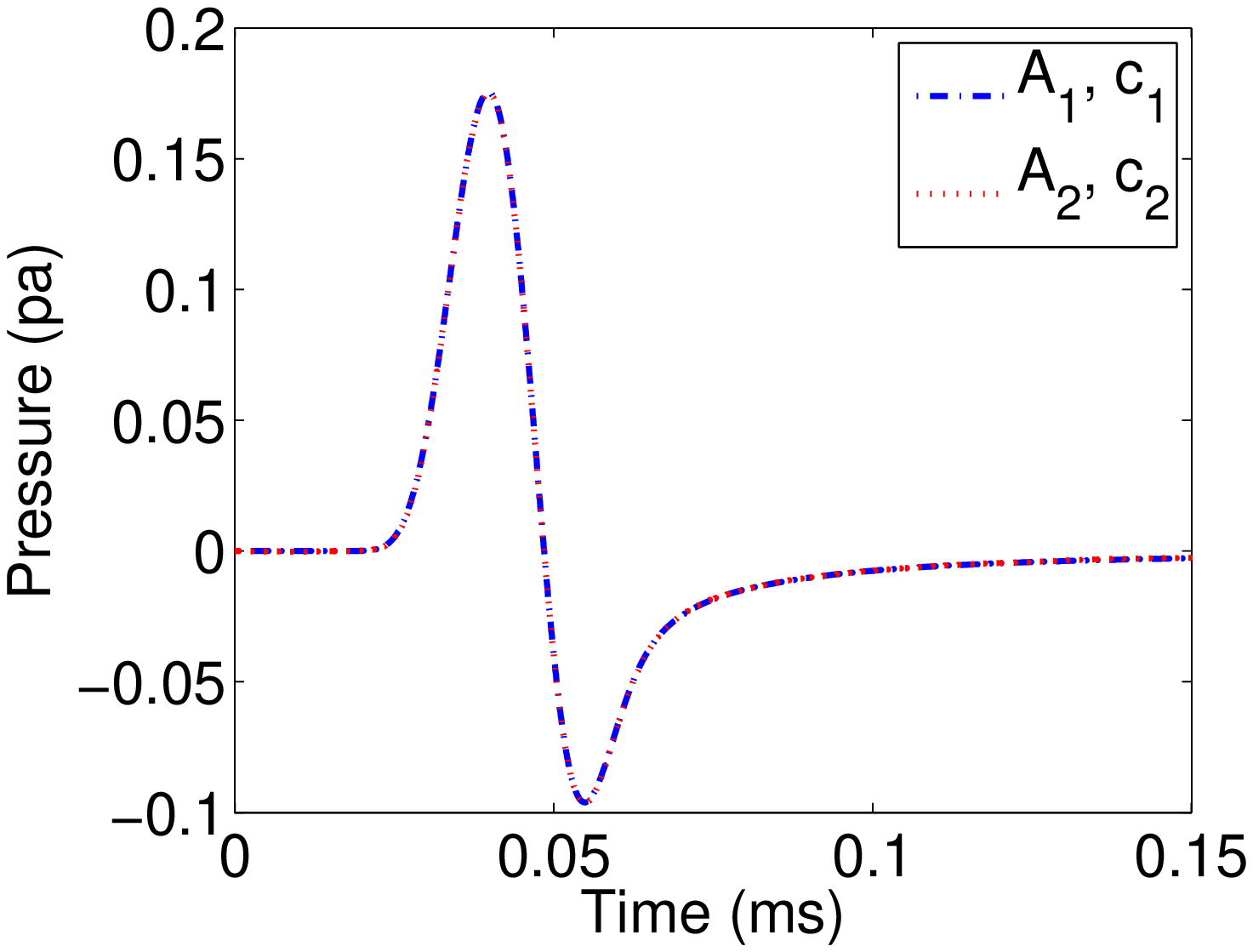}}}
\caption{\label{fig:uniqueP}
Numerical evidence of non-uniqueness of the JR problem:
Simulated PA measurement data were computed from
the $(\mathbf A, \mathbf c)$ pairs shown in
 Figs. \ref{fig:ill-cond}(a) and (b) and
 Figs. \ref{fig:ill-cond}(c) and (d). The two pressure profiles
corresponding to an arbitrary transducer location
are  superimposed in subfigure (a).
Subfigure (b) displays a zoomed-in version of subfigure (a).
Similar agreement between the profiles was observed at
all transducer locations.
}
\end{figure}

It is interesting to note that the ($\mathbf A$, $\mathbf c$) pair in the top 
row of Fig. \ref{fig:ill-cond} produces 
nearly identical PA data, at all transducer locations,
to that produced by the ($\mathbf A$, $\mathbf c$) pair
in the bottom row. 
The simulated noiseless pressure data at an arbitrary transducer location 
produced by the two ($\mathbf A$, $\mathbf c$) pairs is
shown in Fig. \ref{fig:uniqueP}, where the pressure signals are observed
to overlap almost completely.
The RMSE between the two sets of PA data was $\text{RMSE}=3.2           
\times 10^{-4}$.
These results suggest that the solutions of 
the JR problem in PACT may not be unique. 
Consequently, it indicates that accurate 
JR of $\mathbf A$ and $\mathbf c$, 
in general, may not be possible.

\subsection{Feasibility of JR with idealized data}
\label{sect:JR_ideal}

\begin{figure*}[ht]
\centering
  \subfigure[]{\resizebox{1.5in}{!}{\includegraphics{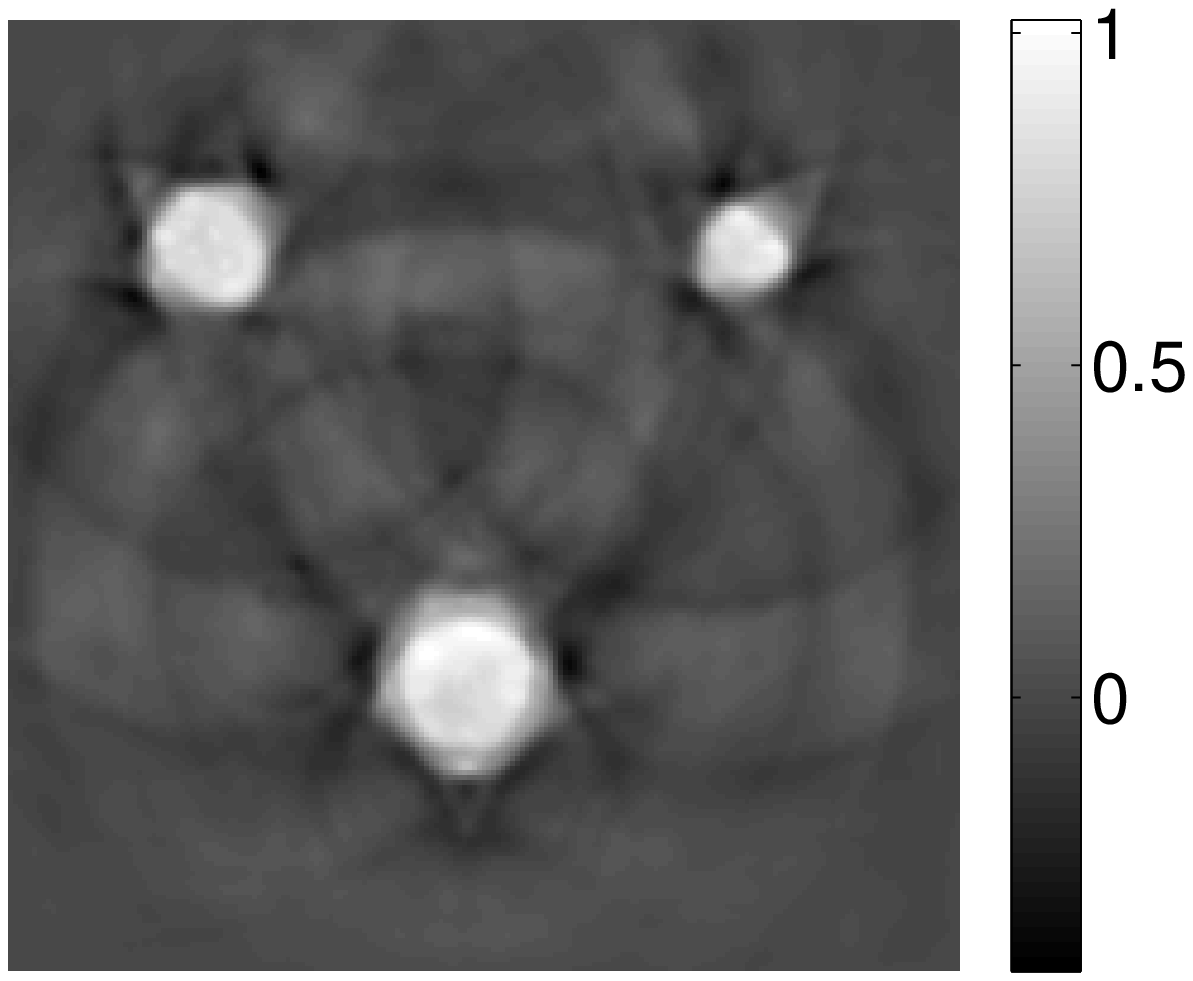}}}
  \subfigure[]{\resizebox{1.5in}{!}{\includegraphics{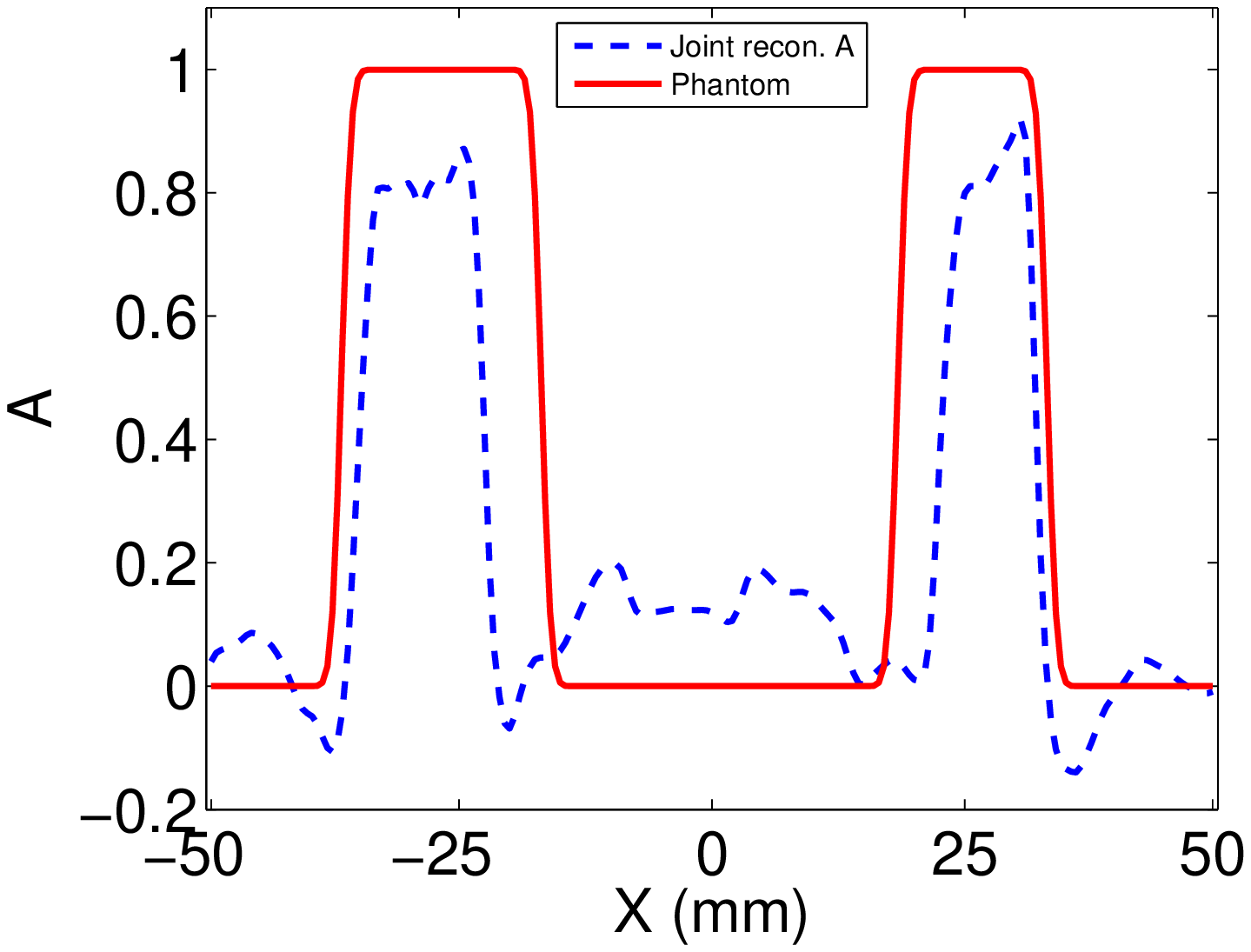}}} 
  \subfigure[]{\resizebox{1.5in}{!}{\includegraphics{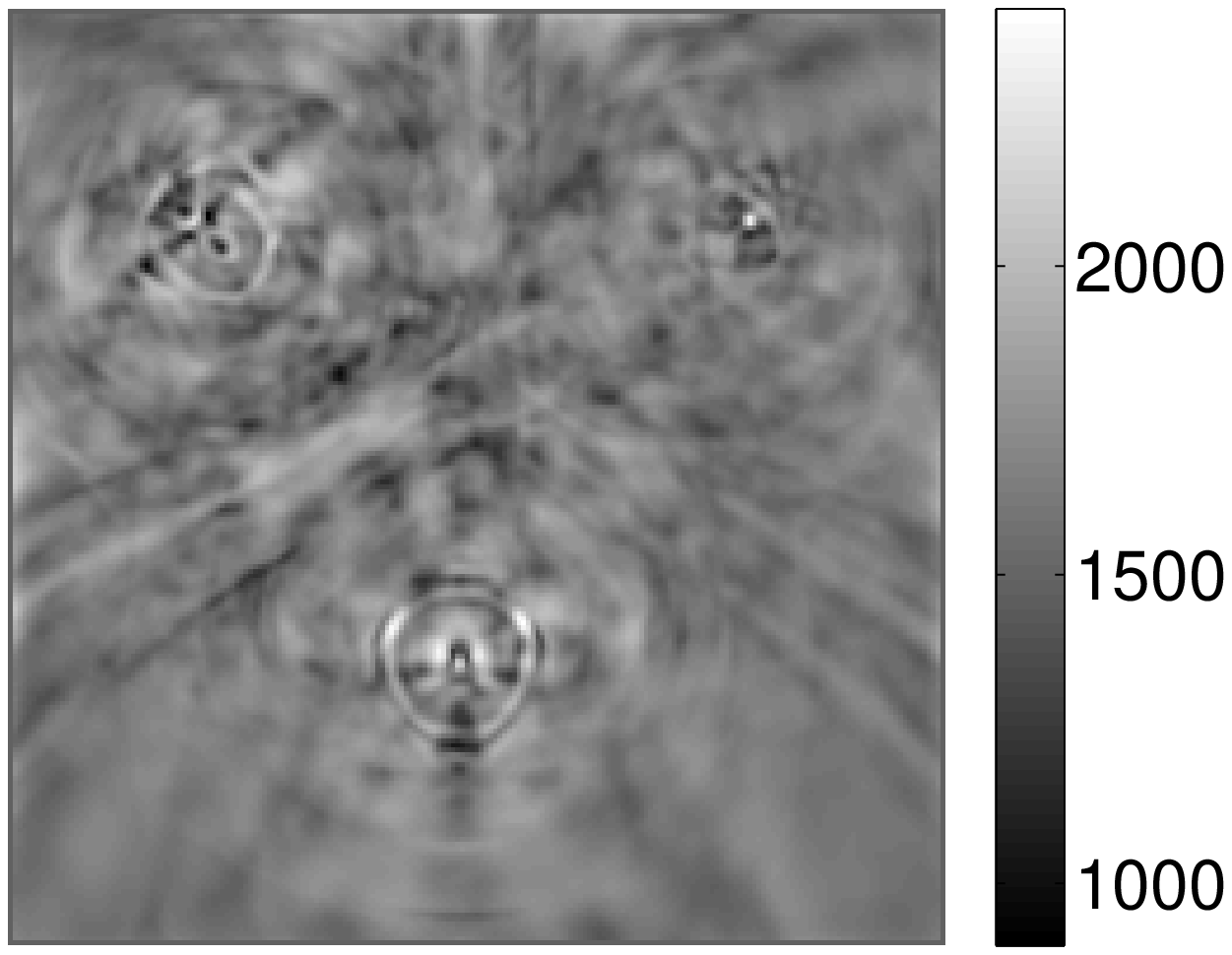}}}
  \subfigure[]{\resizebox{1.5in}{!}{\includegraphics{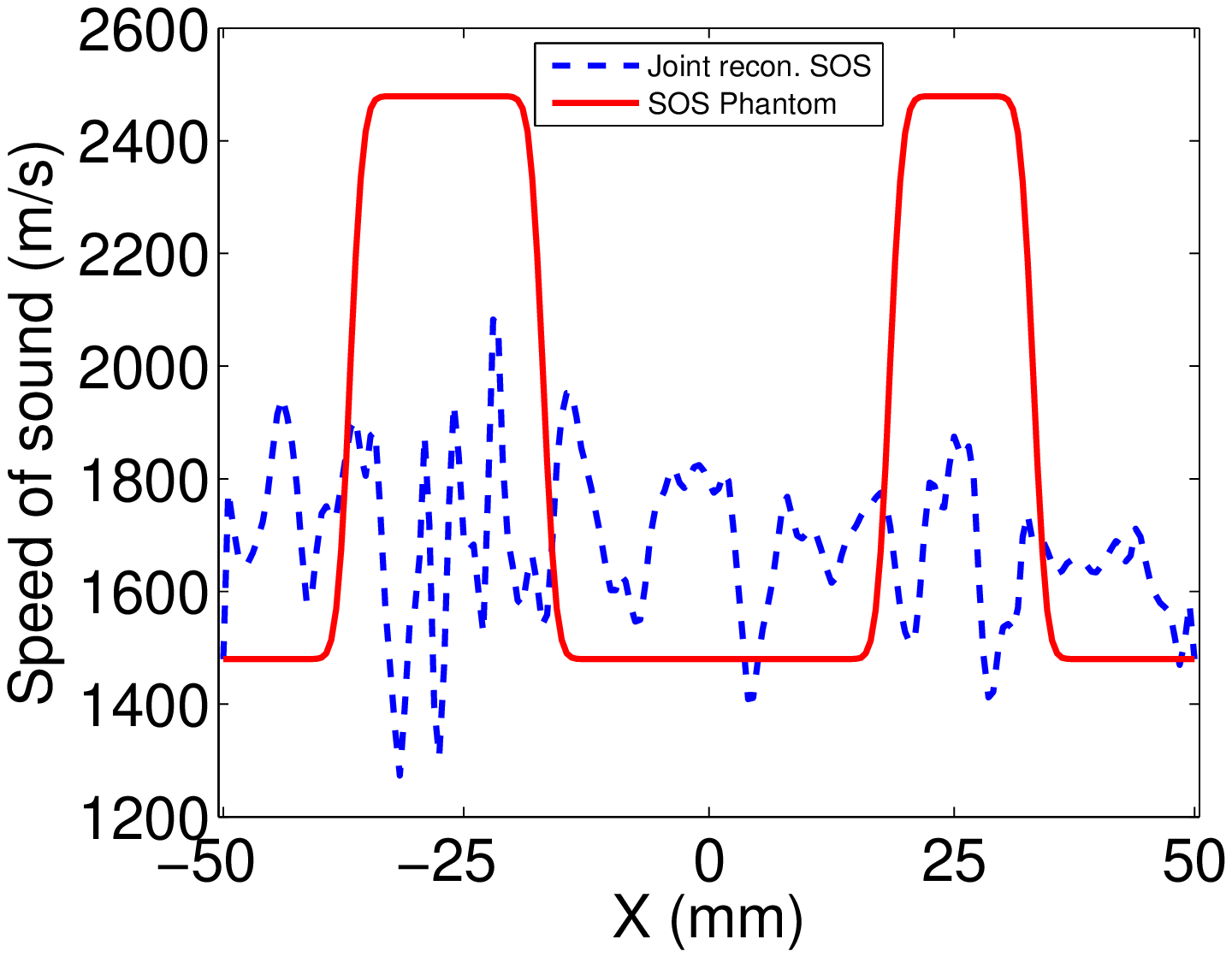}}}\\
\vskip -0.1cm
  \subfigure[]{\resizebox{1.5in}{!}{\includegraphics{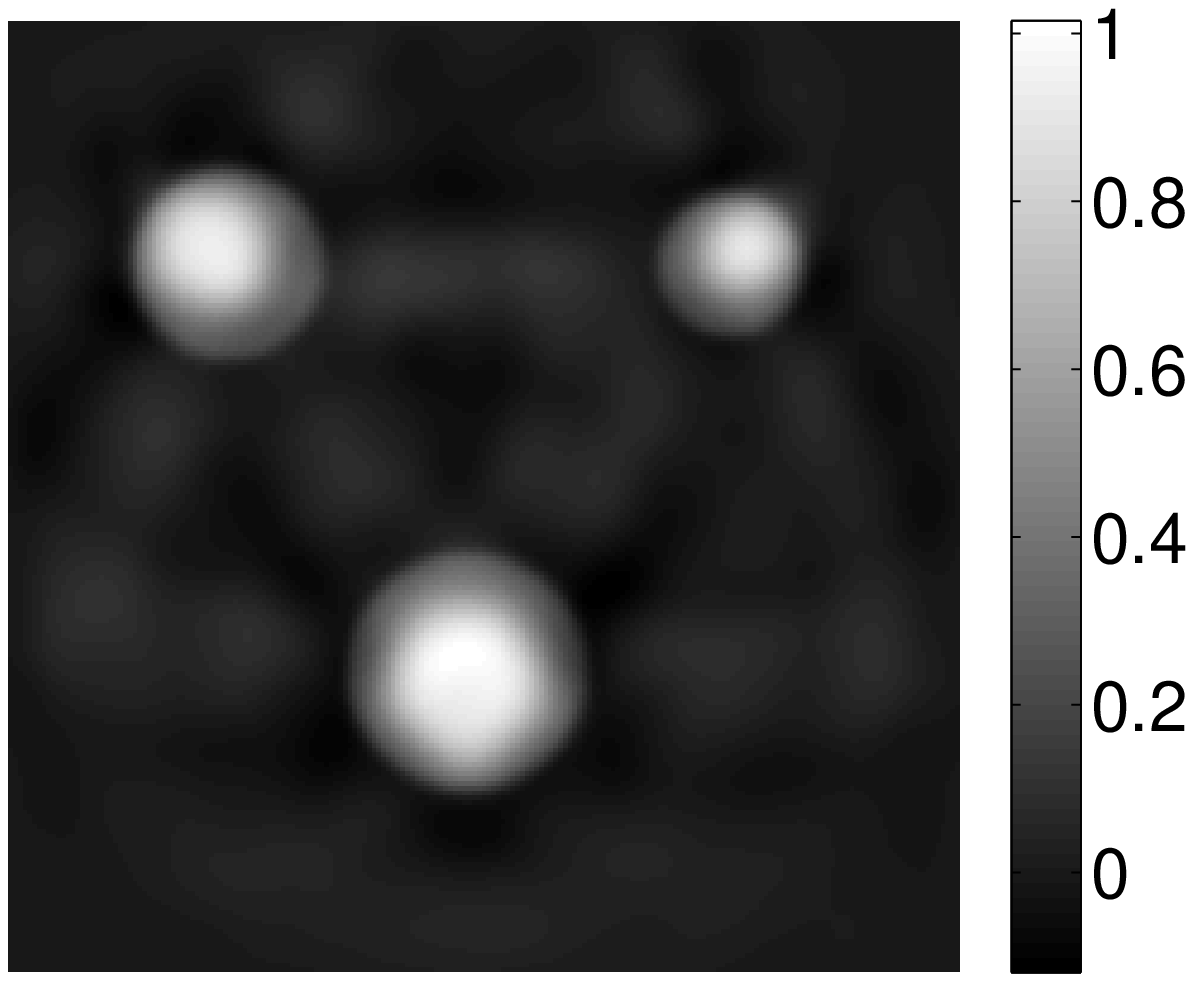}}}
  \subfigure[]{\resizebox{1.5in}{!}{\includegraphics{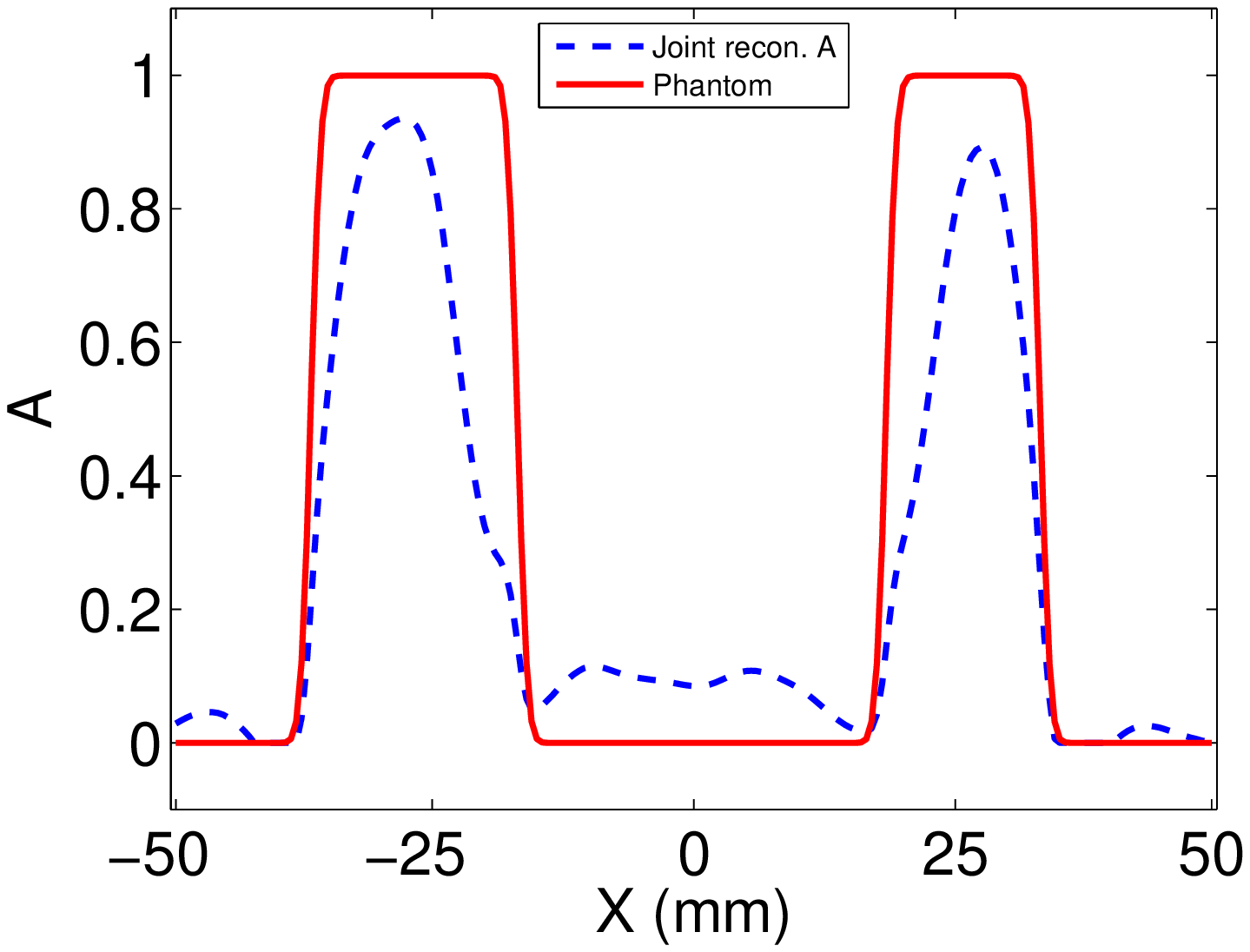}}}
  \subfigure[]{\resizebox{1.5in}{!}{\includegraphics{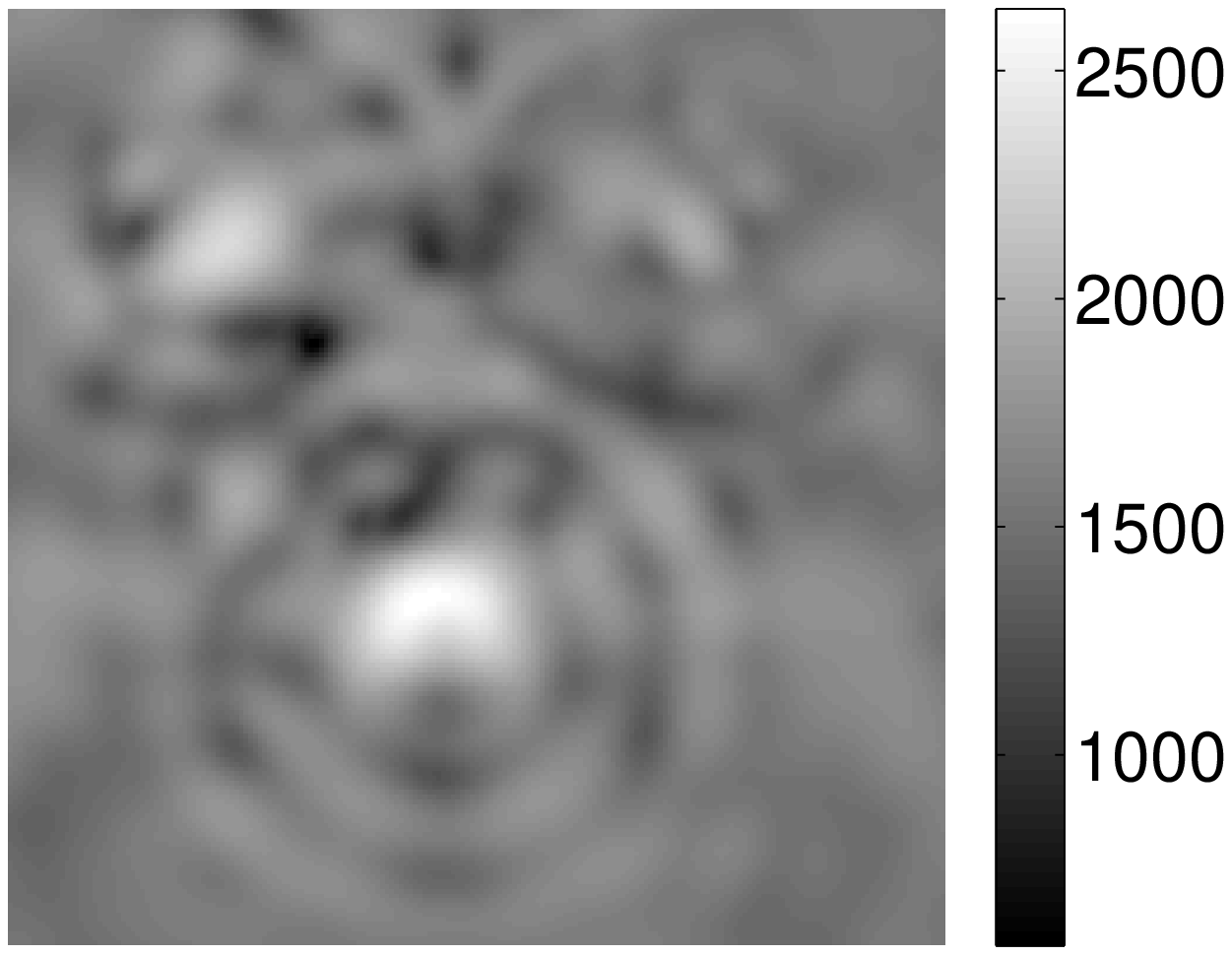}}}
  \subfigure[]{\resizebox{1.5in}{!}{\includegraphics{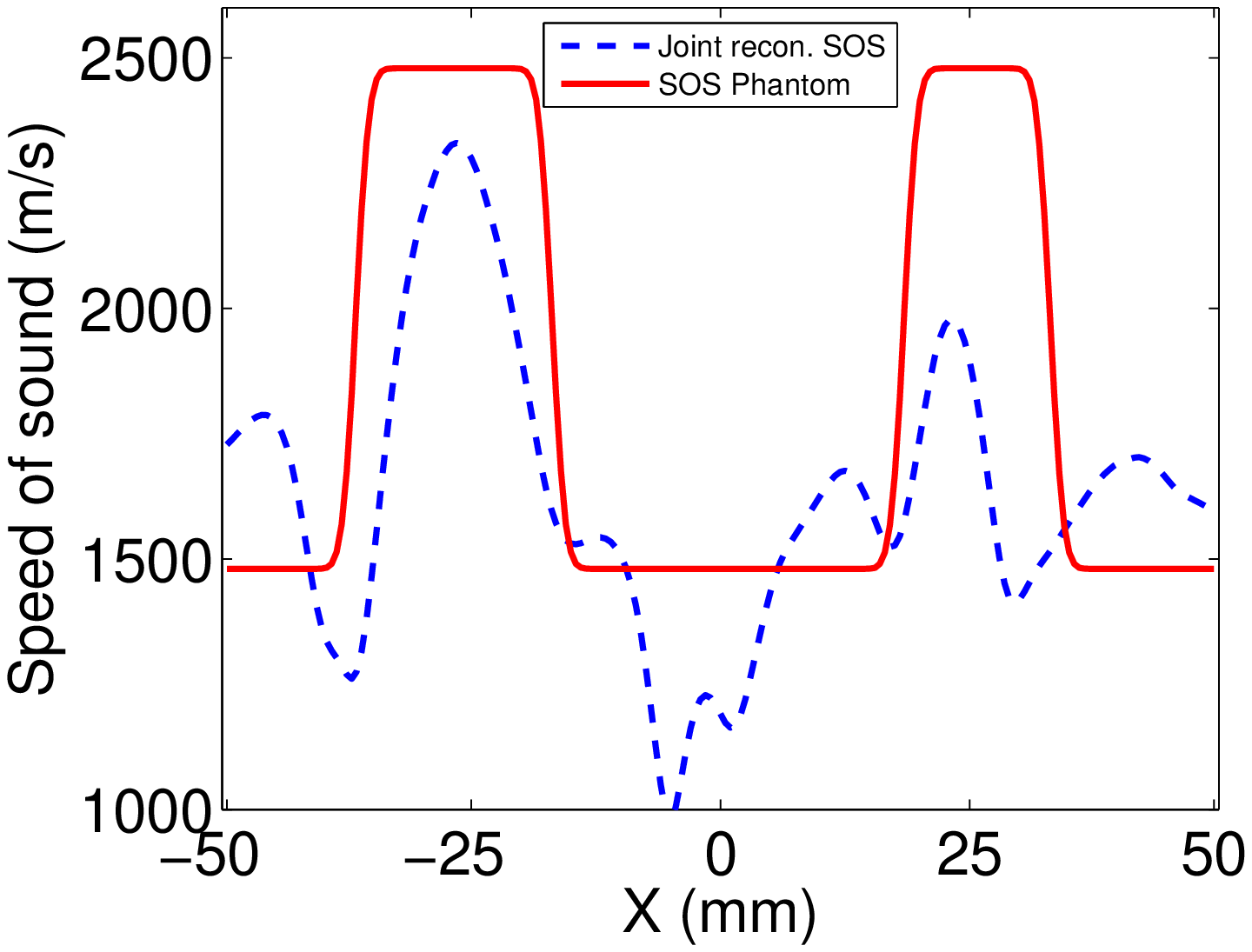}}}\\
\vskip -0.1cm
  \subfigure[]{\resizebox{1.5in}{!}{\includegraphics{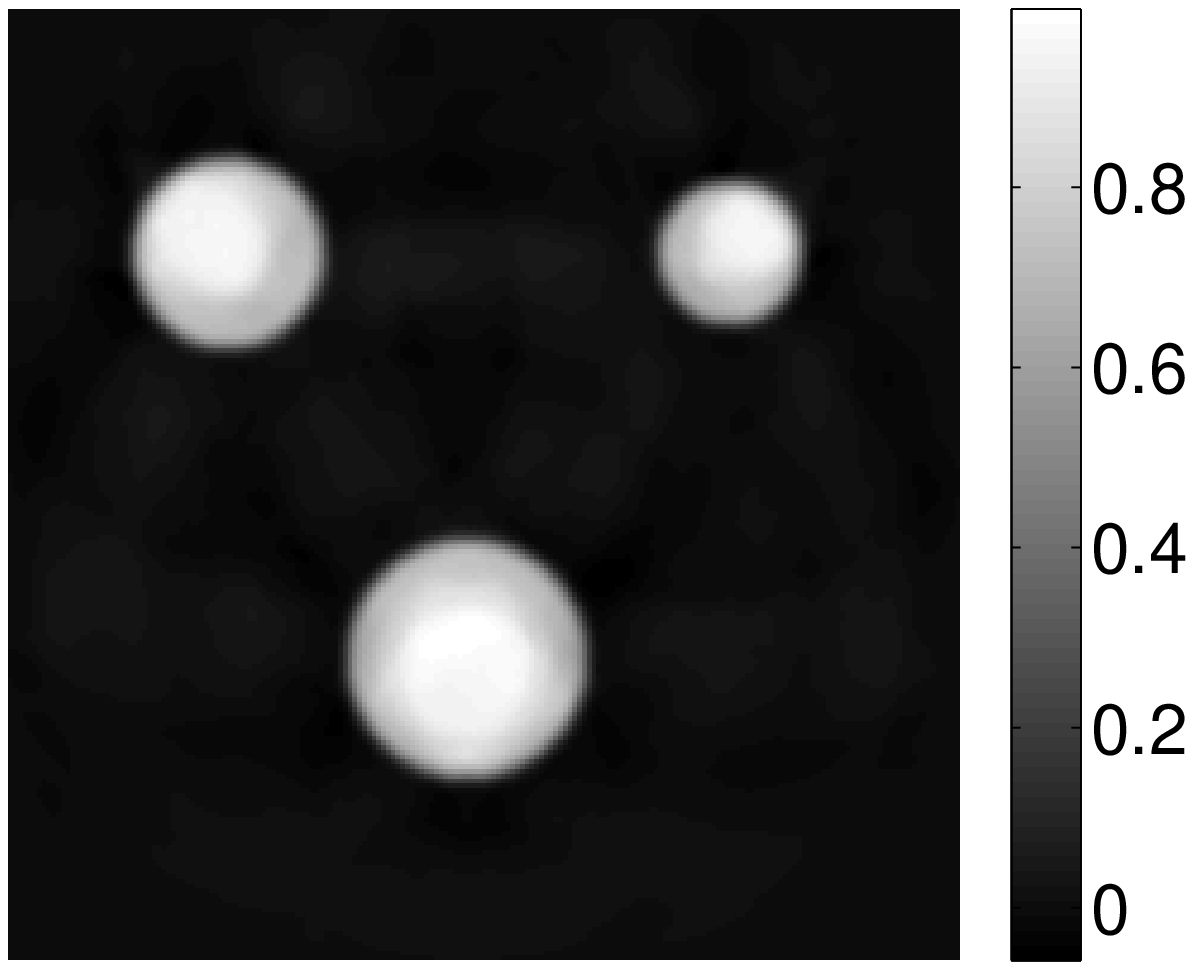}}}
  \subfigure[]{\resizebox{1.5in}{!}{\includegraphics{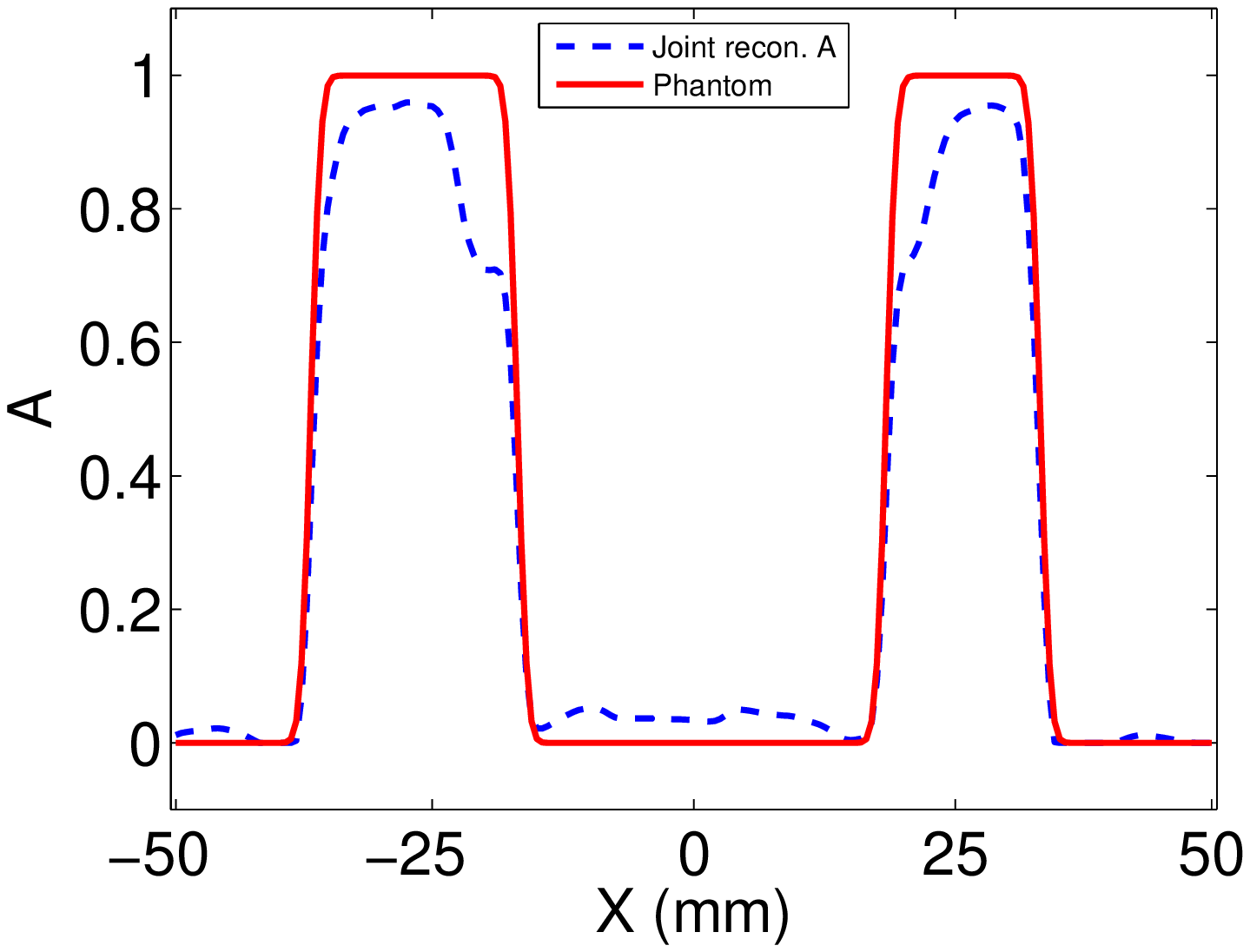}}}
  \subfigure[]{\resizebox{1.5in}{!}{\includegraphics{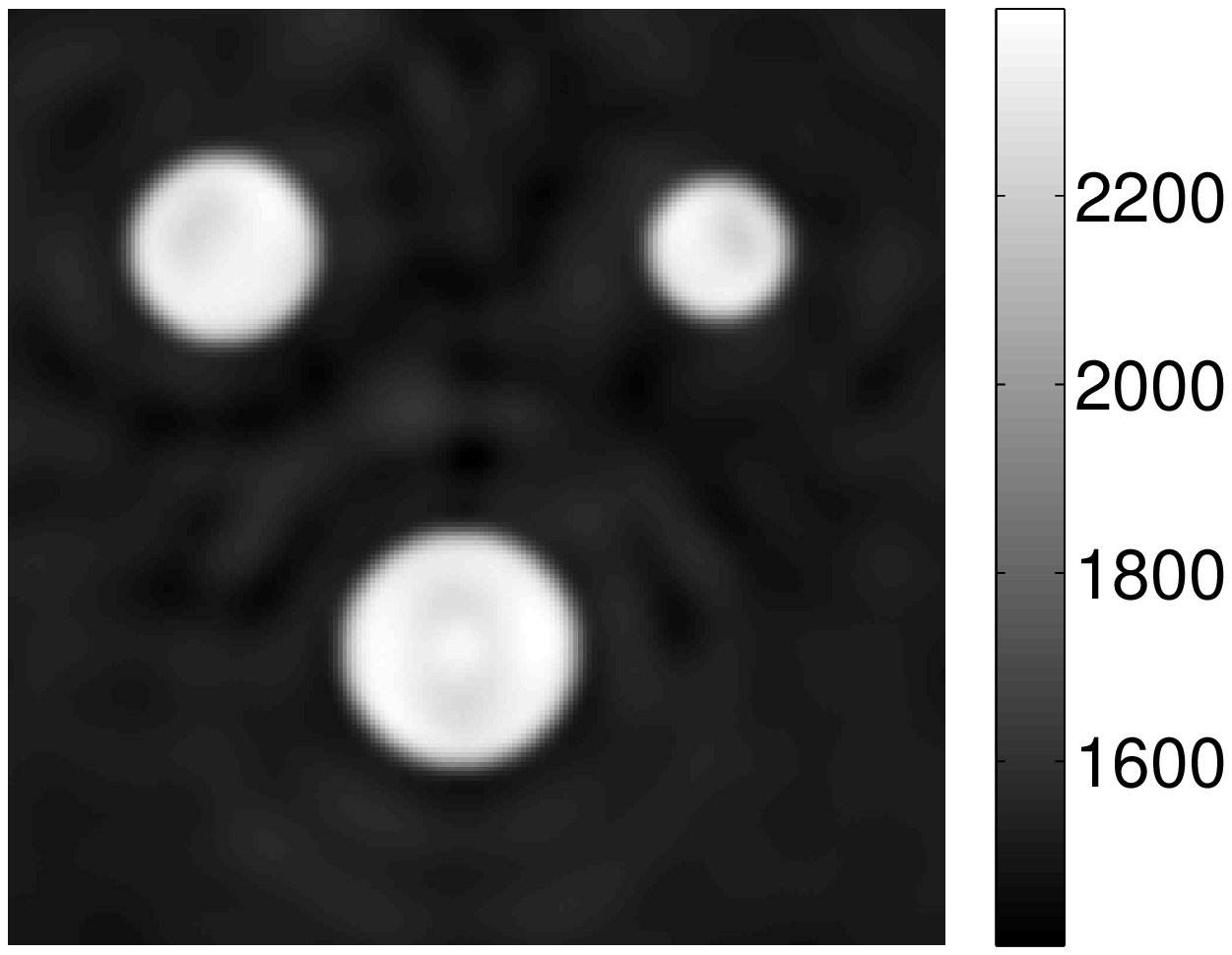}}}
  \subfigure[]{\resizebox{1.5in}{!}{\includegraphics{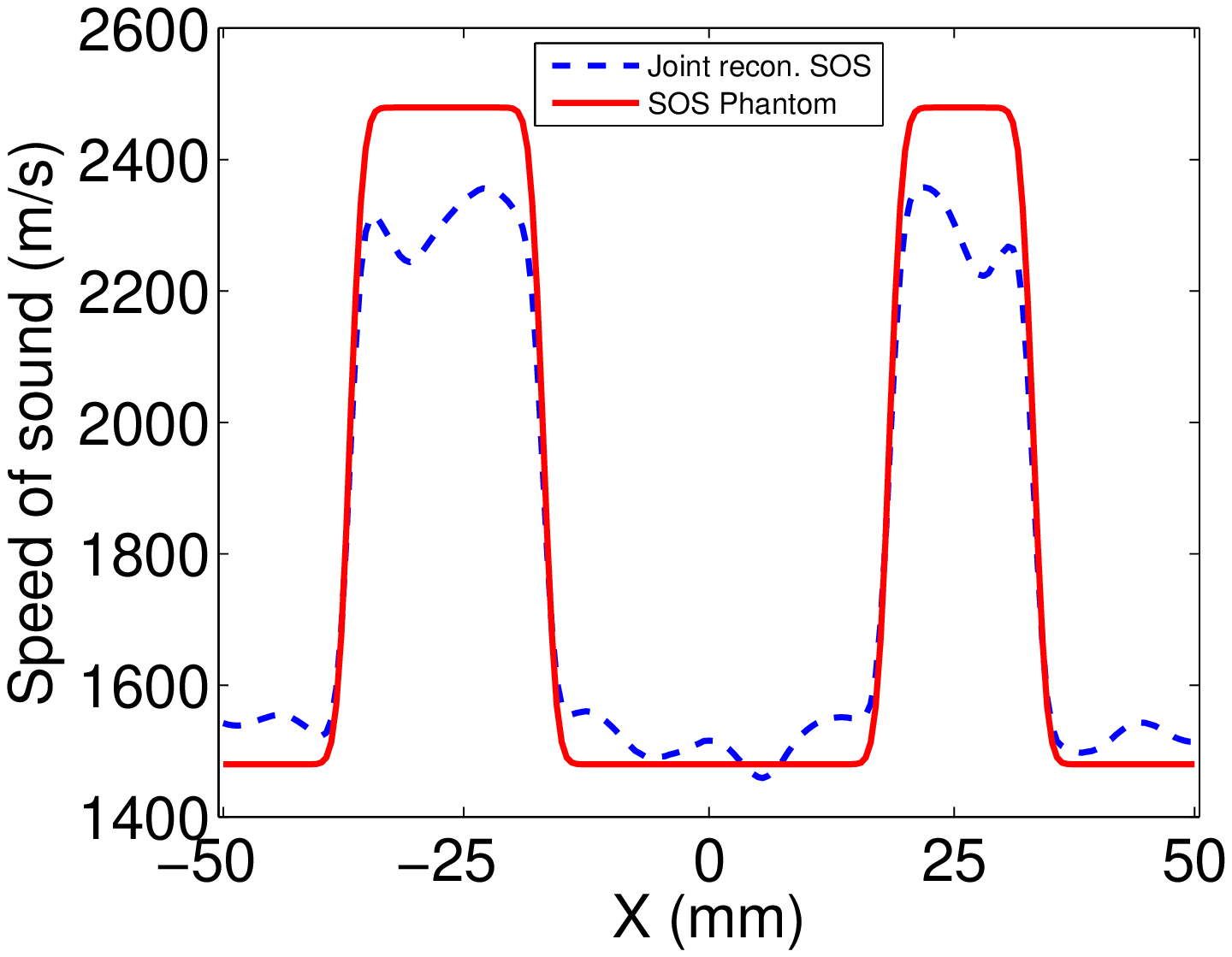}}}\\
\vskip -0.1cm
  \subfigure[]{\resizebox{1.5in}{!}{\includegraphics{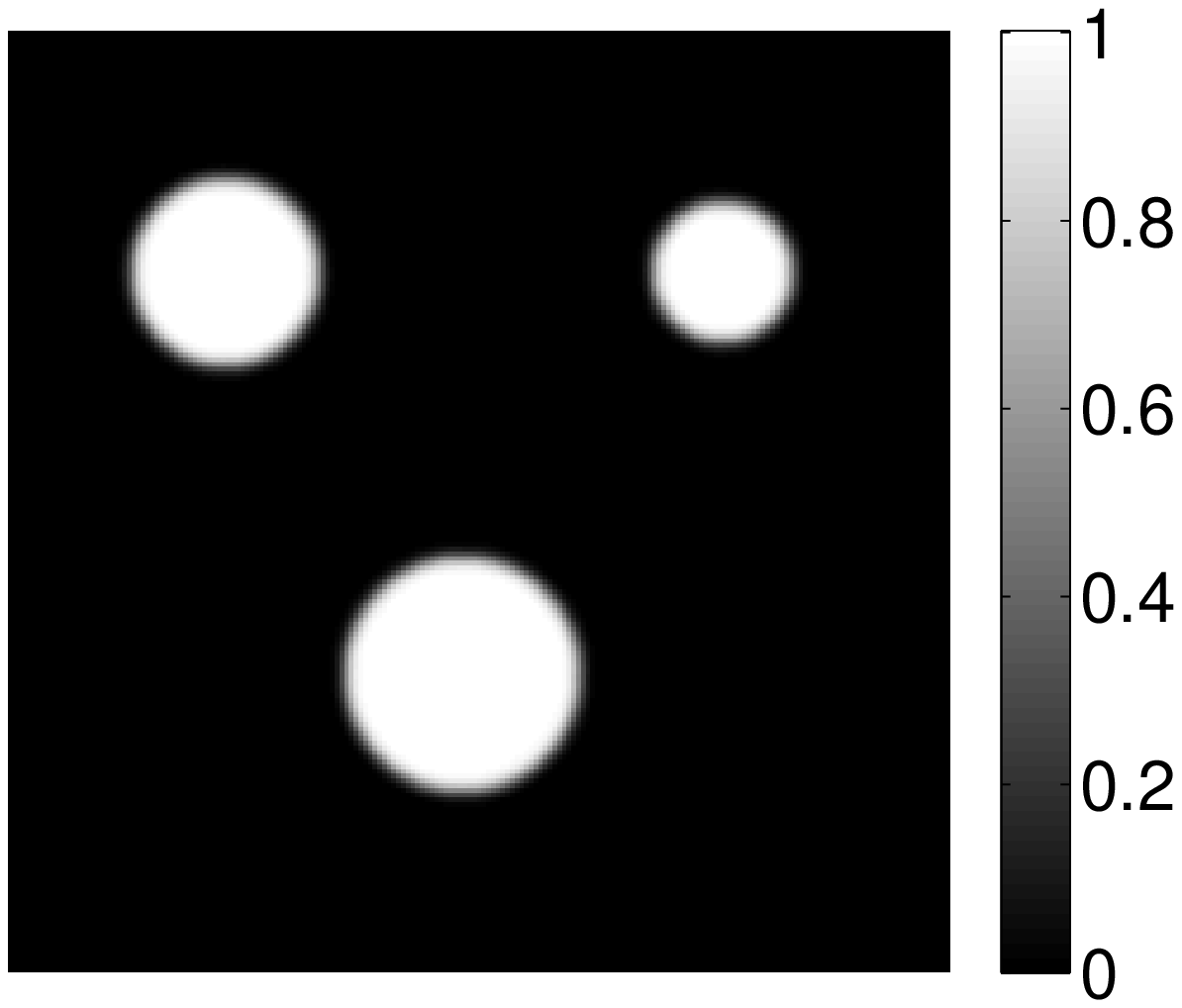}}}
  \subfigure[]{\resizebox{1.5in}{!}{\includegraphics{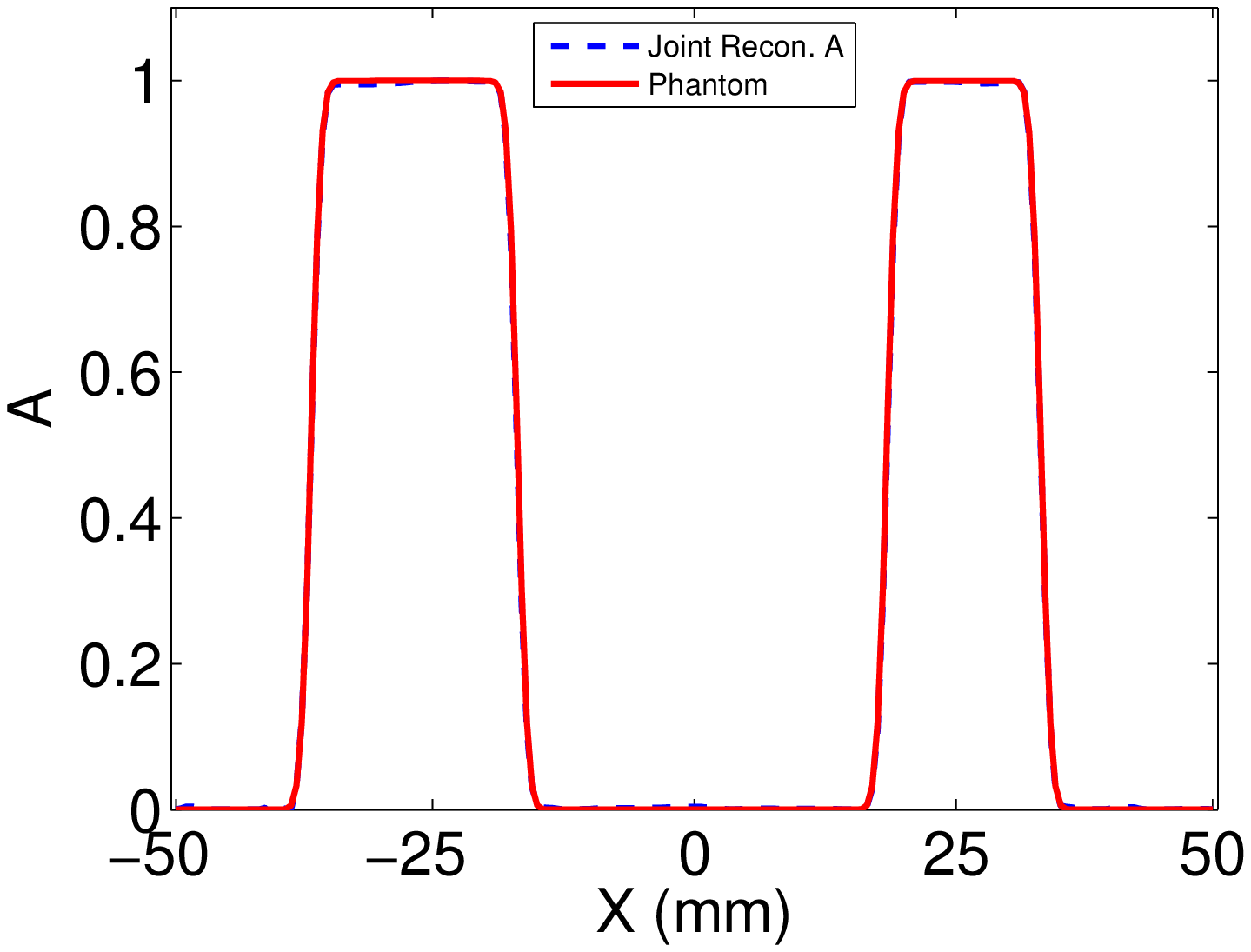}}}
  \subfigure[]{\resizebox{1.5in}{!}{\includegraphics{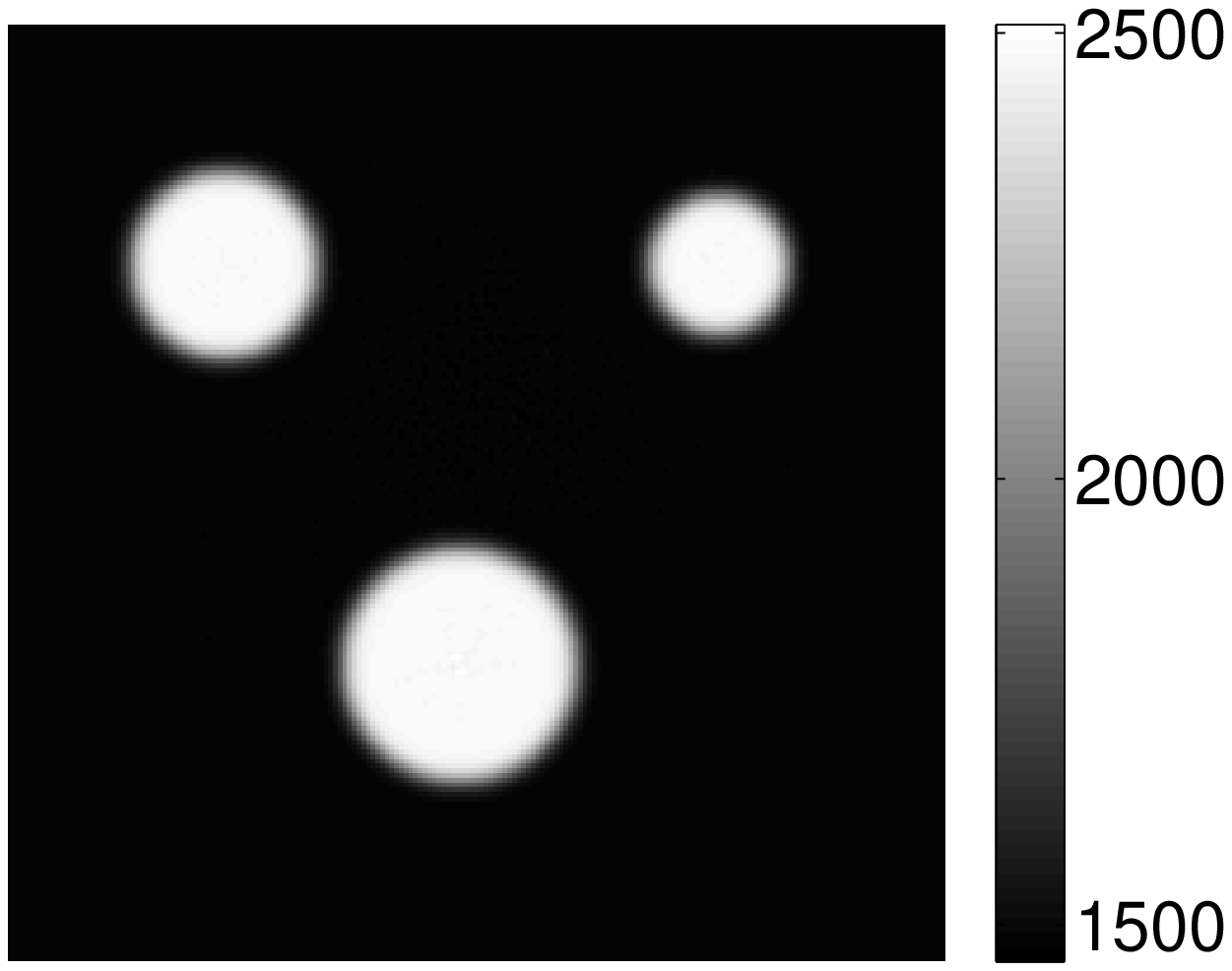}}}
  \subfigure[]{\resizebox{1.5in}{!}{\includegraphics{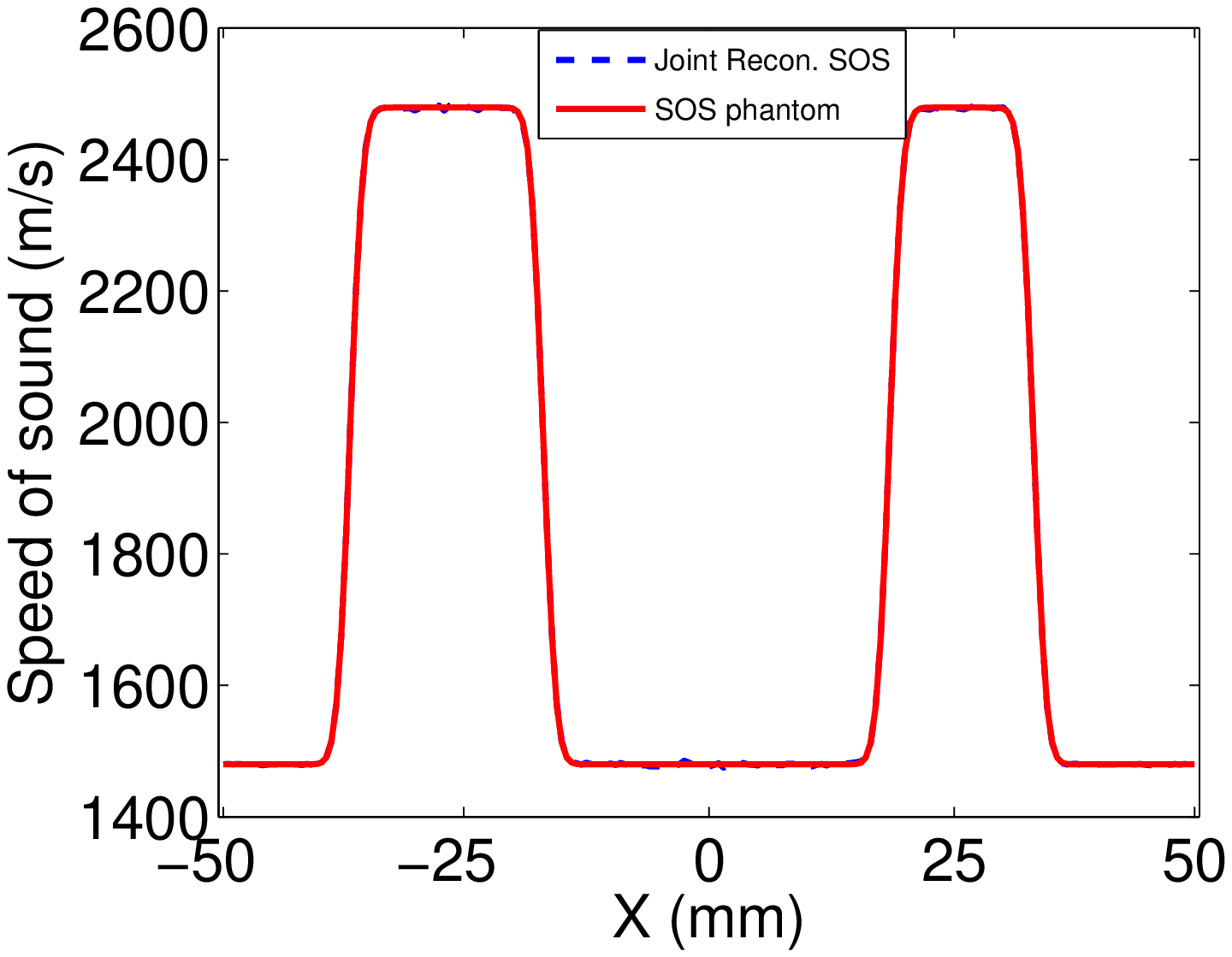}}}
%  \subfigure[]{\resizebox{2.7in}{!}{\includegraphics{iter_diskA.eps}}}
%  \subfigure[]{\resizebox{2.7in}{!}{\includegraphics{iter_diskA_pfl.eps}}}
\caption{\label{fig:jr}
Images obtained via JR from noiseless PA measurements:
The first and third columns display the estimates of $\mathbf A$
and $\mathbf c$, and the corresponding image profiles are displayed
in the second and fourth columns, respectively.
Each row corresponds to  use of
different regularization parameters $\lambda_A$ 
and $\lambda_c$. From top to bottom, $(\lambda_A, \lambda_c)$
 are $(0,10^{-5})$, $(10^{-4}, 10^{-3})$, $(0, 10^{-4})$, and $(10^{-3}, 10^{-2}$), 
respectively.
}
\end{figure*}

The numerical instability of Sub-Problem \#2, as examined
in the previous section,
implies that the solution to the  JR problem is also numerically unstable.
This was confirmed by computing solutions to the JR problem
from perfect measurement data.  The data were perfect in the
sense that they did not contain measurement noise.  Moreover,
`inverse crime' was committed in which the same forward model
and discretization parameters (pixel size $=0.5$ mm) were employed to produce the
simulation data and to conduct image reconstruction.
These data were produced by use of the phantoms shown in Fig.\  \ref{fig:disk_phantoms},
which satisfied both the support  and the k-space conjectures.
Unregularized JR was performed by use of Algorithm 1 with $\lambda_A=\lambda_c=0$
and the results are displayed in the top row of Fig. \ref{fig:jr}.
Despite the use of perfect measurement data and a favorable choice
of $\mathbf A$ and $\mathbf c$,
%where $\mathbf A$ was sufficient and inverse crime was committed, 
neither $\mathbf A$ nor $\mathbf c$ could be
accurately reconstructed.

Additional studies were conducted in which noiseless measurement
data were computed without committing inverse crime.  Namely, the
pixel size employed to simulate the measurement data was 0.5 times
the pixel size employed in the reconstruction algorithm. 
%In order to accurately reconstruct both 
%$\mathbf A$ nor $\mathbf c$, it was necessary 
%to compute regularized solutions.
%The regularized JR results  were generated without inverse
%crime and are displayed in the 2-4 row of
From these data, regularized estimates of $\mathbf A$
and $\mathbf c$ were computed.
In rows 2-4 of Fig. \ref{fig:jr}, the corresponding
regularization parameters are $\lambda_A=\lambda_c=10^{-5}$,
 $10^{-4}$,  and
$10^{-3}$, respectively.
These results 
demonstrate that the numerical instability
of the JR problem can be mitigated by 
incorporating appropriate regularization.
%Since the reconstruction 
%of $\mathbf c$ is more instable than 
%reconstruction of $\mathbf A$, the jointly
%reconstructed $\mathbf c$ is less accurate
%than the reconstructed $\mathbf A$.
However, the salient observation here is that regularization was
required to obtain accurate image estimates, even if no stochastic
measurement noise or
other significant modeling errors other than discretization effects
were introduced.

\subsection{Feasibility of JR with imperfect data}
\label{sect:JR_real}

\subsubsection{Effect of stochastic measurement noise}

\begin{figure}[h]
\centering
  \subfigure[]{\resizebox{1.5in}{!}{\includegraphics{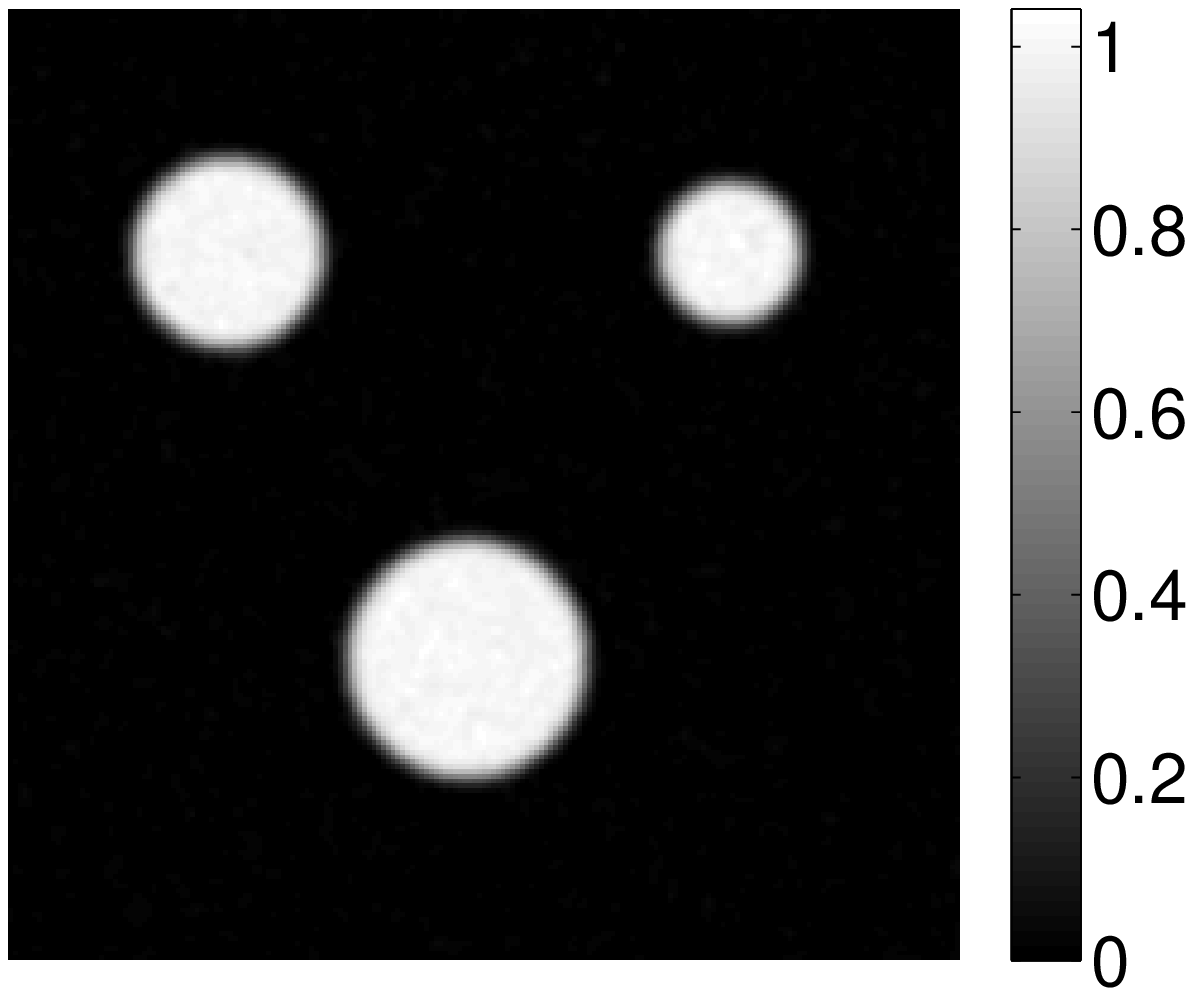}}}
  \subfigure[]{\resizebox{1.5in}{!}{\includegraphics{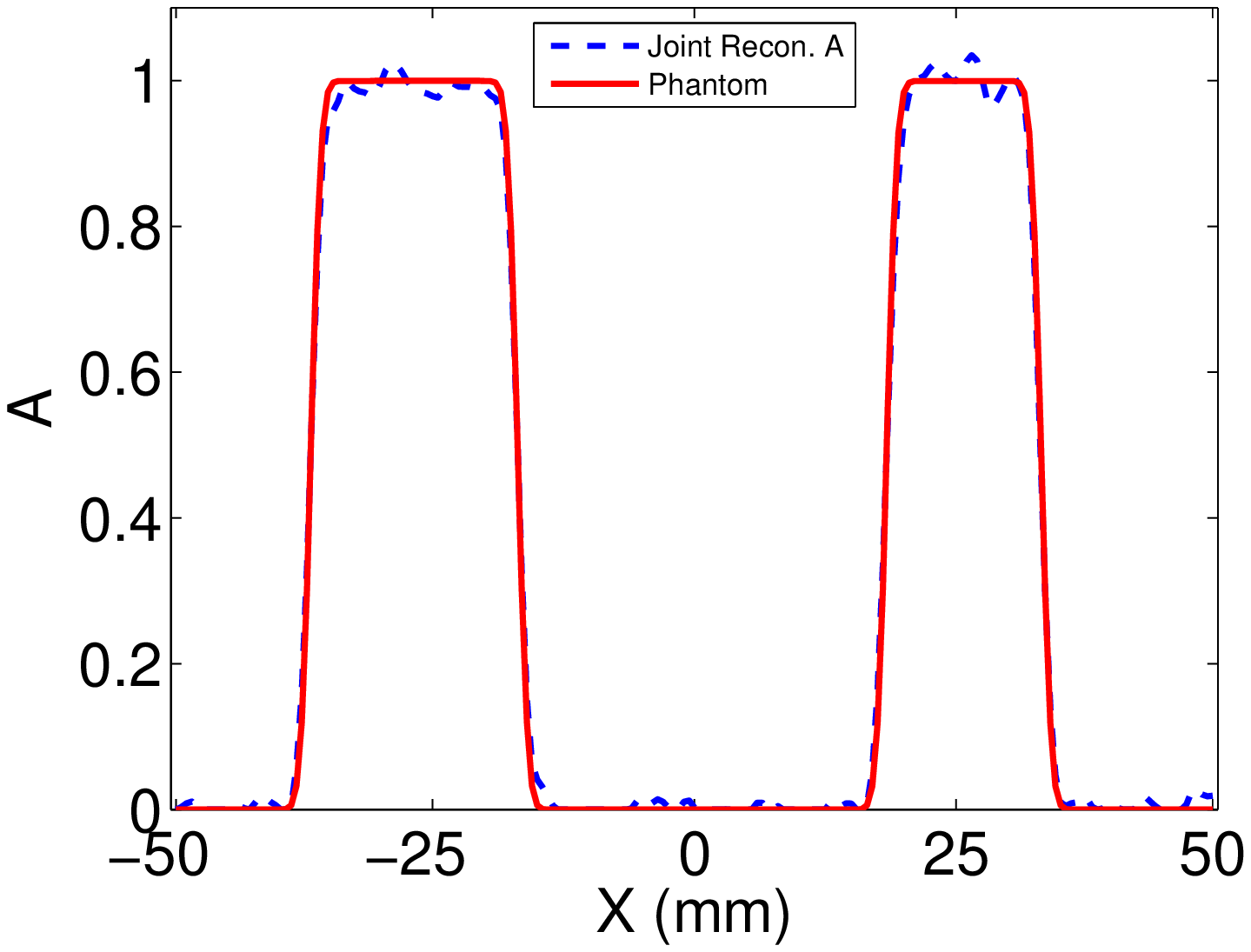}}}\\
\vskip -0.1cm
  \subfigure[]{\resizebox{1.5in}{!}{\includegraphics{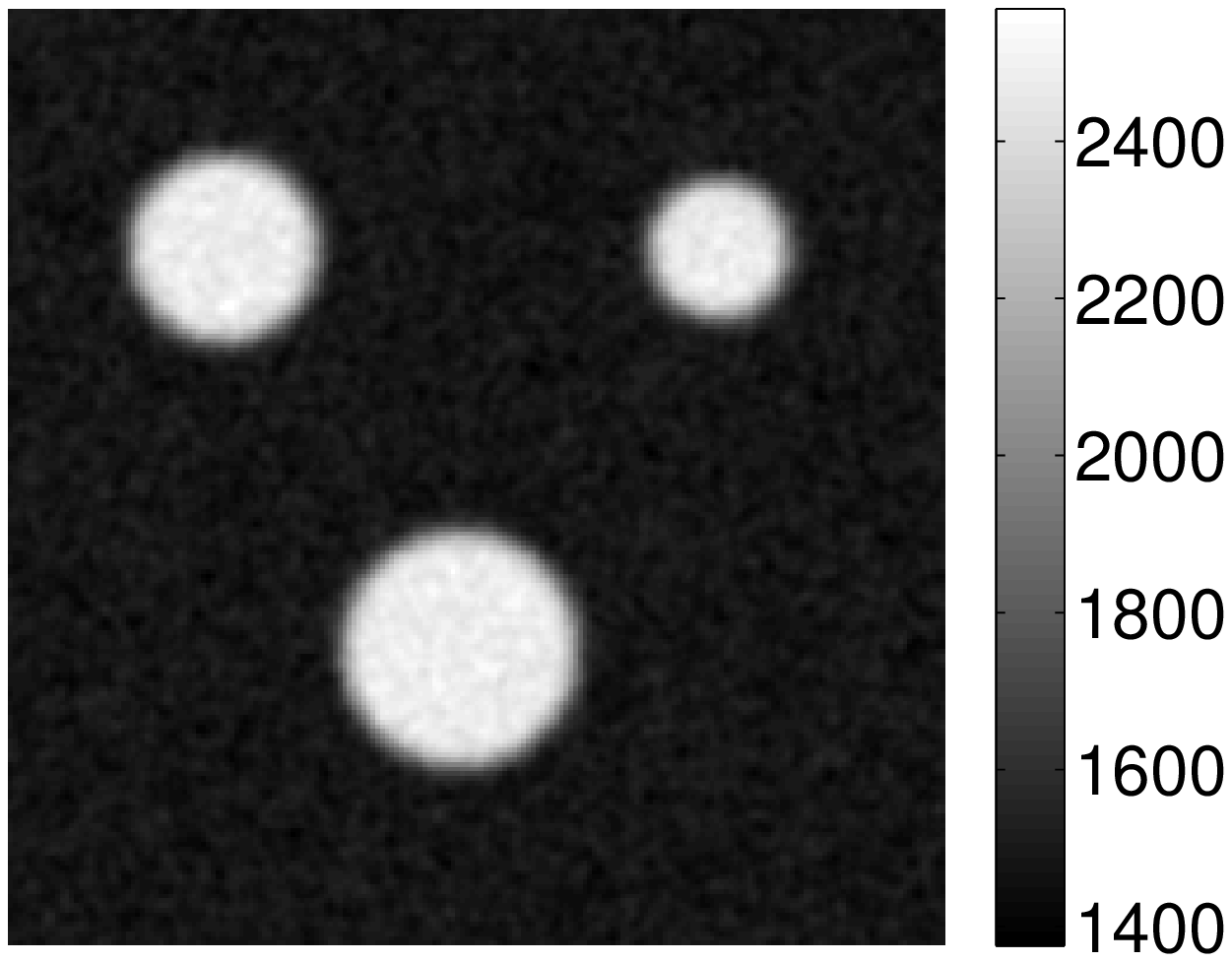}}}
  \subfigure[]{\resizebox{1.5in}{!}{\includegraphics{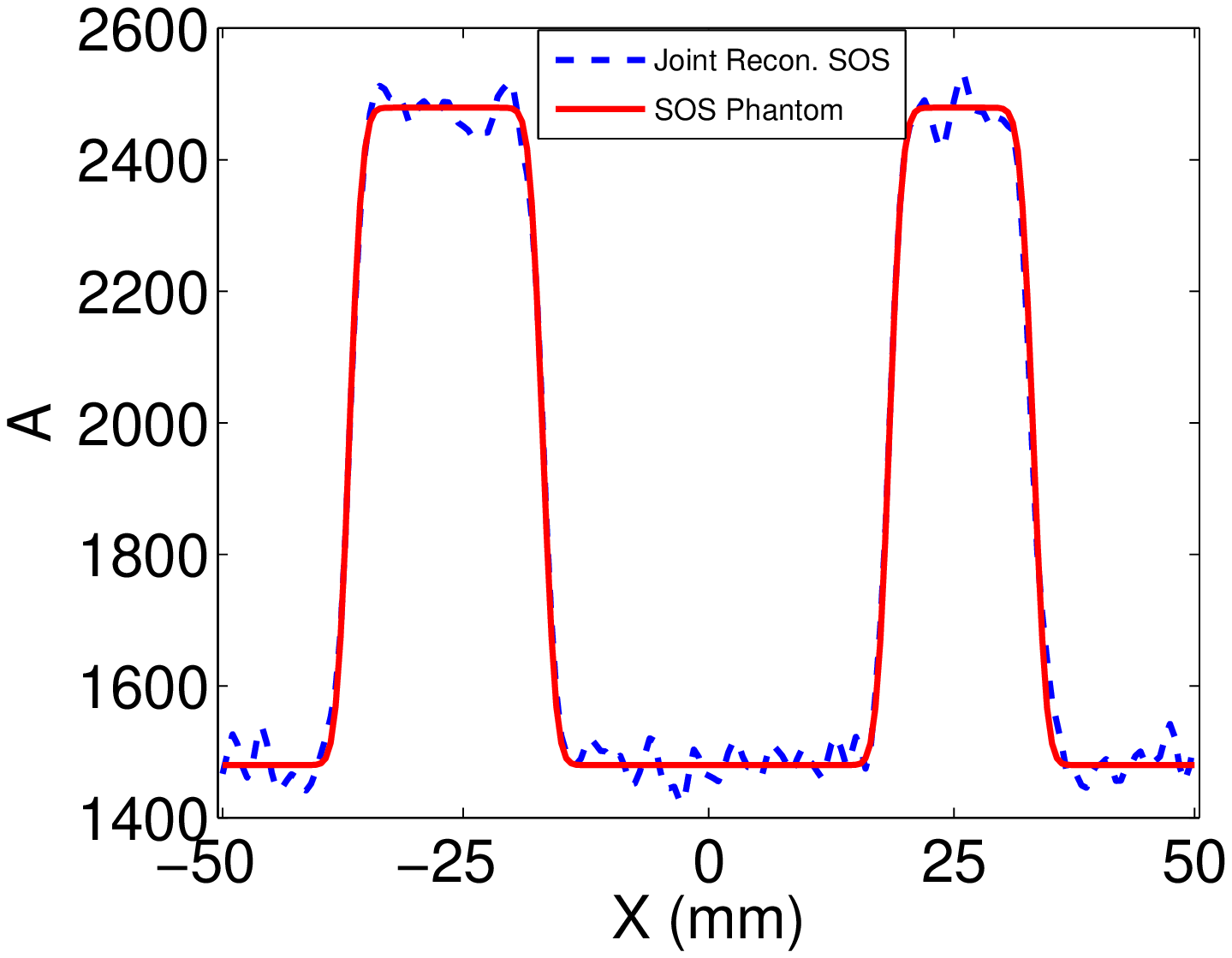}}}\\
\vskip -0.1cm
  \subfigure[]{\resizebox{1.5in}{!}{\includegraphics{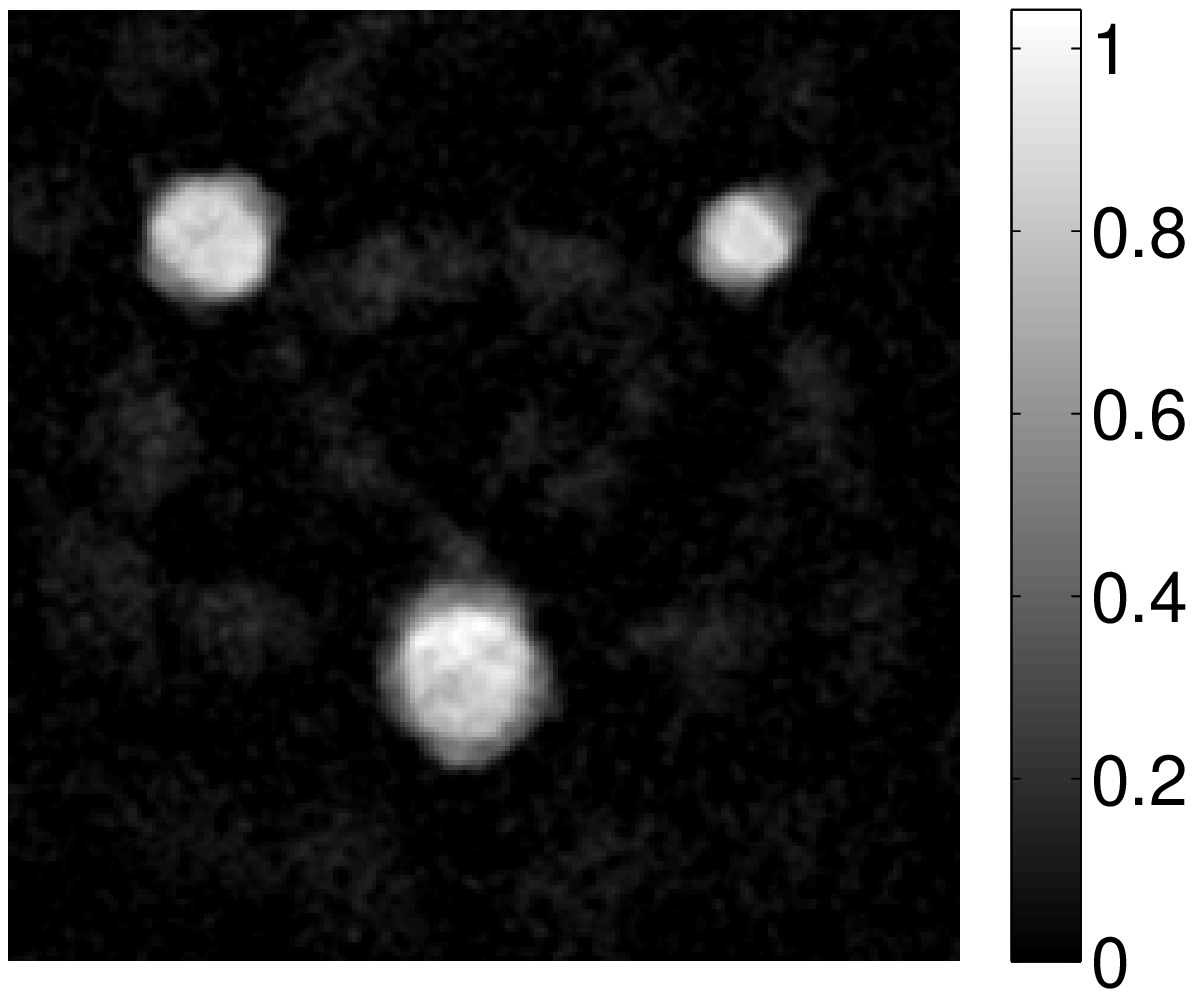}}}
  \subfigure[]{\resizebox{1.5in}{!}{\includegraphics{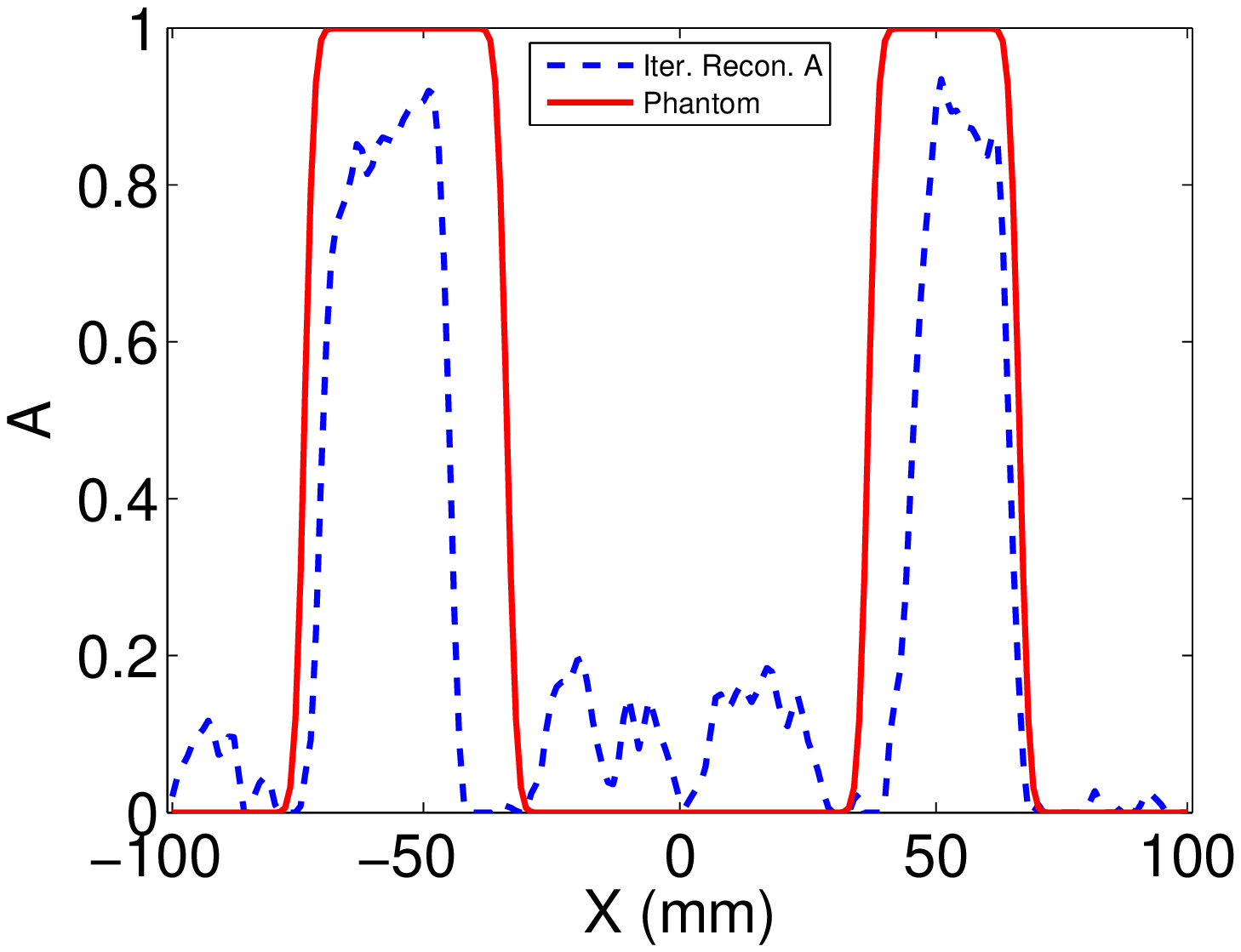}}}
\caption{\label{fig:jrn}
Images obtained via JR from noisy PA measurements:
The top and middle rows display the  reconstructed  estimates of
$\mathbf A$ and $\mathbf c$, respectively, along with the corresponding
image profiles. The bottom 
row displays an estimate of $\mathbf A$ reconstructed by a PACT
reconstruction method that employed  a constant SOS that was manually tuned to minimize RMSE.
}
\end{figure}

%The simulated PA data described above were
%contaminated by 3\% AWGN and JR was performed.
JR was performed by use of noisy versions of the PA data.
 The first and second
rows of Fig. \ref{fig:jrn} display
the estimates of $\mathbf A$ and $\mathbf c$, respectively, along
with corresponding image profiles.
These results were obtained with regularization parameters 
$\lambda_A=10^{-3}$ and $\lambda_c=10^{-2}$. 
There results show that, with appropriate regularization,
JR can be robust to the effects of AWGN.

To compare with the JR results, $\mathbf A$ 
was also reconstructed by use of a full-wave 
PACT reconstruction method \cite{HuangchaoTMIiter}
that neglected acoustic heterogeneity  by employing a constant SOS value.
This estimate of $\mathbf A$ is 
displayed in the third
row of Fig \ref{fig:jrn}.
 The  reconstruction
method assumed a constant SOS value of 1600 m/s,
 which was manually tuned  to
minimize the RMSE of the reconstructed image.
 The RMSE of the reconstructed 
$\mathbf A$ by use of the JR method and 
the PACT method ignoring SOS variations was $0.01$ and $0.21$, 
respectively.
 These results confirm that 
the estimate of $\mathbf A$ obtained by performing JR is 
more accurate than the corresponding estimate reconstructed  by use of a PACT
method assuming a constant SOS.

%\subsection{Feasibility of JR with imperfect data}
%\label{sect:JR_real}

\if 0
The above JR results show that, in an 
idealized scenario where $\mathbf A$
is sufficient and model errors can be 
neglected, it is possible to achieve 
accurate JR of $\mathbf A$ and $\mathbf c$ 
when appropriate regularization is employed. 
However, in practice, $\mathbf A$ may
not be sufficient, and there always exist 
model errors.

 In this section, we will 
investigate the feasibility of accurate 
JR in practice.
\fi

\subsubsection{Effect of other modeling errors}

In practice, besides stochastic measurement noise due to the electronics,
other forms of measurement data inconsistency will be present.
For example, the imaging model utilized in this work neglects the
spatial impulse response (SIR) of the transducers
% assumes the electrical
%impulse response (EIR) of the transducer can be perfectly deconvolved,
and  acoustic attenuation within the object and
coupling medium  \cite{KunTMI2011,rosenthal2011model,sheng2015photoacoustic,PatrickOL2006,BenPMB2011,TreebyIP2010}.
% While these factors can, in principle, be compensated
%for in the imaging model \cite{KunTMI2011,rosenthal2011model,sheng2015photoacoustic,PatrickOL2006,BenPMB2011,TreebyIP2010}, %this is beyond
%the scope of this initial study. 
Additionally, because the pressure data are assumed to represent
the measurable quantity,  the electrical
impulse response (EIR) of the transducer is assumed to be known
exactly and the deconvolution process to obtain the pressure data
is assumed to be error-free.
  These assumptions will be violated in a real-world experiment.
A study was conducted to demonstrate the performance
of JR based on our idealized imaging model in the presence of stochastic
measurement noise and
 these factors that are neglected or only
partially compensated for.

\if 0
 First, we consider the
impact of different model errors on JR
results, including neglecting acoustic
attenuation, point-like transducer assumption
and imperfect EIR deconvolution. We then
show the impact of deficient $\mathbf A$
on the JR results. Finally, we will show
the combining effects of model errors and
deficient $\mathbf A$ on the JR results.
In this section, all the JRs were conducted 
with noisy data, where 3\% AWGN were added 
to the simulated PA data.
\fi

%\emph{Effects of acoustic attenuation}:
%In many applications, acoustic attenuation 
%is not negligible \cite{TreebyIP2010,
%PatrickOL2006,BenPMB2011}. To investigate
%the effects of model error of neglecting 
%acoustic attenuation, the

The simulated PA data containing the effects of these
physical factors was generated as follows.

\begin{enumerate}
\item  Simulated PA data were generated in a lossy medium.
The  $\mathbf A$ and $\mathbf c$ phantoms shown
in Fig.\ \ref{fig:disk_phantoms} were employed.
Acoustic attenuation was introduced by use of an
 acoustic attenuation coefficient
$\alpha$ that was described by a frequency 
power law of the form $\alpha(\mathbf r, f)
=\alpha_0(\mathbf r) f^y$ \cite{SzaboJASA1994}.
The frequency-independent attenuation coefficient
$\alpha_0 = 10$ dB MHz and the power law exponent
$y=2.0$ were employed in the data generation,
which correspond to the values of $\alpha_0$
and $y$ in human kidneys that have the strongest
acoustic attenuation among typical biological
tissues \cite{SzaboBook2004}.  The simulated PA data were
contaminated by 3\% AWGN. 
\item To model the effects of the SIR,
the simulated PA data, described above, were computed on 
a grid with a pitch of 0.1 mm and recorded 
by 4000 transducers that were evenly distributed 
on the sides of a square with side length 
100 mm. The recorded data from every 20 
consecutive transducers were then averaged 
to emulate the SIR of a 2 mm line 
transducer.
% By use of the averaged data, 
%JR was conducted with regularization parameters 
%$\lambda_A=10^{-3}$ and $\lambda_c=10^{-2}$.
%Figure \ref{fig:jr_sir} displays the JR
%results, which show that, when the 
%transducer aperture size is small,
%the proposed JR method is robust to 
%the model error of neglecting SIR.
Note that this simplified model of the SIR accounts only for the
 averaging of the pressure data over the active surface
of a transducer \cite{GHarris1981} and neglects other
factors that may contribute to the SIR of a real-world transducer.

%\emph{Effects of EIR}: In addition to the
%SIRs, ultrasonic transducers also have EIRs
%in practice, which model the the electrical 
%responses of the piezoelectric transducers.
%Before performing JR, EIR needs to deconvolved
%from the measured voltage signals to recover
%the detected pressure data. However, the EIR 
%of a transducer usually cannot be accurately 
%measured.

\item The simulated PA data containing {the acoustic attenuation effects,}
the SIR effects and measurement noise were
 convolved with an EIR of an actual 
transducer \cite{Conjusteau2009, KunTMI2011}. 
The convolved data were then deconvolved by 
use of a curvelet deconvolution technique 
\cite{KunCurvelet}. In the deconvolution,  a perturbed EIR 
was employed that was produced by adding 2\% Gaussian 
noise into the spectrum of the original 
EIR.  Figure \ref{fig:eir}(a)  displays
the perturbed and original EIR  
and Fig.\  \ref{fig:eir}(b) displays the deconvolved and true pressure signal 
for a particular transducer location.
\begin{figure}[h]
\centering
  \subfigure[]{\resizebox{1.5in}{!}{\includegraphics{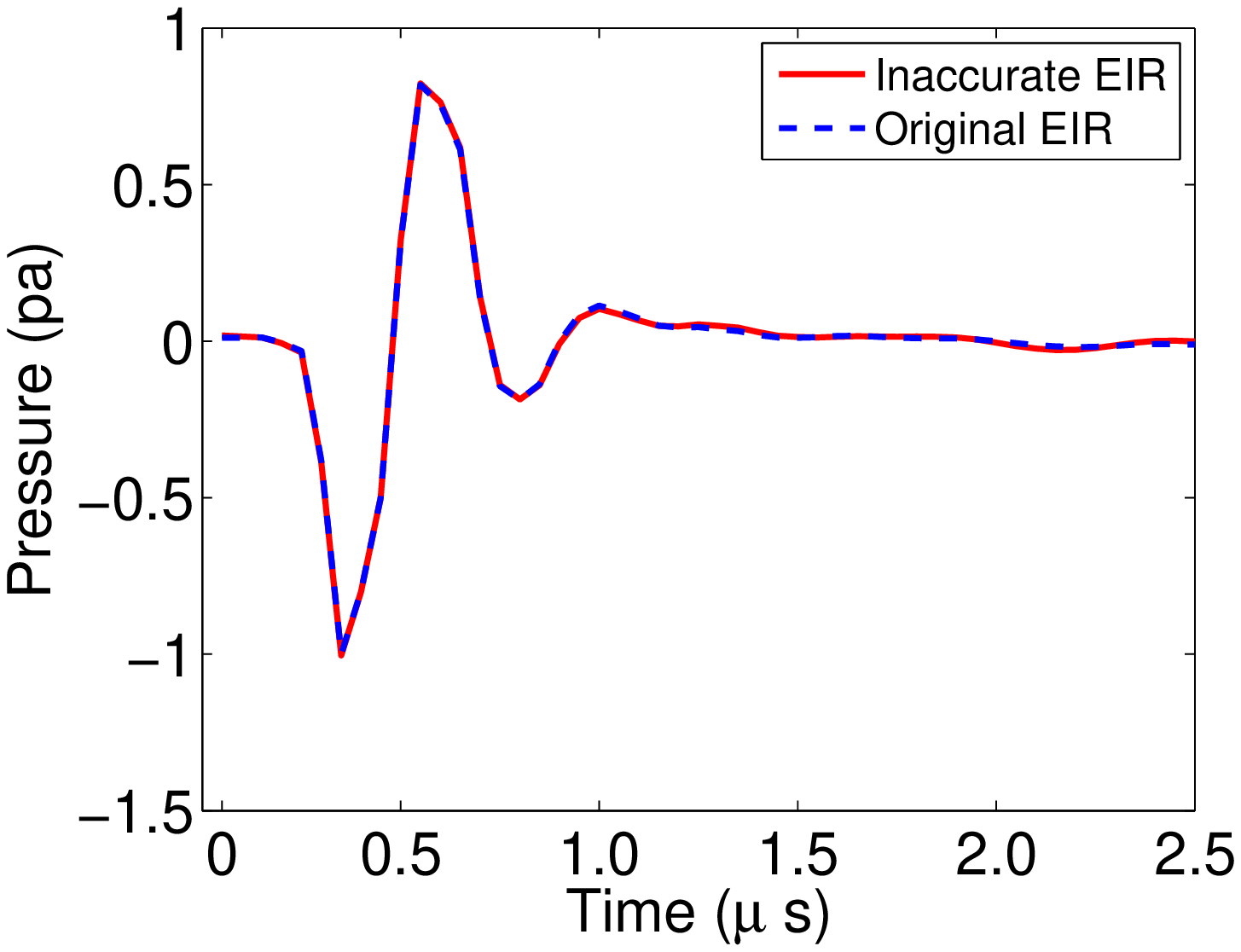}}}
  \subfigure[]{\resizebox{1.5in}{!}{\includegraphics{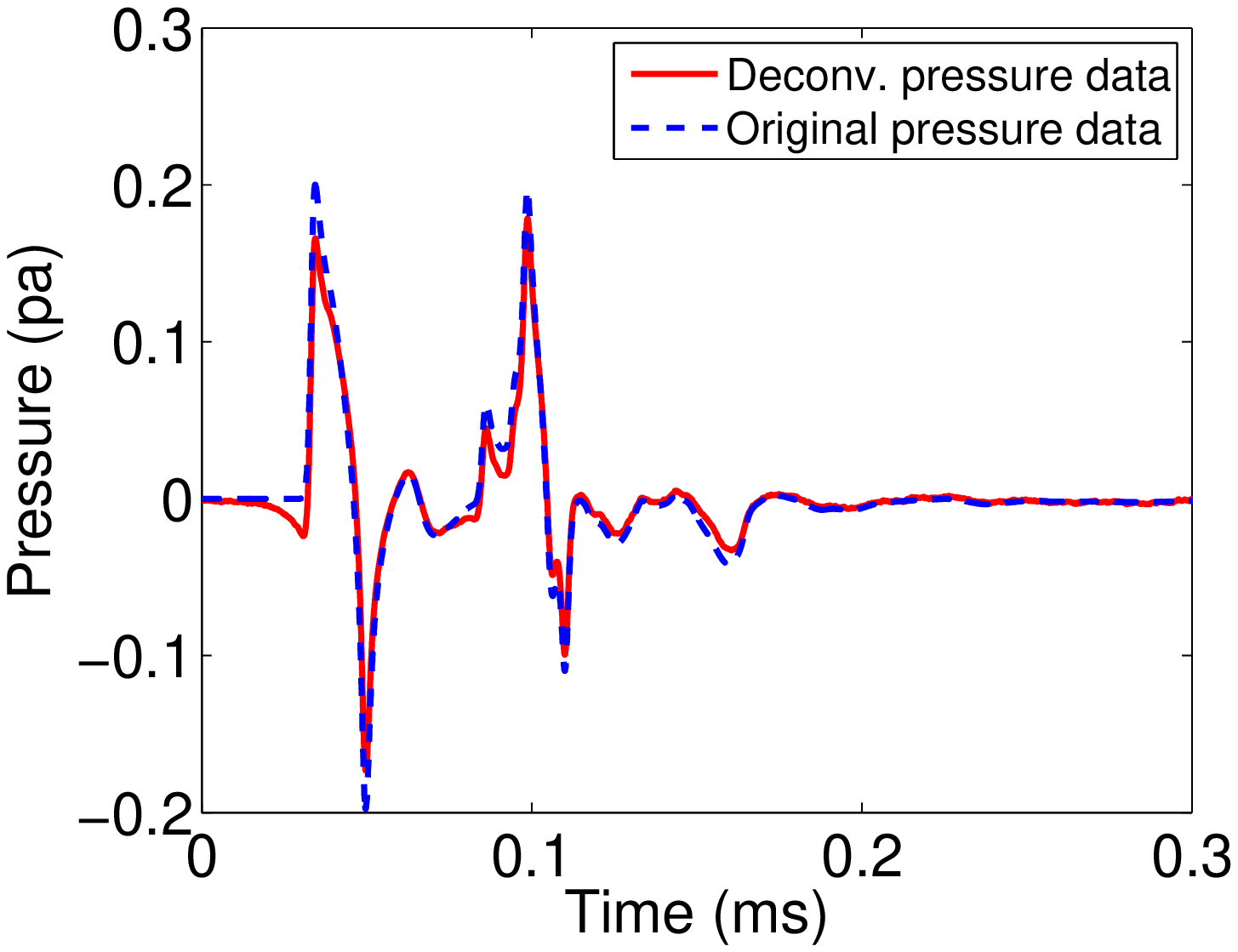}}}
\caption{\label{fig:eir}
Panel (a): inaccurate EIR compared to 
the original EIR. Panel (b): deconvolved 
pressure data by use of the inaccurate EIR
compared to original pressure data.
}
\end{figure}

\if 0 By use of the deconvolved
data, JR was conducted with regularization parameters 
$\lambda_A=10^{-2}$ and $\lambda_c=10^{-1}$.
Figure \ref{fig:jr_eir} displays the JR
results, which show that the model error
of inaccurate EIR deconvolution has larger
impact on the JR results than acoustic
attenuation or SIR effects. This can be
explained by the fact that both the JR and 
EIR deconvolution are ill-conditioned problems, 
so small errors in the EIR measurement
could be significantly amplified in the
final JR results.
\fi
\end{enumerate}

\if 0
\emph{Effects of combined model errors}:
Usually, the model errors of neglecting
acoustic attenuation, neglecting SIR and
inaccurate EIR deconvolution all exist
in practice. To investigate the effects
of the combined model errors, the above 
procedures were repeated to first generate
attenuated data, which were then averaged 
to emulate the SIR effects. The averaged,
attenuated data were used to generate
the inaccurately deconvolved data, which
were employed for
\fi

\begin{figure}[h]
\centering
  \subfigure[]{\resizebox{1.5in}{!}{\includegraphics{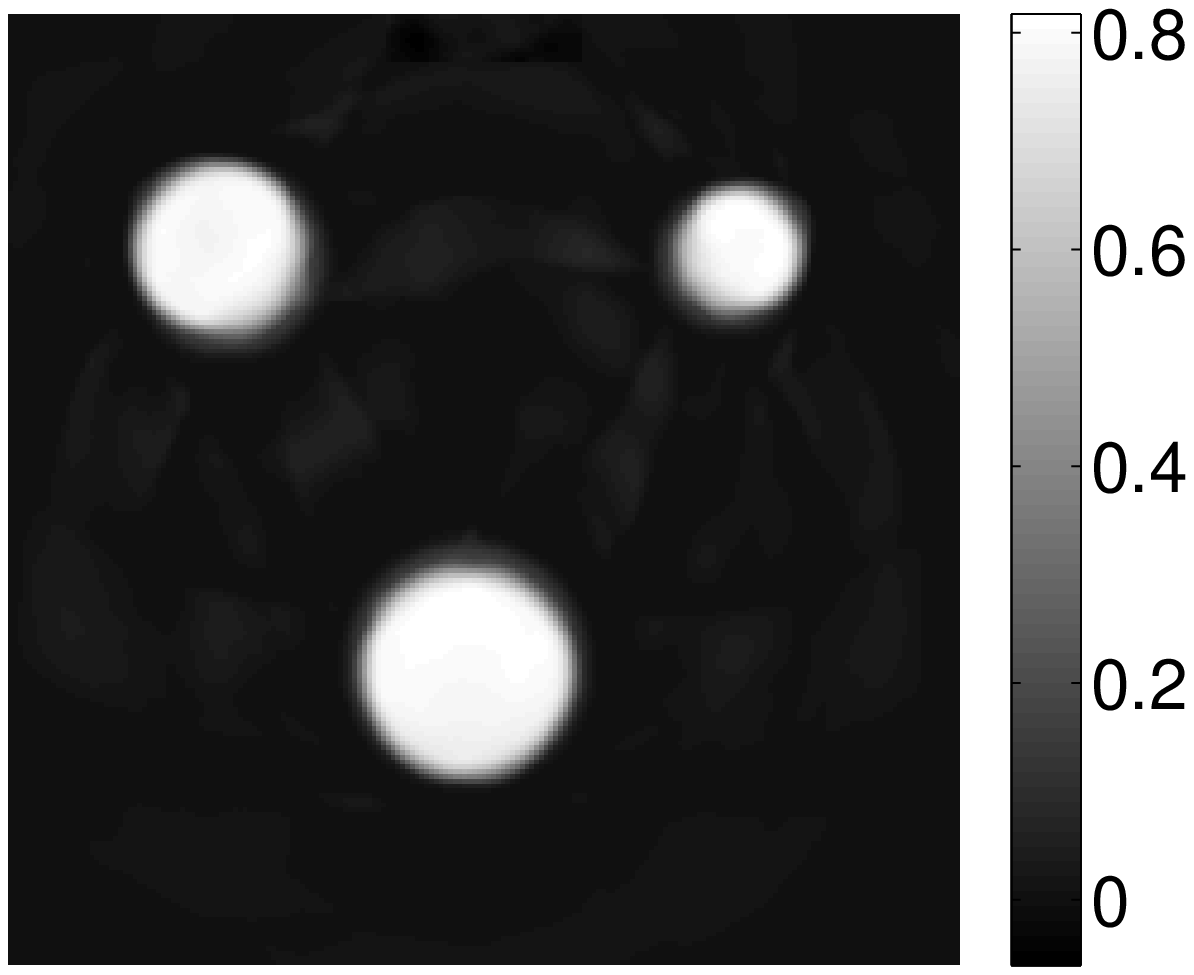}}}
  \subfigure[]{\resizebox{1.5in}{!}{\includegraphics{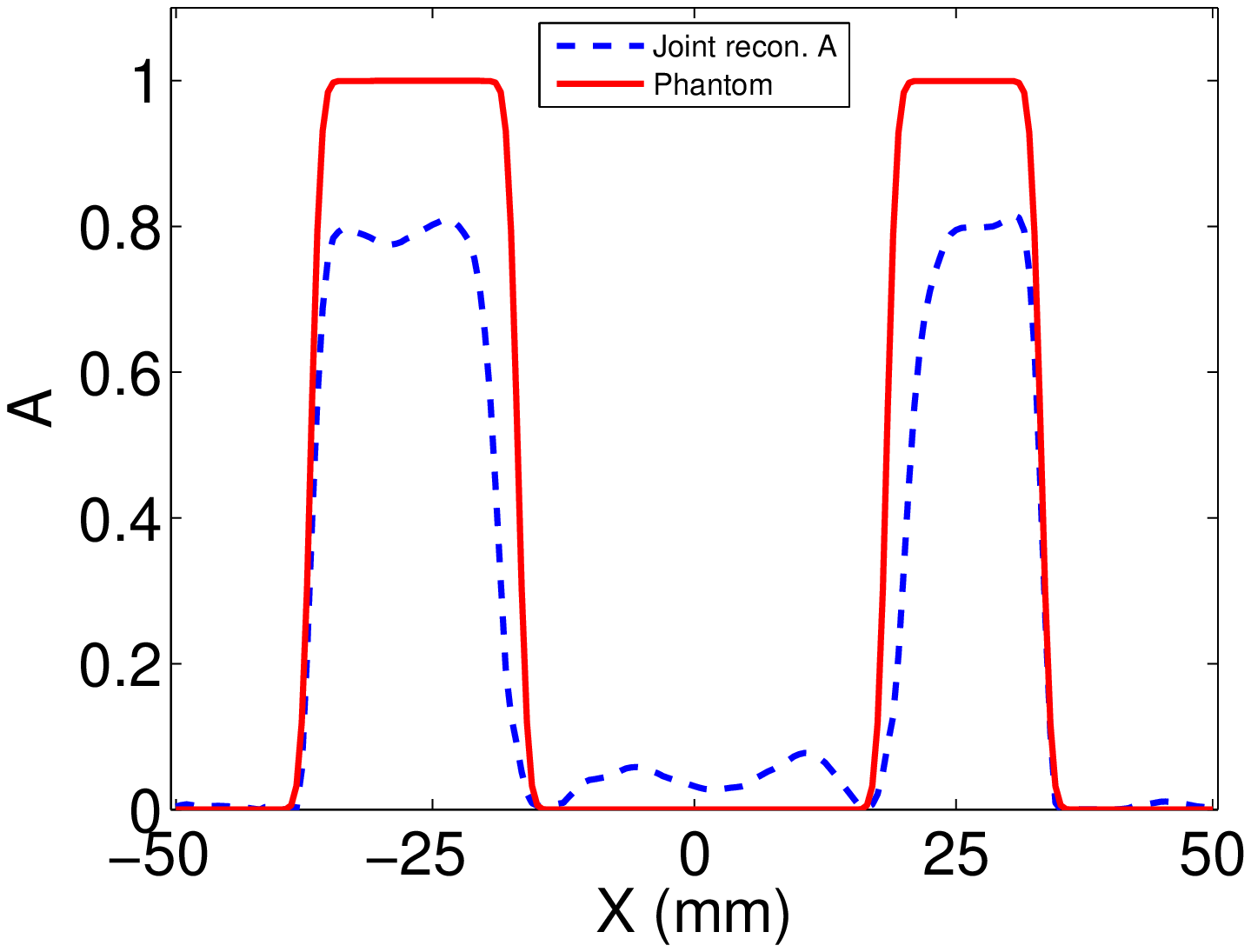}}}\\
\vskip -0.1cm
  \subfigure[]{\resizebox{1.5in}{!}{\includegraphics{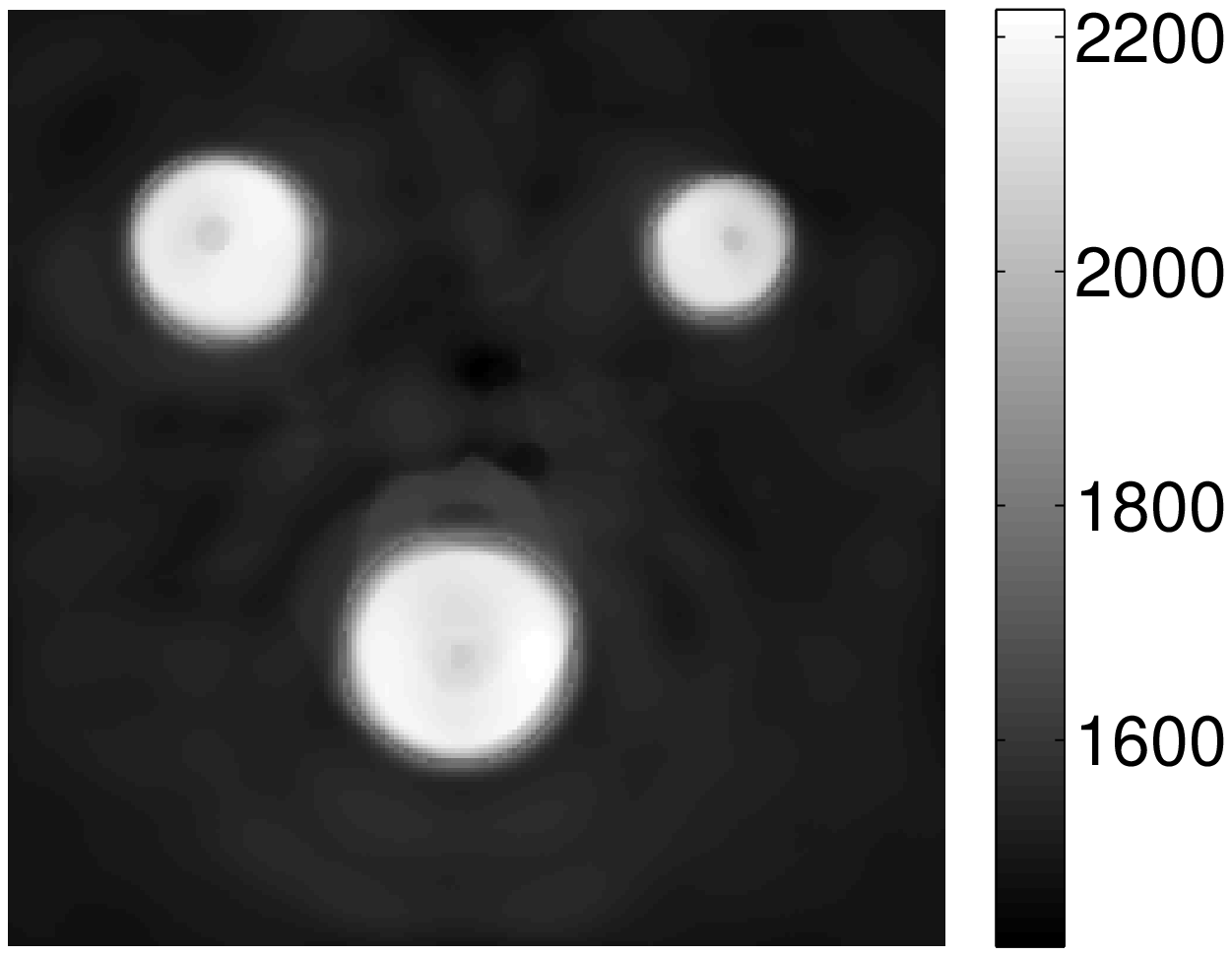}}}
  \subfigure[]{\resizebox{1.5in}{!}{\includegraphics{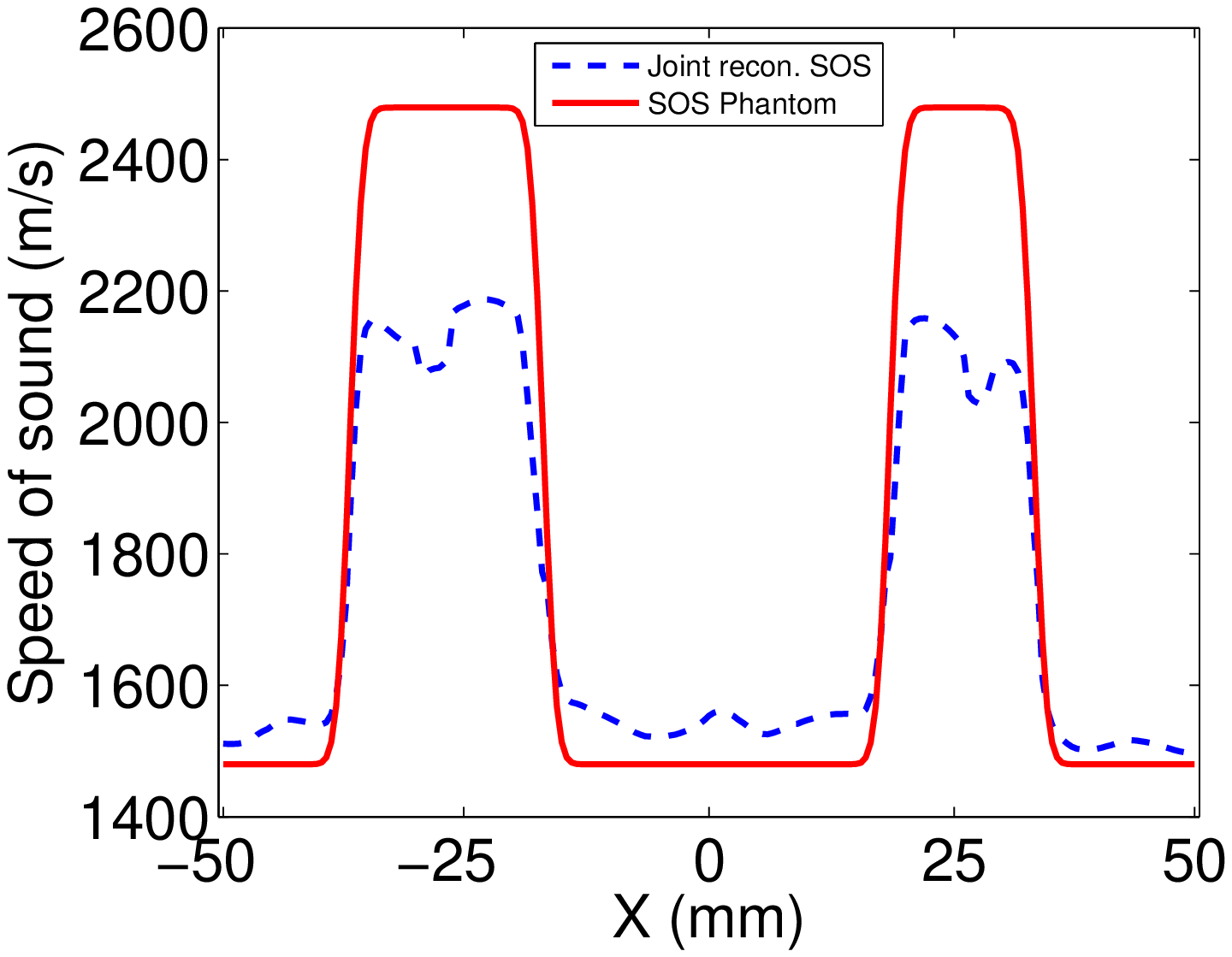}}}\\
\vskip -0.1cm
  \subfigure[]{\resizebox{1.5in}{!}{\includegraphics{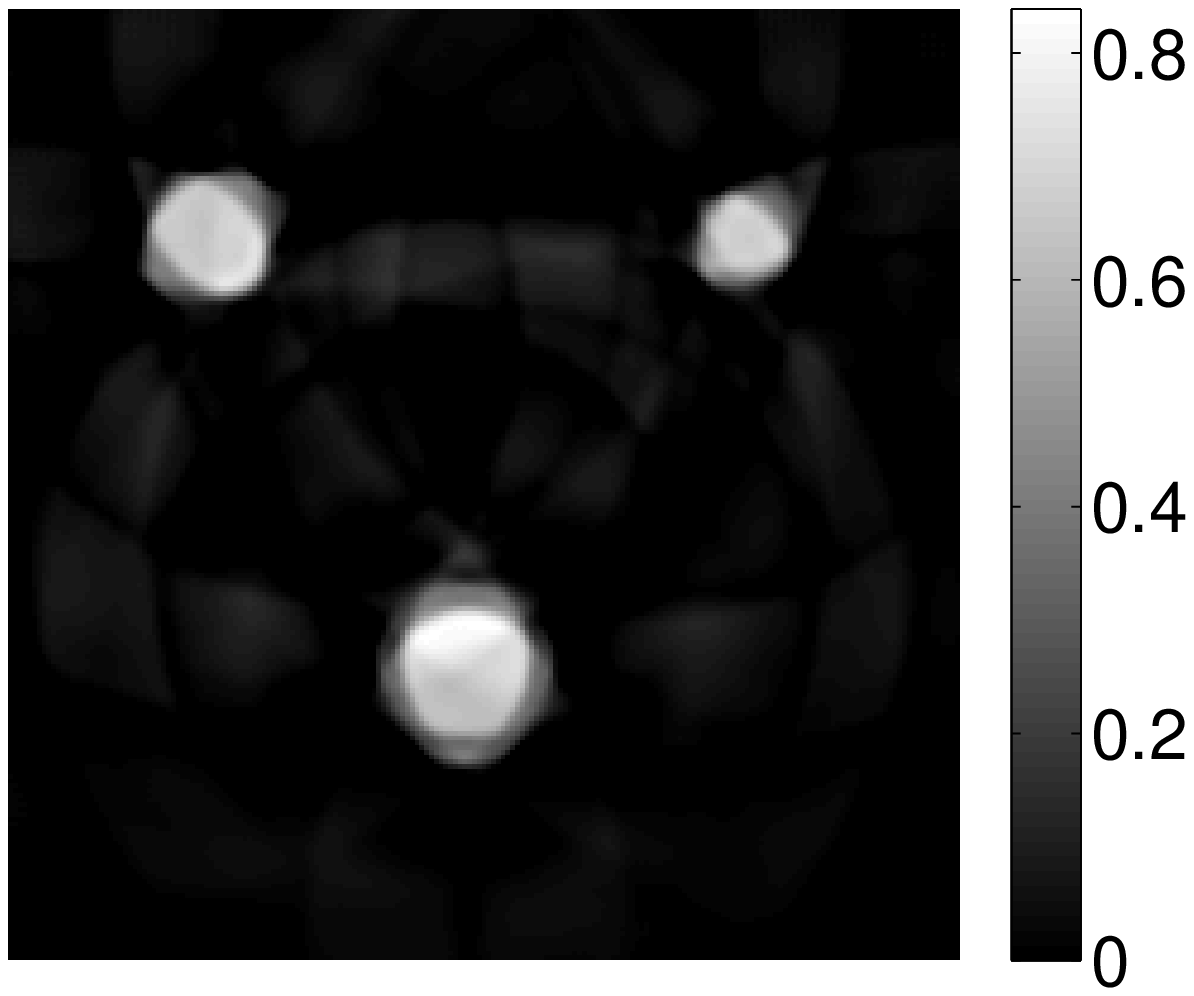}}}
  \subfigure[]{\resizebox{1.5in}{!}{\includegraphics{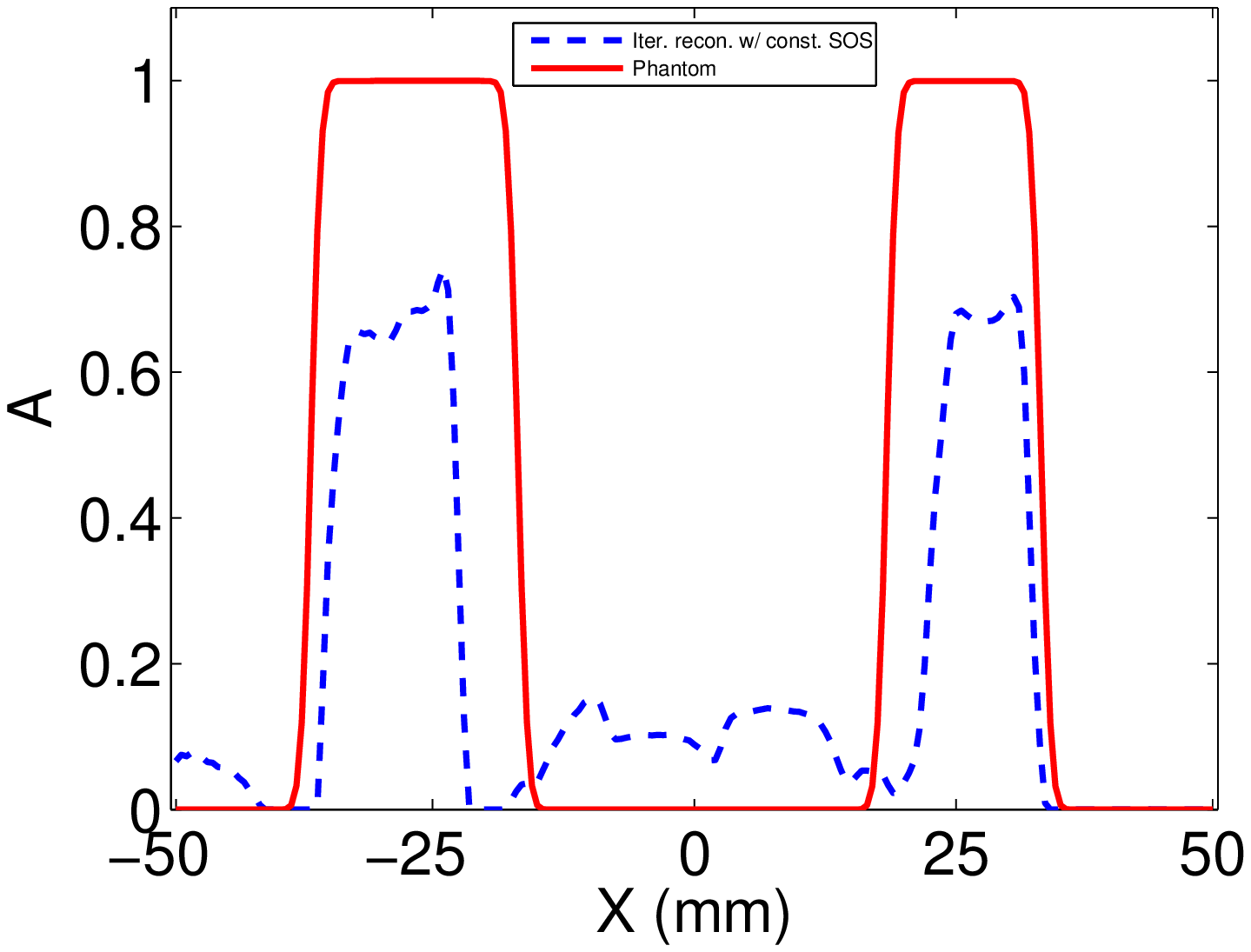}}}
\caption{\label{fig:jr_me}
Images obtained via JR from noisy PA measurements in the presence of model error:
The top and middle rows display the  reconstructed  estimates of
$\mathbf A$ and $\mathbf c$, respectively, along with the corresponding
image profiles. The bottom
row displays an estimate of $\mathbf A$ reconstructed by a PACT
reconstruction method that employed  a constant SOS that was manually tuned to minimize RMSE.
}
\end{figure}

From these data, JR of $\mathbf A$ and $\mathbf c$ was performed
 with regularization 
parameters $\lambda_A=10^{-2}$ and 
$\lambda_c=10^{-1}$. The JR results are
displayed in the top and middle rows of
Fig. \ref{fig:jr_me}. These results suggest
that, even with a sufficient $\mathbf A$ that satisfies
the support and k-space conjectures,
accurate JR may not be feasible in practice
due to its instability  unless
model errors are small. However, the jointly 
reconstructed $\mathbf A$ has smaller 
RMSE = 0.12 compared to the iterative 
result (the bottom row of Fig. \ref{fig:jr_me}) 
that was reconstructed with constant SOS 
of 1600 m/s and has RSME = 0.22. This 
shows that, even though accurate JR may not be 
feasible in practice, the JR method provides
the opportunity to improve the accuracy of 
the reconstructed $\mathbf A$ 
as compared to the use of a 
PACT reconstruction method that assumes  a constant SOS \cite{zhu2014active}.

\section{Conclusion and discussion}
\label{sect:summary}

Because variations in the  SOS distribution
induce aberrations in the measured PA wavefields,
certain information regarding an object's SOS distribution
is encoded in the PACT measurement data.
As such, several investigators have proposed a JR problem
in which the SOS distribution is concurrently estimated along with the sought-after
absorbed optical energy density.
The purpose of this work was to contribute to a broader understanding of the extent to which this problem
  can be accurately and reliably solved under realistic conditions.
This was accomplished by conducting a series of numerical experiments that elucidated some
important numerical properties of the JR problem.
%This was accomplished by use of a novel
%optimization-based reconstruction algorithm.
%Additionally, heuristic analytic insights into the reconstruction problem were also
%provided, which were referred to as support and k-space conjectures.
These studies demonstrated the numerical instability of the
JR problem.
%in practice due to the model errors and deficiency
%of $\mathbf A$. However, we also showed that
%the accuracy of the reconstructed $\mathbf A$ 
%can still be improved by the JR method compared 
%to the image reconstructed with a constant SOS.

\if 0
Note that, although our observations and
conclusions were drawn from the numerical 
results that were obtained by use of an 
alternating optimization algorithm to 
solve a wave-equation-based optimization
problem, their validity may be independent
of the approaches employed to solve the 
JR problem.
\fi

The presented findings are consistent with and corroborate previous theoretical studies of
 the JR problem.  Namely,  Stefanov  \emph{et al} 
proved the instability of the linearized
JR problem, which suggested instability 
of the more general JR problem as well \cite{StefanovArxiv12}. 
%This result corroborates our conclusion
%of the instability of the JR problem. 
In \cite{StefanovArxiv11},  a 
condition for unique reconstruction of 
$c(\mathbf{r})$ given $A(\mathbf{r})$ was provided, 
which is consistent with our support condition 
(see Theorem 3.3 therein). In previous 
work regarding the development of JR 
algorithms, Chen \emph{et al} proposed 
a similar optimization-based approach to 
JR \cite{TRadjoint}. They solved the 
optimization problem by use of an 
optimization algorithm called 
the TR adjoint method. Although their 
algorithm was different from ours, they 
obtained similar results; Accurate JR 
images were not produced when $\mathbf{A}$ 
is deficient, but the jointly reconstructed 
$\mathbf{A}$ could be more accurate than the 
one reconstructed by use of the TR method 
with a constant SOS.

Similar and results can be found in the 
works by Jiang \emph{et al} 
\cite{JiangJOSA06,YuanOE06}, in which the
authors proposed an optimization approach 
to JR that was based on the Helmholtz 
equation instead of the wave equation. 
By use of that method, the authors observed
that the accuracy of JR results was 
affected by the temporal frequency band 
employed in the reconstruction. Specifically,
the frequency ranges covering lower
frequencies gave more accurate JR results
than higher frequencies. This observation
is implicitly contained in our heuristic
k-space conjecture.
%where only low-pass filtered $\mathbf A$ is considered.
 Their 
observations and our k-space  conjecture
can be understood by noting that band-pass 
or high-pass filtered $\mathbf A$ are not 
physical because the non-negativity of 
$\mathbf A$ does not hold in those cases. 
Both results showed that the accuracy of
JR is impacted by the spatial spectrum of
$\mathbf A$.
When $\mathbf A$ and $\mathbf c$  possess sharp boundaries and the same structures 
(i.e., are simply scaled versions of each other),
 the authors also showed that  biased estimates
 of $\mathbf A$ and $\mathbf c$ 
could be jointly reconstructed by incorporating 
Marquardt and Tikhonov regularizations into
the reconstruction method.
Note that in this case,  $\mathbf A$ and $\mathbf c$ satisfied our
support and k-space conjectures.
 By use of regularization,
the authors showed that their algorithm was
insensitive to random noise in the measurement,
which is congruous with our observations.
Although the reconstructed images were 
biased, the authors 
showed that the jointly reconstructed 
$\mathbf A$ was more accurate than the 
image reconstructed with a homogeneous 
SOS, which is consistent with our results. 
In addition, they also observed that the
jointly reconstructed $\mathbf A$ was more
accurate than the jointly reconstructed 
$\mathbf c$, which, again, indicated the
inverse problem of reconstructing $\mathbf c$
is more unstable compared to the reconstruction
of $\mathbf A$.

%However, in the works mentioned above, 
%the authors only showed the JR results
%produced by  their proposed methods; 
However, 
none of these works reported a systematic
investigation of the numerical properties of the JR problem
or provided broad insights that allow one to predict when accurate JR may be possible.
In particular, the impact of the spatial support and spatial
frequency content of $A(\mathbf r)$ relative to that of $c(\mathbf r)$
was not explored.
 In the current study, we have demonstrated that, 
even if the measurement data are perfect,
accurate JR may not be achievable.
%$\mathbf A$ is deficient.
% Furthermore, 
%those works did not investigate the numerical 
%instability of the JR problem and its 
%implication of the feasibility of accurate 
%JR in practice. In this work, we demonstrated 
%the numerical instability of the JR problem, 
%and systematically studied the practical 
%limitations of JR due to its instability 
%and inevitable model errors.

%There remain several important topics for future investigation.
%In addition 
%to computer simulations, the proposed 
%method can be further evaluated through 
%experimental studies,

The investigation of the JR problem
by use of experimental data remains a topic
for future investigation. However,
based on the presented studies, the task of
performing accurate JR under experimental conditions
is likely to be highly challenging and will require
accurate modeling of the imaging operator.
%judicious choice of the penalty functions to regularize
%the solution.
This will generally require the use of the 3D wave equation
instead of 2D wave equation. A line
search is inevitable in any nonlinear 
optimization algorithm that is employed
to solve Sub-Problem \#2, which will create
a very large computational burden in the 3D case.
Additionally, in this 
study, 
a lossless fluid  medium was assumed by the reconstruction method.
%media where the mass density is assumed 
%to be homogenous.
 However, in certain
applications, density variations and/or
acoustic absorption may not be negligible
\cite{HuangchaoJBObrain,HuangchaoJBOatten}.
These assumptions can, in principle, be relaxed when formulating
the JR problem.

%In some cases, for example transcranial
%PACT brain imaging of primates, shear 
%wave mode conversion needs to be taken 
%into account as well.
% The development of 
%the JR method, which is based on the wave 
%equation that describes density variation, 
%acoustic absorption and/or shear wave 
%mode conversion, is another important topic
%for future studies.

 Finally, due to its instability, it will likely be beneficial to 
incorporate additional information into the JR problem.
 One possibility is to augment the PACT measurement data with
a small number of ultrasound computed tomography (USCT) measurements.
The investigation of the JR problem 
by combining PACT and USCT measurements is underway \cite{matthews2015synergistic}.

\section*{acknowledgments}

This work was supported in part by NIH awards CA1744601 and EB01696301.
The authors thank Dr.\ Stephen Norton for insightful discussions regarding
the computation of the  Fr\'echet derivative and Dr. Konstantin Maslov
for informative discussions pertaining to transducer modeling.  The authors
also thank Professor Gunther Uhlmann for numerous discussions
and guidance regarding the mathematical properties of the 
JR problem.

\section*{Appendix: calculating the gradient of \eqref{eq:cost_c}}

The gradient of the first term in
Eq.\ \eqref{eq:cost_c} can be calculated by
discretizing the the Fr\'echet derivative
\eqref{eq:fd}
\begin{equation} 
\label{eq:fdd}
\begin{split}
\frac{\partial \| \mathbf{H}(\mathbf c) 
\mathbf{A} - \hat{\mathbf p}\|^2}
{\partial \mathbf c} 
= -4 \mathbf{C}^{-3} \circ
\big \{ \sum_{l=1}^{L-2} 
\frac{\mathbf{p}_{l+1} - \mathbf{p}_{l-1}}{2} \\
\circ
\frac{\mathbf{q}_{l+1} - \mathbf{q}_{l-1}}{2} 
 +
(\mathbf{p}_{1} - \mathbf{p}_{0})
\circ
(\mathbf{q}_{1} - \mathbf{q}_{0})\\
+
(\mathbf{p}_{L-1} - \mathbf{p}_{L-2})
\circ
(\mathbf{q}_{L-1} - \mathbf{q}_{L-2})
\big \}
\end{split}
\end{equation}
where $\circ$ denotes Hadamard product,
$\mathbf{C}^{-3}$ is defined as
\begin{equation}
\label{eq:c3}
\mathbf{C}^{-3} \equiv 
[c(\mathbf{r}_1)^{-3},\cdots, c(\mathbf{r}_N)^{-3}]^{\rm T},
\end{equation}
$\mathbf{p}_l$ and $\mathbf{q}_l$ ($l=0,\cdots,L-1$)
are defined as
\begin{equation}
\label{eq:p}
\mathbf{p}_l \equiv [p(\mathbf{r}_1,l\Delta t), \cdots, p(\mathbf{r}_N,l\Delta t)]^{\rm T},
\end{equation}
and
\begin{equation}
\label{eq:q}
\mathbf{q}_l \equiv [q(\mathbf{r}_1,l\Delta t), \cdots, q(\mathbf{r}_N,l\Delta t)]^{\rm T},
\end{equation}
representing the PA wavefield and the adjoint wavefield
sampled at the 3D Cartesian grid vertices $\mathbf{r}_n$
($n=1,\cdots,N$) and at time $t=l \Delta t$, respectively.

If TV-penalty is adopted, the gradient of the second term in
\eqref{eq:cost_c} is given by \cite{SidkyArxiv09}
\begin{equation}
\label{eq:tvd0}
\frac{\partial \lambda_c |\mathbf{c}|_{\text{TV}}}
{\partial \mathbf c} = \lambda_c
[\dot c_1, \cdots, \dot c_n, \cdots, \dot c_N]^{\rm T},
\end{equation}
and
\begin{equation}
\label{eq:tvd1}
\begin{split}
&\dot c_n \equiv 
 { (3 [\mathbf c]_n - 
\sum_{i=1}^3
[\mathbf c]_{n_i^-} % - [\mathbf c]_{n_2^-} - [\mathbf c]_{n_3^-}
) }
{ \{\epsilon + \sum_{i=1}^3 ([\mathbf c]_n - [\mathbf c]_{n_i^-})^2 
\}^{-\frac{1}{2}} }  \\
&- \sum_{i=1}^3{( [\mathbf c]_{n_i^+} - [\mathbf c]_{n}) }
{\{ \epsilon + 
\sum_{j=1}^3 ([\mathbf c]_{n_i^+} - [\mathbf c]_{{(n_i^+)}_j^-})^2 
\}^{-\frac{1}{2}} } \\
% &-  {( [\mathbf c]_{n_2^+} - [\mathbf c]_{n} )}
% { \{\epsilon + 
% \sum_{i=1}^3
% ([\mathbf c]_{n_2^+} - [\mathbf c]_{{(n_2^+)}_i^-})^2 
% \}^{-\frac{1}{2}} } \\
% &-  {( [\mathbf c]_{n_3^+} - [\mathbf c]_{n} )}
% { \{\epsilon + 
% \sum_{i=1}^3 ([\mathbf c]_{n_3^+} - [\mathbf c]_{{(n_3^+)}_i^-})^2 
% \}^{-\frac{1}{2}} },
\end{split}
\end{equation}
where $\epsilon$ is a small positive number
to prevent the denominators being zeros,
and $[\mathbf c]_n$ denotes the $n$-th
grid node of $\mathbf c$, and $[\mathbf c]_{n_i^-}$
and $[\mathbf c]_{n_i^+}$
are neighboring nodes before and after the $n$-th node
along the $i$-th dimension ($i=1,2,3$), respectively.
Likewise, $[\mathbf c]_{{(n_i^+)}_j^-}$ denotes
the neighboring node that is after the $n$-th node
along the $i$-th dimension and before
the $n$-th node along the $j$-th dimension.

The gradient of the objective function
in \eqref{eq:cost_c} is then given by
the sum of Eqs.\  \eqref{eq:fdd} and \eqref{eq:tvd0}.

\bibliographystyle{IEEEtran}

\bibliography{reference}
%%%%%%%%%%%%%%%%%%%%%%%%%%

\end{document}